\documentclass[journal=jpclcd,manuscript=letter]{achemso}
\usepackage[version=3]{mhchem} 
\usepackage[T1]{fontenc}       
\usepackage{amsmath}
\usepackage{braket}
\usepackage{booktabs}
\usepackage{siunitx}
\usepackage{ragged2e}
\usepackage{array}
\usepackage{color}
\usepackage{multirow}
\usepackage{xr}
\usepackage{subcaption}
\usepackage{lscape}

\newcolumntype{?}[1]{!{\vrule width #1}}

\newcommand{\onlinecite}[1]{\hspace{-1 ex} \nocite{#1}\citenum{#1}} 
\SectionNumbersOn
\SectionsOn

\title{Regio-Connectivity and Torsional Angle Effects on Singlet Fission and SOCT-ISC in Aza-BODIPY Dimers}

\author{Sophiya Goyal}
\affiliation{School of Chemical Sciences and Pharmacy,
	Central University of Rajasthan, Ajmer, India}

\author{S. Rajagopala Reddy}
\email{rajagopala.seelam@curaj.ac.in}
\affiliation{School of Chemical Sciences and Pharmacy,
	Central University of Rajasthan, Ajmer, India}
	
\begin{document}
\begin{abstract}
	Aza-BODIPY dimers represent promising molecular systems for efficient triplet-state generation through either intramolecular-singlet fission (iSF) or spin–orbit charge transfer intersystem crossing (SOCT-ISC). In this work, we investigate the role of molecular geometry in governing these mechanisms across four regioisomeric aza-BODIPY dimers (D[1,1], D[1,3], D[3,3], and D[2,2]) using multireference quantum-chemical calculations. Ground- and excited-state properties were analyzed at the MP2 and SA-XMCQDPT levels of theory, while diabatic couplings and spin–orbit matrix elements were evaluated to estimate iSF and SOCT-ISC rate constants, respectively. Our results reveal that triplet formation is strongly governed by the torsional angle ($\Phi$) between monomer units, with regio-connectivity exerting a secondary influence. Dimers D[1,1] and D[1,3] exhibit favorable iSF energetics and coupling magnitudes, whereas D[2,2] displays low iSF rate constant ($k_\text{SF}$) but enhanced SOCT-ISC activity. The D[3,3] dimer shows exothermic multiexciton formation but reduced iSF efficiency due to destructive coupling interactions. The dominant ISC channel proceeds through the $\textrm{S}_1$–$\textrm{T}_3$  transition with large spin–orbit coupling and a small energy gap. These findings provide critical mechanistic insights into geometry-dependent triplet generation in aza-BODIPY dimers.
\end{abstract}
\section{Introduction}

Triplet photosensitizers (PS) are molecules capable of efficiently undergoing intersystem crossing (ISC) to form long-lived triplet states upon photoexcitation. These triplet states play a pivotal role in a wide array of applications, including photocatalysis\cite{Welin2017,Kim2018,Strieth-Kalthoff2018}, photovoltaics\cite{Bessette,Kolemen2011,Miao2022}, bioimaging\cite{Fernandez-Moreira2010,Zhao2015}, photodynamic therapy (PDT),\cite{Zhao2013,Li2017a,Majumdar2014} and photothermal therapy (PTT)\cite{Guo2022}. Traditionally, triplet state generation relies on ISC promoted by the incorporation of heavy atoms, typically transition metals, that strengthen spin–orbit coupling (SOC) and enable effective triplet energy transfer from the organometallic sensitizer. However, the use of heavy atoms often raises concerns regarding toxicity, cost, and photostability. To overcome these limitations, heavy-atom-free strategies for promoting ISC have gained considerable attention. One of the most promising among them is spin–orbit charge transfer intersystem crossing (SOCT-ISC), which leverages twisted donor–acceptor geometries to enable efficient ISC in purely organic molecules. In these systems, orbital angular momentum changes during the charge-transfer (CT) transition can mimic the role of SOC, facilitating spin-flip transitions without requiring heavy atoms.\cite{Bassan2021,Rana2022}

Within this context, BODIPY derivatives have attracted considerable attention as promising heavy-atom-free systems capable of generating triplet states.\cite{Adarsh2012,Zhao2015,Ge2016,Chen2019} BODIPY\cite{Treibs1968} (4,4'-difluoro-4-bora-3a,4a-diaza-s-indacene) and its aza-analog Aza-BODIPY\cite{Killoran2002} (4,4'-difluoro-4-bora-3a,4a,8-triaza-s-indacene), obtained by nitrogen substitution at the meso position of the BODIPY core, exhibit intense visible absorption and high fluorescence efficiency,\cite{Ziessel2007} excellent photo- and thermal stability,\cite{Di2023} and tunable photophysical properties.\cite{Loudet2007} These attributes make them highly suitable for designing next-generation triplet sensitizers.

The mechanism underlying triplet formation in BODIPY-based dimers has been the subject of considerable debate. Montero et al. proposed that orthogonal ($\Phi\simeq$ 90°) BODIPY dimers undergo singlet fission (SF), based on transient absorption spectroscopy (TAS) experiments and positive $\Delta_\textrm{SF}=E(\textrm{S}_1)-2E(\textrm{T}_1)$ values calculated using TDDFT.\cite{Montero2018} However, Michl and co-workers revisited these systems using a multireference MS-CASPT2(16,16) approach and arrived at a different conclusion.\cite{Michl2018} Unlike TDDFT, which is a single-reference method, MS-CASPT2 can accurately describe systems with strong multiconfigurational character. Their group demonstrated that the BODIPY ground-state wave function is inherently multiconfigurational, meaning that a multireference treatment is essential for a reliable description of its electronic structure. Their calculations revealed that $\Delta_\textrm{SF}$ is in fact unfavorable, thereby ruling out iSF in orthogonal BODIPY dimers and instead supporting SOCT-ISC as the dominant triplet formation pathway. Subsequent experimental and theoretical investigations, including solvent polarity studies, further reinforced this interpretation, consistently favoring SOCT-ISC over iSF in orthogonal BODIPY systems.\cite{Kandrashkin2019,Casanova2022}

Most recently, Kandrashkin et al. (2024) employed time-resolved electron paramagnetic resonance (TREPR) techniques examine BODIPY dimers with non-orthogonal ($\Phi\neq$ 90$^\circ$) geometries, focusing on how solvent polarity influences the fluorescence quantum yield ($\Phi_{f}$) and the efficiency of singlet oxygen production ($\Phi_{\Delta}$). They reported that the sum of  $\Phi_{f}$+$\Phi_{\Delta}$ lies in the range $\sim$0.16–0.6, indicating that non-radiative pathways also contribute to deactivation in these systems.\cite{Kandrashkin2024} Collectively, these findings highlight that torsional geometry critically governs the triplet generation pathway: orthogonal systems favor SOCT-ISC, while non-orthogonal arrangements may enable multiexciton (ME) formation and possible iSF.

 In our recent study, we systematically examined a series of covalently linked BODIPY dimers to assess their suitability for SF.\cite{Goyal2024} Although iSF rate constants ($k_\text{SF}$) were calculated for these systems, the values were consistently low, pointing to intrinsic limitations in their ability to support efficient SF. To gain deeper mechanistic insight, we investigated the effect of inter-monomer torsional angles (0°–180°) in two representative systems, B[3,3] and B[2,2]. Our analysis revealed that B[3,3] is capable of populating a multiexcitonic (ME) state under non-orthogonal geometries. This ME state could, in principle, serve as an intermediate for SF-mediated triplet generation; however, in the absence of an energetically accessible quintet state, we concluded that the ME state instead decays predominantly through non-radiative relaxation to the ground state. In orthogonal geometries, no ME formation was observed, in line with previous findings that orthogonality disfavors ME pathways and instead promotes SOCT-ISC. For B[2,2], no ME formation was detected in either orthogonal or non-orthogonal conformations, reinforcing the sensitivity of iSF efficiency to connectivity and torsional geometry. Taken together, these results highlight that the substantial energy gap between the locally excited (LE) and the ME states represents a key bottleneck that suppresses efficient iSF in BODIPY dimers. To overcome this limitation, we introduced targeted substitutions on both BODIPY and Aza-BODIPY monomers with the aim of tuning their electronic structure and narrowing the energy spacing between the LE and ME states. Within this series, the Aza-BODIPY derivatives consistently exhibited smaller $\Delta_\textrm{SF}$ values and enhanced diradical character compared to their BODIPY counterparts. Importantly, these features correlated with higher  $k_\text{SF}$ values, suggesting the possibility of a singlet-fission-like process in the aza-BODIPY based systems.
 
 A review of the reported literature on aza-BODIPY dyes shows a pronounced decrease in fluorescence quantum yield ($\Phi_f \approx 0.01$) upon dimer formation. This substantial fluorescence quenching indicates the presence of efficient non-radiative relaxation channels and motivates the exploration of iSF as a possible excited-state deactivation pathway. For instance, the aza-BODIPY dimer connected through a direct C–C linkage at the 2,2-position, D[2,2] displays an extremely low $\Phi_f$ of less than 0.01\cite{Nepomnyashchii2011} and 0.008\cite{Tian2021}, in contrast to the much higher values observed for its corresponding monomer.\cite{Liu2018} Likewise, an ethylene-bridged dimer of D[2,2] shows a markedly reduced fluorescence yield of 0.006 compared with 0.161 for the monomeric unit.\cite{Guo2022}
 
These experimental observations, combined with our theoretical predictions of favorable electronic characteristics in aza-BODIPY systems, motivated a detailed investigation into the mechanism of excited-state relaxation and triplet formation in such dimers. Although triplet state generation has been widely reported in BODIPY-based systems, studies exploring the mechanism of triplet state generation in Aza-BODIPY systems, particularly in covalently linked dimeric structures, remain limited. In our recent work, we investigated the D[1,3] Aza-BODIPY dimer, where monomer units are connected via the 1 and 3 positions.\cite{Goyal2026} Introducing asymmetry in the electronic distribution is a well-established approach for modulating photophysical behavior.\cite{Filatov2025} In this context, the aza-BODIPY dimer D[1,3], featuring an asymmetric linkage between the monomer units and a relatively short interchromophoric separation, emerges as a promising scaffold for facilitating triplet state formation. To probe the mechanism of triplet generation, we performed quantum nuclear dynamics simulations on this dimer. The adiabatic states obtained from the calculations exhibited extensive mixing among the underlying diabatic configurations, necessitating a description within a super-exchange-like framework. Within this representation, configurations associated with multiexcitonic (ME) character contributed substantially to several adiabatic states, such as $^\textrm{1}\left(\textrm{M}_\textrm{1}\right)$, $^\textrm{1}\left(\textrm{M}_\textrm{2}\right)$ and $^\textrm{1}\left(\textrm{M}_\textrm{5}\right)$, though none of them were dominated by a purely ME configuration. Quantum dynamical population analyses revealed an ultrafast redistribution of populations across the electronic manifold, leading to the rapid occupation of the lowest-lying singlet state, $^\textrm{1}\left(\textrm{M}_\textrm{1}\right)$. This state exhibits a mixed electronic character, comprising roughly 30\% charge-transfer (CT), 20\% multiexcitonic (ME) and 20\% locally excited (LE) contributions. The population evolution in both delocalized and super-exchange bases indicated transient but significant ME state involvement. Furthermore, SOC calculations and the evaluated ISC rate constants confirm that the $^\textrm{1}\left(\textrm{M}_\textrm{1}\right)$ efficiently channels population toward the triplet manifold via the SOCT-ISC mechanism.

In the present study, we extend our investigation to four Aza-BODIPY dimers, D[1,1], D[1,3], D[3,3], and D[2,2] to systematically examine the effects of monomer linkage positions and inter-monomer torsion angles on triplet state formation. Given that orthogonal arrangements are known to enhance SOCT-ISC efficiency, our aim is to understand how variations in molecular geometry and electronic coupling influence the competition between SOCT-ISC and iSF in these systems. Through this analysis, we seek to provide a deeper understanding of the photophysical mechanisms that govern triplet generation in Aza-BODIPY-based systems, ultimately guiding the rational design of heavy-atom-free triplet photosensitizers.

\begin{figure}
	\centering
	\includegraphics[width=0.5\textwidth]{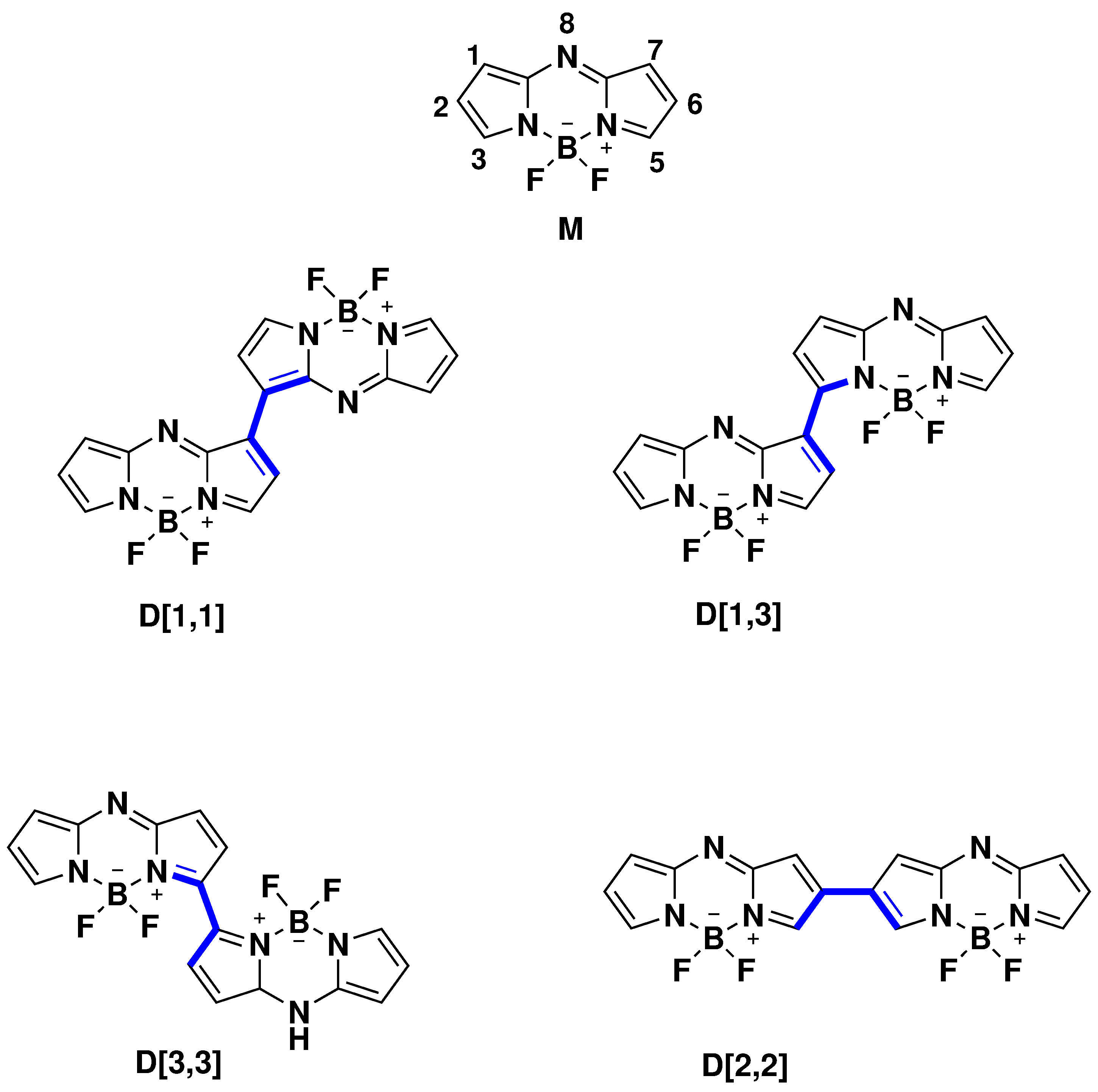}
	\caption{Structure of the aza-BODIPY monomer (M) with atom numbering following IUPAC conventions.  The dimer systems D[x,y] are generated by linking the x$^\textrm{th}$ atom of one monomer to the y$^\textrm{th}$ atom of another. In this study, the investigated dimers include D[1,1], D[1,3], D[3,3], and D[2,2]. The torsional angle, highlighted in blue, is 0.0°, 14.9°, 23.0°, and 0.0° for each dimer, respectively.}
	\label{fig:aza} 
\end{figure}

\section{Methods}
\label{sec:methods}
BODIPY dyes, known for their stable diradical character, consist of two pyrrole units joined through a BF$_2$ group that connects their nitrogen atoms. This character arises from the unique structure where a N-B–N framework replaces the typical C-C–C framework.\cite{Michl2018,Casanova2021} Aza-BODIPY, a derivative of BODIPY, has a nitrogen atom at the meso position in place of a CH group, further diversifying its photophysical properties. These dyes have garnered attention due to their unique chemical structure and identified to support SF.\cite{Goyal2024} Due to the radical nature of ground electronic state, the electronic states of BODIPYs and aza-BODIPYs require multireference methods.\cite{Brown2016,Casanova2021,Michl2018,Goyal2024} We have employed correlation consistent methods through out this work to correctly describe the multireference nature of the electronic states.

We optimized the geometries of aza-BODIPY dimers in their ground-level singlet state using Second-order Møller–Plesset perturbation (MP2) theory\cite{Moller1934} with Dunning's correlation-consistent polarized valence double-zeta (cc-pVDZ) basis set.\cite{Jr1989} All the optimized geometries represent a local minimum on the account of all positive frequencies. 15 lowest lying singlet and triplet electronic states are calculated by using SA-XMCQDPT (state-averaged extended multiconfigurational quasi-degenerate perturbation theory) method\cite{Granovsky2011},  and cc-pVDZ\cite{Dunning2013} basis set. In addition to the singlet and triplet states, we have also calculated 3 lowest lying quintet states at the same level of theory. A shift of 0.02 a.u. was applied to mitigate intruder state effects on energy calculations. Exploratory calculations in terms of active space and basis set are performed to find a compromise between accuracy and computational load. An active space (8,8) and cc-pVDZ basis set found to be optimal for the current problem (See SI for details)

The theoretical analysis of iSF follows a two-part framework. In the first part, a spin-independent Hamiltonian, which involves pertinent diabatic state configurations (See Sec.\ref{sec:diab}) is constructed. By including the vibrational motions in the spin-independent Hamiltonian, the formation of ME state is successfully discussed.\cite{Reddy2018,Reddy2019,Reddy2022,Zeng2016,Beljonne2015,Lan2016,Granucci2019} Then, in the second part, a spin-dependent Hamiltonian featuring singlet $^1\left(\textrm{T}_\textrm{1}\textrm{T}_\textrm{1}\right)_{0}$, three triplet $^3\left(\textrm{T}_\textrm{1}\textrm{T}_\textrm{1}\right)_{0,\pm 1}$, and five quintet  $^5\left(\textrm{T}_\textrm{1}\textrm{T}_\textrm{1}\right)_{0,\pm 1,\pm 2}$ sublevels devised and the diffusion of ME state to two triplet states is discussed \cite{Hartzler2018,Matsuda2020,Taffet2020,Abraham2021,Kathir2024} The contrasting photophysical properties of aza-BODIPY dimers can be explained by examining the static component of the spin-free Hamiltonian.

To gain deeper insights into the iSF mechanism, the coupling of diabatic electronic states among the relevant states are evaluated \cite{Zirzlmeier2016,Basel2017,Reddy2018,Reddy2019} employing Truhlar and Nakamura’s fourfold diabatization approach\cite{Nakamura2001,Nakamura2002}. We have calculated 8 diabatic states of relevance to iSF namely the ground-state  $^1\left(\textrm{S}_\textrm{0} \textrm{S}_\textrm{0}\right)$, a ME state  $^1\left(\textrm{T}_\textrm{1} \textrm{T}_\textrm{1}\right)$, two states corresponding to local excitation (LE) in which one monomer is in its ground state and the other monomer in its first excited state ($^1\left(\textrm{S}_\textrm{1} \textrm{S}_\textrm{0}\right)$ and  $^1\left(\textrm{S}_\textrm{0} \textrm{S}_\textrm{1}\right)$), two charge-transfer (CT) configurations where an electron from the HOMO of given monomer transfer to the LUMO of its paired monomer ($^{1}(\textrm{C}\textrm{A})$ and $^{1}(\textrm{A}\textrm{C})$) and two double excitation (DE) states $^{1}\left(\textrm{DE}_\textrm{1}\right)$ and $^{1}\left(\textrm{DE}_\textrm{2}\right)$. The effective coupling $V_\text{eff}$ between these diabatic states was computed using the following expression. \cite{Berkelbach2013,Reddy2018a}
\begin{eqnarray}
		V_{eff} \approx \bra{^{1}(\textrm{S}_\textrm{1}\textrm{S}_\textrm{0})}V\ket{^{1}(\textrm{T}_\textrm{1}\textrm{T}_\textrm{1})}
		-2\tfrac{\bra{^{1}(\textrm{S}_\textrm{1}\textrm{S}_\textrm{0})}V\ket{^{1}(\textrm{CA})}\bra{^{1}(\textrm{CA})}V\ket{^{1}(\textrm{T}_\textrm{1}\textrm{T}_\textrm{1})}+\bra{^{1}(\textrm{S}_\textrm{1}\textrm{S}_\textrm{0})}V\ket{^{1}(\textrm{AC})}\bra{^{1}(\textrm{AC})}V\ket{^{1}(\textrm{T}_\textrm{1}\textrm{T}_\textrm{1})}}{[E(^{1}(\textrm{CT})) - E(^{1}(\textrm{T}_\textrm{1}\textrm{T}_\textrm{1}))]+[E(^{1}(\textrm{CT})) - E(^{1}(\textrm{S}_\textrm{1}\textrm{S}_\textrm{0}))]}
		\label{eq:eff_coup}
\end{eqnarray}
\noindent In this equation, $\bra{\Psi}V\ket{\Phi}$ denotes the coupling between the diabatic states $\Psi$ and $\Phi$. The terms $E(^{1}(\textrm{T}_\textrm{1}\textrm{T}_\textrm{1}))$, $E(^{1}(\textrm{S}_\textrm{1}\textrm{S}_\textrm{0}))$ reflect the  energies associated with the diabatic states $^{1}(\textrm{T}_\textrm{1}\textrm{T}_\textrm{1})$  and $^{1}(\textrm{S}_\textrm{1}\textrm{S}_\textrm{0})$ respectively, evaluated at the geometry optimized in the ground state. The first term in the expression corresponds to the direct mechanism, and the second term describes the mediated mechanism. The rate constant ($k_\text{SF}$) for LE to ME state conversion, is calculated using:
\begin{equation}
	k_{SF} = \frac{2 \pi}{\sqrt{4 \pi \hbar^2 \lambda k_B T}}  \vert V_{eff} \vert ^2  \exp{\left ({-\frac{[E({^{1}(\textrm{T}_\textrm{1}\textrm{T}_\textrm{1})})-(E({^{1}(\textrm{S}_\textrm{1}\textrm{S}_\textrm{0})})+\lambda)]^2}{4\lambda k_{B} T}}\right )},
	\label{eq:sf_rate}
\end{equation}
\noindent Here, $\lambda$ represents the reorganization energy.\cite{Nitzan2006} Using $V_\text{eff}$ and assigning $\lambda$ a value of 100 meV allowed the
determination of the rate constant $k_\text{SF}$ in order to generate the ME state within the studied dimers.

 To further examine the SOCT-ISC mechanism, we computed spin–orbit couplings within the Breit–Pauli framework. The intersystem crossing (ISC) rate from a singlet state $\Psi_n^{1}(0)$ to any accessible triplet state $\Psi_m^{3}(0)$ can be expressed as
 
 \begin{eqnarray}
 	k_\textrm{ISC}^n & = & \sum_m k_\textrm{ISC}^{nm} \nonumber \\
 	k_\textrm{ISC}^{nm} & = & \frac{2 \pi}{\hbar} |\braket{\Psi_n^{1}(0)|\hat{H}_{S0}|\Psi_m^{3}(0)}|^2 \times FCWD \label{eq:rates_triplet}
 \end{eqnarray}
 Where $|\braket{\Psi_n^{1}(0)|\hat{H}_{S0}|\Psi_m^{3}(0)}|$ represents the spin–orbit coupling matrix element between the singlet and triplet states. The Franck–Condon weighted density of states (FCWD) is defined as
 \begin{eqnarray} 
 	FCWD = \frac{1}{\sqrt{4 \pi \lambda k_B T}} \sum_{i=0}^{\infty} \exp{(-S)} \frac{S^i}{i!} \exp{\left[\frac{-(\Delta E + i \hbar \omega + \lambda)^2}{4 \lambda K_B T}\right]}
 \end{eqnarray}
 
 Where $\Delta E$ denotes the energy gap between the singlet S$_\textrm{n}$ and T$_\textrm{m}$ states; $\lambda$ is the Marcus reorganization energy, encompassing both intramolecular and intermolecular vibrational contributions; $k_B$ and $T$ are the Boltzmann constant and temperature, respectively. The parameter $\hbar \omega$ is the effective energy of a mode representing the nonclassical high-frequency intramolecular vibrations ($\frac{\hbar \omega}{k_B T} >> 1$); and $S$ is the effective Huang-Rhys factor associated with effective mode.\cite{Samanta2017,Karin2007}
 
\section{Results and Discussion}
In this section, the methodologies outlined in Sec.~\ref{sec:methods} are employed to investigate the photophysical properties of aza-BODIPY dimers. We have established the excited states of aza-BODIPY monomer in our previous studies.\cite{Goyal2024} Further we proceed to discuss the effect of regional isomerism on the excitation energies (singlet, triplet and quintet states) and diabatic electronic couplings of aza-BODIPY dimers (See Fig. \ref{fig:aza}). With the help of the diabatic energies and couplings, we show that ME state generation is slightly an endothermic process.

\subsection{Excited states of aza-BODIPY dimers}
\label{sec:regio}

Based on the singlet and triplet excitation energies of the aza-BODIPY monomer reported in our previous study,\cite{Goyal2024} the thermodynamic driving force for iSF is estimated to be $\Delta_\textrm{SF} \approx 0.4$ eV. This represents a significant improvement over BODIPY systems, for which $\Delta_\textrm{SF}$ is approximately 1.0 eV. Although the process remains slightly endothermic at the monomer level, molecular vibrations and structural modifications introduced through covalent linking of the monomer units may help reduce this energetic barrier.

Previous studies on acene systems such as tetracene and pentacene have shown that the efficiency of iSF is highly sensitive to intermolecular packing and $\pi$--$\pi$ coupling.\cite{Korovina2016,Zirzlmeier2015} In covalently linked dimers, the electronic coupling between excited states can be tuned by the linkage position. To explore this effect in aza-BODIPY systems, we constructed four dimers, D[1,1], D[1,3], D[3,3], and D[2,2], by connecting two monomer units through a single C--C bond (Fig.~\ref{fig:aza}). The optimized torsional angles of D[1,1], D[1,3], D[3,3], and D[2,2] were found to be 0.0$^\circ$, 14.9$^\circ$, 23.0$^\circ$, and 0.0$^\circ$, respectively. For these dimers, singlet, triplet, and quintet states were calculated at the SA15-XMCQDPT(8,8)/cc-pVDZ level of theory. Additional exploratory calculations with respect to active space and basis set effects are provided in the Supporting Information.

Tables \ref{tab:vees-regio} and \ref{tab:vees-regio-ortho} summarize the vertical excitation energies (in eV), oscillator strengths ($f$), dipole moments ($\mu$, in Debye), and corresponding state assignments for the four dimers in their non-orthogonal and orthogonal conformations, respectively. The optimized frontier orbitals of these dimers are largely delocalized over both monomer units. To obtain a chemically intuitive description of the excited states, the electronic configurations were further analyzed using a (4,4) active space constructed from localized orbitals. The resulting state characterizations are illustrated in Figs.~\ref{fig:dimer11-localized}--\ref{fig:dimer22-localized} for the non-orthogonal geometries and in Figs.~\ref{fig:dimer11-ortho-localized}--\ref{fig:dimer22-ortho-localized} for the orthogonal geometries.

Analysis of the seven lowest singlet excited states reveals several characteristic excitation motifs. Two types of local excitations (LE) localized on a single monomer are identified: the HOMO$\rightarrow$LUMO transition (denoted LE$_1$) and the HOMO$-1\rightarrow$LUMO transition (LE$_2$). In addition, two charge-transfer excitations are observed, corresponding to inter-monomer electron transfer, namely HOMO$\rightarrow$LUMO (CT$_1$) and HOMO$-1\rightarrow$LUMO (CT$_2$). Alongside these LE and CT configurations, multiexcitonic (ME) states are also present in the low-lying excited-state manifold. In practice, the adiabatic excited states exhibit substantial mixing among LE, CT, and ME diabatic configurations, as reflected in the diabatic analysis summarized in Tables \ref{tab:diab11}--\ref{tab:diab22}.

The strong configuration interaction among LE, CT, and ME diabatic states gives rise to adiabatic states of highly mixed character. To indicate this mixing, a superscript star ($*$) is used in the state assignments to denote adiabatic states with significant contributions from multiple diabatic configurations. For D[1,1], D[1,3], and D[3,3], the lowest excited singlet state (S$_1$) is optically dark due to its negligible oscillator strength, whereas S$_2$ is optically bright and exhibits dominant CT character. This behavior is consistent with previous observations for non-orthogonal tetracene dimers linked at the 5,5$^{\prime}$ positions, where CT states acquire appreciable oscillator strength due to intermolecular electronic coupling.\cite{Alvertis2019} 

To further assess the iSF potential of these systems, we evaluated the singlet fission driving force ($\Delta_\textrm{SF}$) for all four dimers. The calculated $\Delta_\textrm{SF}$ values for D[1,1], D[1,3], D[3,3], and D[2,2] are 0.33, 0.26, $-$0.15, and 0.75 eV, respectively. Thus, ME state formation is exothermic for D[3,3], slightly endothermic for D[1,1] and D[1,3], and strongly endothermic for D[2,2]. In addition, the feasibility of iSF depends on the energetic relationship between the ME and quintet states.\cite{Kathir2024} In the present systems, the quintet state lies below the ME state for D[1,1], D[1,3], and D[3,3], whereas it lies above the ME state for D[2,2], suggesting that iSF is thermodynamically more plausible in the former three dimers.

Another important factor is the super-exchange nature of the adiabatic states in the non-orthogonal dimers.  In case of D[1,1], the $\textrm{S}_{1}$ contains 22\% , 21\%, 29\%, 27\% and 0\% of $^1\textrm{(CA)}$, $^1\textrm{(AC)}$, $^1(\textrm{S}_{1}\textrm{S}_{0})$ and $^1(\textrm{S}_{0}\textrm{S}_{1})$, $^1(\textrm{T}_{1}\textrm{T}_{1})$
 while in case of D[1,3] the $\textrm{S}_{1}$ contains 48\%, 21\%, 2\%, 13\% and 15\% of $^1\textrm{(CA)}$, 
  $^1(\textrm{S}_{1}\textrm{S}_{0})$, $^1\textrm{(AC)}$, $^1(\textrm{S}_{0}\textrm{S}_{1})$ and $^1(\textrm{T}_{1}\textrm{T}_{1})$, in case of D[3,3] 25\%, 25\%, 26\%, 24\%, 0\% of $^1\textrm{(CA)}$, $^1\textrm{(AC)}$, $^1(\textrm{S}_{1}\textrm{S}_{0})$ and $^1(\textrm{S}_{0}\textrm{S}_{1})$, $^1(\textrm{T}_{1}\textrm{T}_{1})$ and for D[2,2], it contains 38\%, 35\%, 0\%, 14\% and 14\% of  $^1\textrm{(CA)}$, $^1\textrm{(AC)}$, $^1(\textrm{T}_{1}\textrm{T}_{1})$, $^1(\textrm{S}_{1}\textrm{S}_{0})$ and $^1(\textrm{S}_{0}\textrm{S}_{1})$ states. Interestingly the optically bright $\textrm{S}_{2}$ state is a linear combination of 27\%, 27\%, 5\%, 5\% and 34\% of $^1\textrm{(CA)}$, $^1\textrm{(AC)}$, $^1(\textrm{S}_{1}\textrm{S}_{0})$ and $^1(\textrm{S}_{0}\textrm{S}_{1})$, $^1(\textrm{T}_{1}\textrm{T}_{1})$
  while in case of D[1,3] the 4\%, 4\%, 58\%, 12\%, and 22\% of $^1\textrm{(CA)}$, 
  $^1(\textrm{S}_{1}\textrm{S}_{0})$, $^1\textrm{(AC)}$, $^1(\textrm{S}_{0}\textrm{S}_{1})$ and $^1(\textrm{T}_{1}\textrm{T}_{1})$, in case of D[3,3] 34\%, 34\%, 0\%, 0\% and 28\% for $^1\textrm{(CA)}$, $^1\textrm{(AC)}$, $^1(\textrm{S}_{1}\textrm{S}_{0})$ and $^1(\textrm{S}_{0}\textrm{S}_{1})$, $^1(\textrm{T}_{1}\textrm{T}_{1})$ and for D[2,2], it contains 11\%, 12\%, 40\%, 18\% and 18\% for $^1\textrm{(CA)}$, $^1\textrm{(AC)}$, $^1(\textrm{T}_{1}\textrm{T}_{1})$, $^1(\textrm{S}_{1}\textrm{S}_{0})$ and $^1(\textrm{S}_{0}\textrm{S}_{1})$ states.

  Owing to the strong mixing among LE, CT, and ME configurations, it is generally difficult to identify an adiabatic state with dominant ME character.\cite{Goyal2026} However, D[3,3] stands out in this respect: the ME configuration becomes dominant in the S$_7$ state, indicating a more clearly defined multiexcitonic state and thereby suggesting a more favorable iSF pathway. In contrast, for the other dimers, the absence of a clearly dominant ME state within the adiabatic manifold reduces the likelihood of an efficient iSF mechanism. Instead, the spin-orbit charge-transfer intersystem crossing (SOCT-ISC) pathway appears to be the more likely route for triplet-state generation.

Motivated by this observation, we further examined the effect of torsionality between the two monomer units on the excited-state properties by considering orthogonal dimers. In these structures, the mixing among LE, CT, and ME configurations is significantly reduced, leading to more well-defined adiabatic state characters (see Table \ref{tab:eff}). Although the ME state becomes dominant in one of the adiabatic states for the orthogonal dimers, the large energy separation among the LE, CT, and ME diabatic states still limits the possibility of efficient iSF (see Sec.~\ref{sec:diab}).

Therefore, in the present study, we evaluate and compare the rate constants for iSF and SOCT-ISC in order to establish the dominant triplet-generation pathway in these aza-BODIPY dimers. A detailed discussion is provided in the following sections (Secs.~\ref{sec:diab} and \ref{sec:rot}).

\begin{landscape}
	\begin{table}[!ht]
		\caption{Computed vertical excitation energies (VEEs, eV)$^a$, corresponding oscillator strengths ($f$, dimensionless)$^b$, dipole moments ($\mu$, Debye)$^b$, and the nature of the electronic states of non-orthogonal aza-BODIPY dimers D[1,1], D[1,3], D[3,3], and D[2,2].}
		\small
		\begin{tabular}{rrrrrrrrrrrrrrrrrrrr}
			\toprule
			\multirow{2}{*}{State}&\multicolumn{4}{c}{D[1,1]} && \multicolumn{4}{c}{D[1,3]} && \multicolumn{4}{c}{D[3,3]}  && \multicolumn{4}{c}{D[2,2]}\\
			\cmidrule[0.5pt]{2-5} \cmidrule[0.5pt]{7-10} \cmidrule[0.5pt]{12-15} \cmidrule[0.5pt]{17-20}
			& VEEs & $f$ & $\mu$ &char &&  VEEs & $f$ & $\mu$ &char && VEEs & $f$ & $\mu$  &char && VEEs & $f$ & $\mu$ &char \\
			\toprule
			S$_\text{0}$ & 0.000 & 0.000 &  0.76& GS&& 0.000 & 0.000 & 4.46 &GS&& 0.000 & 0.000 & 1.25 &GS&& 0.000  & 0.000  &4.97 &GS\\
			S$_\text{1}$ & 1.847 & $<$0.001&  0.88 &$\textrm{LE}_1$&& 1.626 & 0.010 & 4.88 &$\textrm{CT}_1$$^*$&& 1.538 &$<$0.001& 1.32 &$\textrm{LE}_1$&& 1.761 & 0.756&5.91 &$\textrm{LE}_1$ \\ 		
			S$_\text{2}$ & 1.898 & 1.021& 5.26 &$\textrm{CT}_1$$^*$&& 1.811 & 0.573&5.47 &$\textrm{CT}_1$&&1.727 & 1.066 &0.07 &$\textrm{CT}_1$$^*$&& 2.099 & 0.617  &9.09&$\textrm{ME}$$^*$ \\  			
			S$_\text{3}$ &2.410 & 0.224& 2.70 &$\textrm{LE}_1$&& 2.400 &0.054 & 4.17&$\textrm{LE}_1$&&2.713 & $<$0.001 &0.66  &$\textrm{LE}_1$&& 2.399 & 0.001 & 3.75 &$\textrm{LE}_1$$^*$\\
			S$_\text{4}$ & 2.475 &0.164& 3.68 &$\textrm{LE}_2$&& 2.719&0.267 &4.93 &$\textrm{LE}_1$$^*$&& 3.035 & 0.466 &0.75  &$\textrm{LE}_2$&& 2.721 & 0.826 &5.61 &$\textrm{CT}_1$\\
			S$_\text{5}$ & 2.815& 0.018 &4.01 &$\textrm{CT}_1$$^*$&& 2.893 & 0.052 & 6.55 &$\textrm{LE}_2$&&  3.063 & 0.065 &3.13  &$\textrm{LE}_2$&&2.819 & 0.092 &8.35&$\textrm{CT}_1$ \\
			S$_\text{6}$ & 2.870 &0.169 & 2.68&$\textrm{ME}$&& 3.112 & 0.090& 5.36&$\textrm{LE}_2$&&  3.273 & 0.038&1.03 &$\textrm{CT}_1$&&2.969 & 0.070 &5.08&$\textrm{LE}_2$\\
			S$_\text{7}$ & 3.043 & 0.424& 2.82  &$\textrm{LE}_2$&& 3.319 & 0.032 & 1.56 &$\textrm{ME}$$^*$&& 3.604 & 0.170 &0.86 &$\textrm{ME}$&& 3.116 & 0.045 &6.78&$\textrm{LE}_2$\\	
			T$_\text{1}$ & 1.090 &0.000& $<$0.01&$\textrm{LE}_1$ &&0.941 &0.000&5.49 &$\textrm{LE}_1$ &&0.692 &0.000&1.47 &$\textrm{LE}_1$ &&1.256&0.000 &4.83 &$\textrm{LE}_1$ \\
			T$_\text{2}$ & 1.325 &0.000 & $<$0.01&$\textrm{LE}_1$ &&1.362 &0.000&3.49&$\textrm{LE}_1$ &&1.643 &0.000 &1.48 &$\textrm{LE}_1$ &&1.388 &0.000 &4.48&$\textrm{LE}_1$ \\
		    T$_\text{3}$ & 2.047 &0.000  &$<$0.01  &$\textrm{CT}_1$&&2.069 &0.000 &3.88  &$\textrm{CT}_1$&&2.375  &0.000  &1.58  &$\textrm{CT}_1$&&2.212  &0.000   &  8.41 &$\textrm{CT}_1$  \\
		    Q$_\text{1}$ &2.857   &0.000 & 0.60 &&&2.898 &0.000 &3.67&&&3.173  &0.000  &1.42 &&& 2.867 &0.000   & 4.33 &  \\
			\bottomrule
		\end{tabular}\\
		\label{tab:vees-regio}
		$^a$ Calculated at SA15-XMCQDPT(8,8)/cc-pVDZ level of theory $^b$ Calculated at SA15-CASSCF(8,8)/cc-pVDZ level of theory $^c$Character of the excited state: GS = Ground state, LE$_\textrm{1}$ = optically active states corresponding to symmetric and antisymmetric combinations of locally excited states on each aza-BODIPY monomer (HOMO $\rightarrow$ LUMO transitions within the same monomer), LE$_\textrm{2}$ = optically active states corresponding to symmetric and antisymmetric combinations of locally excited states involving HOMO-1 $\rightarrow$ LUMO transitions within the same monomer., ME = multiexcitonic state, and CT$_\textrm{1}$ = charge transfer states involving electron transfer from the HOMO of one monomer to the LUMO of the other monomer. (See Figs.  \ref{fig:dimer11-localized}-\ref{fig:dimer22-localized} for illustration.) 
	\end{table}
\end{landscape}

\begin{landscape}
	\begin{table}[!ht]
		\caption{Computed vertical excitation energies (VEEs, eV)$^a$, corresponding oscillator strengths ($f$, dimensionless)$^b$, dipole moments ($\mu$, Debye)$^b$, and the nature of the electronic states of orthogonal aza-BODIPY dimers D[1,1], D[1,3], D[3,3], and D[2,2].}
		\small
		\begin{tabular}{rrrrrrrrrrrrrrrrrrrr}
			\toprule
			\multirow{2}{*}{State}&\multicolumn{4}{c}{D[1,1]} && \multicolumn{4}{c}{D[1,3]} && \multicolumn{4}{c}{D[3,3]}  && \multicolumn{4}{c}{D[2,2]}\\
			\cmidrule[0.5pt]{2-5} \cmidrule[0.5pt]{7-10} \cmidrule[0.5pt]{12-15} \cmidrule[0.5pt]{17-20}
			& VEEs & $f$ & $\mu$ &char &&  VEEs & $f$ & $\mu$ &char && VEEs & $f$ & $\mu$  &char && VEEs & $f$ & $\mu$ &char \\
			\toprule
			S$_\text{0}$ & 0.000 & 0.000    &3.33 & GS     && 0.000 &0.000     &4.48  &GS     &&0.000  & 0.000   &1.23  &GS     &&0.000  & 0.000   &1.77  &GS    \\
			S$_\text{1}$ & 2.070 & 1.884    &2.88 & LE$_1$ && 1.967 &1.087     &5.01  &LE$_1$ &&1.948  & 0.956   &1.40  &LE$_1$ &&1.996  &0.742    &1.07  &LE$_1$\\ 		
			S$_\text{2}$ & 2.298 & $<$0.001 &2.77 & LE$_1$ && 2.277 &1.031     &6.15  &LE$_1$ &&2.349  &1.027    &1.80  &LE$_1$ &&2.350  &1.035    &0.42  &LE$_1$\\  			
			S$_\text{3}$ & 2.481 & $<$0.001 &3.09 & ME     && 2.441 &$<$0.001  &4.04  &ME     &&2.409  &0.000    &1.39  &ME     &&2.510  &$<$0.001 &1.88  &ME    \\
			S$_\text{4}$ & 2.883 & 0.224    &2.40 & CT$_1$ && 2.627 &0.003     &22.12 & CT$_1$&&2.510  & 0.051   &4.20  &CT$_1$ &&3.051  &0.352    &1.86  &CT$_1$ \\
			S$_\text{5}$ & 2.885 & $<$0.001 &2.49 & CT$_1$ && 2.735 &0.002     &16.29 &CT$_1$ &&2.533  &$<$0.001 &3.48  &CT$_1$ &&3.084  &0.004    &2.12  &CT$_1$\\
			S$_\text{6}$ & 3.012 & 0.035    &3.47 & LE$_2$ && 3.069 &0.121     &5.24  & LE$_2$&&3.132  &0.220    &0.47  &LE$_2$ &&3.128  &0.122    &0.31  &LE$_2$ \\
			S$_\text{7}$ & 3.034 & 0.324    &3.22 & LE$_2$ && 3.090 &0.111     &5.30  & LE$_2$&&3.143  &0.012    &0.24  &LE$_2$ &&3.138  &0.114    & 0.22 &LE$_2$ \\	
			T$_\text{1}$ & 1.238 & 0.000    &2.67 &        && 1.257 &0.000     &4.50  &       &&1.280  &0.000    &1.34  &       &&1.360  & 0.000   & 3.41 &       \\
			T$_\text{2}$ & 1.238 & 0.000    &2.67 &        && 1.307 &0.000     & 4.69 &       &&1.359  &0.000    &1.35  &       &&1.364  & 0.000   & 3.39 &        \\
			T$_\text{3}$ & 2.515 & 0.000    &3.03 &        && 2.533 &0.000     &4.46  &       &&2.538  &0.000    &1.36  &       &&2.628  & 0.000   &3.37  &         \\
			Q$_\text{1}$ &2.839 &0.000 & 2.13 & &  &2.839& 0.000 &4.33 &  &  &2.828  & 0.000 & 1.20 &  &  &2.885  &0.000  & 3.16 &\\
		\bottomrule
		\end{tabular}\\
		\label{tab:vees-regio-ortho}
		$^a$ Calculated at SA15-XMCQDPT(8,8)/cc-pVDZ level of theory $^b$ Calculated at SA15-CASSCF(8,8)/cc-pVDZ level of theory $^c$Character of the excited state: GS = Ground state, LE$_\textrm{1}$ = optically active states corresponding to symmetric and antisymmetric combinations of locally excited states on each aza-BODIPY monomer (HOMO $\rightarrow$ LUMO transitions within the same monomer), LE$_\textrm{2}$ = optically active states corresponding to symmetric and antisymmetric combinations of locally excited states involving HOMO-1 $\rightarrow$ LUMO transitions within the same monomer., ME = multiexcitonic state, and CT$_\textrm{1}$ = charge transfer states involving electron transfer from the HOMO of one monomer to the LUMO of the other monomer. (See Figs.  \ref{fig:dimer11-ortho-localized}-\ref{fig:dimer22-ortho-localized} for illustration.) 
	\end{table}
\end{landscape}

\subsection{Diabatic couplings}
\label{sec:diab}
With the aim of understanding the iSF energetics in aza-BODIPY dimers, the diabatic states and their electronic couplings were computed at the SA8-XMCQDPT(4,4)/cc-pVDZ level of theory. A careful comparison of the vertical excitation energies (VEEs) obtained using the SA15-XMCQDPT(8,8) and SA8-XMCQDPT(4,4) methods (see Tables \ref{tab:11-excited}–\ref{tab:22-excited} for the non-orthogonal dimers and Tables \ref{tab:11-ortho}–\ref{tab:22-ortho} for the orthogonal dimers) indicates that the (4,4) active space provides a reasonable compromise between computational efficiency and accuracy. Owing to the reduced active space, the LE$_\textrm{2}$-type excitations involving the HOMO$-1$ and LUMO$+1$ orbitals of the monomer are not included in the diabatization procedure.

Using the four-fold diabatization scheme proposed by Truhlar and Nakamura,\cite{Nakamura2001,Nakamura2002} eight diabatic states, together with their corresponding energies and electronic couplings, were obtained from eight adiabatic electronic states. The resulting diabatic Hamiltonians for the dimers D[1,1], D[1,3], D[3,3], and D[2,2] are reported in Tables \ref{tab:diab11}–\ref{tab:diab22} for the non-orthogonal configurations and in Tables \ref{tab:diab11-90}–\ref{tab:diab22-90} for the orthogonal configurations. For clarity, the most relevant couplings with respect to iSF are summarized in Table \ref{tab:eff}. As evident from Table \ref{tab:eff}, the diabatic couplings in the non-orthogonal dimers are substantially larger than those in the orthogonal dimers. The diabatic-to-adiabatic transformation matrices for the four dimers are provided in Eqs.~\ref{eq:rot11}–\ref{eq:rot22-90}. In all the non-orthogonal frameworks, the adiabatic states exhibit significant mixing of diabatic states, which leads to more intense diabatic couplings. 

The diabatic energies (particularly CT and ME states) and couplings are strongly influenced by the torsional angle and the relative proximity of $\textrm{BF}_\textrm{2}$ groups in the dimer (See Table \ref{tab:bond-angle-distance}). The order in which the distance between the $\textrm{BF}_\textrm{2}$ groups is as follows D[3,3] < D[1,3] < D[1,1] < D[2,2]. In case of non-orthogonal dimers, the energies of the diabatic states $^{1}(\textrm{C}\textrm{A})$ and $^{1}(\textrm{A}\textrm{C})$ of D[3,3] are the lowest among the four dimers studied. Also the  $^1\left(\textrm{T}_\textrm{1} \textrm{T}_\textrm{1}\right)$ is the highest among the four dimers. The energies of  $^1\left(\textrm{S}_\textrm{1} \textrm{S}_\textrm{0}\right)$ and  $^1\left(\textrm{S}_\textrm{0} \textrm{S}_\textrm{1}\right)$ are relatively stable and did not change much. As the distance between the $\textrm{BF}_\textrm{2}$ units increases, the energies of the $^{1}(\textrm{C}\textrm{A})$ and $^{1}(\textrm{A}\textrm{C})$ states increase, whereas the energy of the $^1\left(\textrm{T}_\textrm{1}\textrm{T}_\textrm{1}\right)$ state decreases. As a result, the energetic ordering of these states changes. In D[3,3] and D[1,1] the first two states are charge transfer states $^{1}(\textrm{C}\textrm{A})$ and $^{1}(\textrm{A}\textrm{C})$ followed by the locally excited states  $^1\left(\textrm{S}_\textrm{1} \textrm{S}_\textrm{0}\right)$, $^1\left(\textrm{S}_\textrm{0} \textrm{S}_\textrm{1}\right)$ and multi-excitonic state  $^1\left(\textrm{T}_\textrm{1} \textrm{T}_\textrm{1}\right)$. In dimer D[1,3], the state ordering have changed by having the state  $^1\left(\textrm{S}_\textrm{1} \textrm{S}_\textrm{0}\right)$ between the two CT states. For the dimer D[2,2], for which the distance between the $\textrm{BF}_\textrm{2}$ units is farthest, the energy of the CT states is higher than the   $^1\left(\textrm{S}_\textrm{1} \textrm{S}_\textrm{0}\right)$ and  $^1\left(\textrm{S}_\textrm{0} \textrm{S}_\textrm{1}\right)$ and multi-excitonic state  $^1\left(\textrm{T}_\textrm{1} \textrm{T}_\textrm{1}\right)$. The system D[1,3] presents an interesting scenario where CT states (as they rise in energy with respect to the $\textrm{BF}_\textrm{2}$ unit distance) strongly mix with the  $^1\left(\textrm{S}_\textrm{1} \textrm{S}_\textrm{0}\right)$ and  $^1\left(\textrm{S}_\textrm{0} \textrm{S}_\textrm{1}\right)$ and multi-excitonic state  $^1\left(\textrm{T}_\textrm{1} \textrm{T}_\textrm{1}\right)$ states, paving way for super-exchange like mechanism for SF. It is also noted that the diabatic energies LE diabatic states are degenerate for D[1,1], D[3,3] and D[2,2]. Despite of our best efforts, the LE diabatic states are not degenerate in D[1,3] (See Table \ref{tab:diab11}-\ref{tab:diab22}). This observation holds good for CT diabatic states as well. We attribute strong mixing (super-exchange like) among the diabatic states for this behaviour reported in our study.\cite{Goyal2026}

Concerning the diabatic couplings, except the dimer D[1,3], the direct couplings   $\bra{^{1}(\textrm{S}_\textrm{1}\textrm{S}_\textrm{0})} V \ket{^{1}(\textrm{T}_\textrm{1}\textrm{T}_\textrm{1})}$ for the remaining dimers are negligible. The direct coupling for D[1,3] is -189 meV. For all the dimers, the mediated couplings (i.e $\bra{^{1}(\textrm{S}_\textrm{1}\textrm{S}_\textrm{0})} V \ket{^{1}(\textrm{C}\textrm{A})}$ and  $\bra{^{1}(\textrm{T}_\textrm{1}\textrm{T}_\textrm{1})} V \ket{^{1}(\textrm{C}\textrm{A})}$) are very strong (See Table \ref{tab:eff}). The excitonic (diabatic) coupling between the  $^1\left(\textrm{S}_\textrm{0} \textrm{S}_\textrm{1}\right)$ and  $^1\left(\textrm{S}_\textrm{1} \textrm{S}_\textrm{0}\right)$ states was calculated as -117, 101, -134, and -189 for D[1,1], D[1,3], D[3,3], and D[2,2], respectively, consistent with previous studies on BODIPY dimers. All the relevant diabatic states lie in the range of $\sim$0.4 eV. The effective coupling calculated by Eq. \ref{eq:eff_coup} and the rate of iSF calculated by using Eq. \ref{eq:sf_rate} are shown in Table \ref{tab:veff}. The iSF rates for the dimers D[1,1] and D[1,3] are as high as cross-conjugated pentacene dimers ($\left(0.63 \pm 0.1\right) \times 10^{12}$).\cite{Zirzlmeier2016} Low iSF rates for dimer D[3,3] despite the exothermic iSF energy condition is due to the small effective couplings. The mediated couplings for dimer D[3,3] are equal in magnitude and are opposite in sign, as a result the effective couplings are close to zero.  

\begin{table}[!ht]
	\caption{Diabatic electronic couplings among the states $^1\left(\textrm{S}_\textrm{1} \textrm{S}_\textrm{0}\right)$,  $^1\left(\textrm{S}_\textrm{0} \textrm{S}_\textrm{1}\right)$,  $^1\left(\textrm{T}_\textrm{1} \textrm{T}_\textrm{1}\right)$, $^{1}(\textrm{C}\textrm{A})$ and $^{1}(\textrm{A}\textrm{C})$ for the dimers D[1,1], D[1,3], D[3,3] and D[2,2] expressed in meV, as obtained via Truhlar’s fourfold diabatization approach.}
	\small
	\begin{center}
		\begin{tabular}{rrrrrrrrrr}
			\toprule
			& \multicolumn{4}{c}{Non-orthogonal} && \multicolumn{4}{c}{Orthogonal}\\
			\cmidrule[0.5pt]{2-5} \cmidrule[0.5pt]{7-10}
			& D[1,1]&  D[1,3] & D[3,3] &D[2,2] && D[1,1]&  D[1,3] & D[3,3] &D[2,2]\\
			\toprule
			$\bra{^{1}(\textrm{S}_\textrm{1}\textrm{S}_\textrm{0})} V \ket{^{1}(\textrm{S}_\textrm{0}\textrm{S}_\textrm{1})}$    &-117 & 101&-134 &-189 && -3 & 160 & 196 & -183\\
			$\bra{^{1}(\textrm{S}_\textrm{1}\textrm{S}_\textrm{0})} V \ket{^{1}(\textrm{T}_\textrm{1}\textrm{T}_\textrm{1})}$    &22 & -189&-9 &-12 && -6 & 0 & -3 & 0\\
			$\bra{^{1}(\textrm{S}_\textrm{1}\textrm{S}_\textrm{0})} V \ket{^{1}(\textrm{C}\textrm{A})}$            &-144 & -381&-571&189 && 6 & -29 & -14 & 67\\
			$\bra{^{1}(\textrm{S}_\textrm{1}\textrm{S}_\textrm{0})} V \ket{^{1}(\textrm{A}\textrm{C})}$            &438 & -335&542& 389 && 41 & 41 & 47 & -15 \\
			$\bra{^{1}(\textrm{S}_\textrm{0}\textrm{S}_\textrm{1})} V \ket{^{1}(\textrm{T}_\textrm{1}\textrm{T}_\textrm{1})}$    &-27 & 196&9& 13 && 8 & -4 & 2 & 2\\
			$\bra{^{1}(\textrm{S}_\textrm{0}\textrm{S}_\textrm{1})} V \ket{^{1}(\textrm{C}\textrm{A})}$            &-432 & 243&-519&383 && 47 & -3 & 16 & 33\\
			$\bra{^{1}(\textrm{S}_\textrm{0}\textrm{S}_\textrm{1})} V \ket{^{1}(\textrm{A}\textrm{C})}$            &132 & 502& 550& 173 && 15 & -6 & -50 & -61 \\
			$\bra{^{1}(\textrm{T}_\textrm{1}\textrm{T}_\textrm{1})} V \ket{^{1}(\textrm{C}\textrm{A})}$            &320 & -440&650 &268 && 2 & 27 & -49 & -58\\
			$\bra{^{1}(\textrm{T}_\textrm{1}\textrm{T}_\textrm{1})} V \ket{^{1}(\textrm{A}\textrm{C})}$            &314 & 491& 650&-267 && -2 & 16 & 76 & 75\\
			\bottomrule
		\end{tabular}
	\end{center}
	\label{tab:eff}
\end{table}

\begin{table}[!ht]
	\small
	\caption{Diabatic direct, mediated, and effective couplings (in meV) for the investigated non-orthogonal and orthogonal dimers. The iSF rate constant is estimated assuming a reorganization energy of 100 meV for the $\ket{^{1}(\textrm{S}_\textrm{1}\textrm{S}_\textrm{0})}$ state.}
	\begin{center}
		\setlength{\tabcolsep}{4pt}
		\begin{tabular}{crrrcccccccc}
			\toprule
			& \multicolumn{4}{c}{Non-orthogonal} && \multicolumn{4}{c}{Orthogonal} &&\\
			\cmidrule[0.5pt]{2-5} \cmidrule[0.5pt]{7-10}
			System & Direct & Mediated & V$_\textrm{eff}$ & k$_\textrm{SF}^a$ && Direct & Mediated & V$_\textrm{eff}$ & k$_\textrm{SF}^a$&& k$_\textrm{SF}^b$\\
			& (meV)  & (meV)    &   (meV)          & (s$^\textrm{-1}$)&& (meV)  & (meV)    &   (meV)          & (s$^\textrm{-1}$)&&(s$^\textrm{-1}$) \\
			\toprule
			D[1,1] & 21.7 &-8332.8&-8310.8 & 7.10 $\times$ 10$^\textrm{16}$  && -5.7 & 0.1544 & -5.5456 & 5.09 $\times$ 10$^\textrm{7}$ && 2.38 $\times$ 10$^\textrm{-18}$ \\
			D[1,3] & -189 & -26.636& -215.64   &1.43 $\times$ 10$^\textrm{13}$ && 0.3  & 0.3499& 0.6499   &6.19 $\times$ 10$^\textrm{2}$ &&8.62 $\times$ 10$^\textrm{-29}$\\
			D[3,3] & -9.0& 35.399& 26.399 &3.82 $\times$ 10$^\textrm{3}$   && -3.1 & 9.1677 & 8.8677 &1.10 $\times$ 10$^\textrm{3}$ && 6.13 $\times$ 10$^\textrm{-23}$\\
			D[2,2] &-12.0 &172.78 & 160.78   &4.85 $\times$ 10$^\textrm{7}$ && 0.0 &-6.0557 & -6.0557   &9.71 $\times$ 10$^\textrm{1}$ &&3.50 $\times$ 10$^\textrm{-31}$\\
			\bottomrule
		\end{tabular} \\
		$^a$ $E(\ket{^{1}(\textrm{T}_\textrm{1}\textrm{T}_\textrm{1})})-E(\ket{^{1}(\textrm{S}_\textrm{1}\textrm{S}_\textrm{0})})$ taken from Tables \ref{tab:diab11} - \ref{tab:diab22}\\ $^b$  Reproduced from Ref. \onlinecite{Goyal2024} for BODIPY dimers B[1,1], B[1,3], B[3,3] and B[2,2].
	\end{center}
	\label{tab:veff}
\end{table}

\subsection{Spin-Orbit coupling}
\label{sec:rot}
To investigate the mechanism of triplet state generation we calculated the spin–orbit coupling constant ($\lambda_\textrm{SOC}$) between low lying singlet and triplet states. 
For efficient intersystem crossing, large $\lambda_\textrm{SOC}$ and small $\Delta{E}_{ST}$ are desirable. The computed $\lambda_\textrm{SOC}$ values for all four dimers are relatively small (see Tables \ref{soc-11}–\ref{soc-22}) as expected for organic molecules. Moreover, the energy gaps ($\Delta{E}_{ST}$ in cm$^{-1}$) between $\textrm{S}_\textrm{1}$ and $\textrm{T}_\textrm{1}$ are significantly large, while their respective SOC constants are small. This combination leads to inefficient intersystem crossing.
To explore whether these gaps can be reduced, we performed rigid potential energy scans along the torsional angle between the monomer units from 0° to 180° with an interval of 5° using the Gaussian09 software package\cite{Frisch2009}. Subsequently, vertical excitation energies were computed for the eight lowest singlet and triplet excited states. Among these, seven singlet states ($\textrm{S}_1$–$\textrm{S}_7$) and three triplet states ($\textrm{T}_1$–$\textrm{T}_3$) were selected for analysis due to their relevance in triplet state generation. For the D[3,3] dimer, optimization beyond a torsional angle of 165° was not achieved, likely due to the onset of significant torsional strain arising from the close proximity of the monomer units.

The potential energy curves show multiple avoided crossings among adiabatic states, suggesting possible population transfer routes that may enhance ISC efficiency. An avoided crossing between $\textrm{S}_2$ and $\textrm{S}_3$ occurs near 90° in all dimers, and $\textrm{S}_1$–$\textrm{S}_2$ avoided crossings appear near the axial conformations (except in D[2,2]). These features suggest that electronic coupling and state mixing are highly dependent on the torsional angle, directly affecting ISC probability.
 
To quantify this effect, ISC rate constants ($k_\textrm{ISC}$) were calculated using Eq.~\ref{eq:rates_triplet} for each dimer across the $\textrm{S}_1$–$\textrm{T}_1$, $\textrm{S}_1$–$\textrm{T}_2$, and $\textrm{S}_1$–$\textrm{T}_3$ transitions (Tables \ref{tab:rates_triplet11}–\ref{tab:rates_triplet22}). The rate constants were evaluated for both axial and orthogonal orientations by varying the parameters $\lambda$, $S$, and $\hbar\omega$ in the rate expression. In all dimers, $k_\textrm{ISC}$ is lowest for $\textrm{S}_1$–$\textrm{T}_1$ due to the large energy gap, slightly higher for $\textrm{S}_1$–$\textrm{T}_2$, and reaches a maximum for $\textrm{S}_1$–$\textrm{T}_3$. This trend clearly supports the $\textrm{S}_1\to\textrm{T}_3$ transition as the most probable SOCT-ISC pathway. Table~\ref{kisc} summarizes the comparative $k_\textrm{ISC}$ values for axial and orthogonal geometries. 

These rate trends can be directly correlated with the torsional-scan results. In the axial geometry, all dimers exhibit smaller $\textrm{S}_1$–$\textrm{T}_2$ gaps, which may slightly increase $k_\textrm{ISC}$ for this transition. Moreover, across all dimers, the $\textrm{S}_1$–$\textrm{T}_3$ gap becomes minimal at the orthogonal geometry, supporting the feasibility of SOCT-ISC in this orientation. This observation is consistent with previous reports that orthogonal orientations promote SOCT-ISC in donor–acceptor systems due to enhanced CT character and SOC efficiency. In the table \ref{kisc} we have compared the rate constants for axial and orthogonal arrangement of dimers for all the three transitions. 

\begin{figure}[!ht]
	\caption{Scan calculation results of all the four dimers D[1,1], D[1,3], D[3,3] and D[2,2]}
	\begin{subfigure}{0.50\textwidth}
		\includegraphics[width=0.95\textwidth]{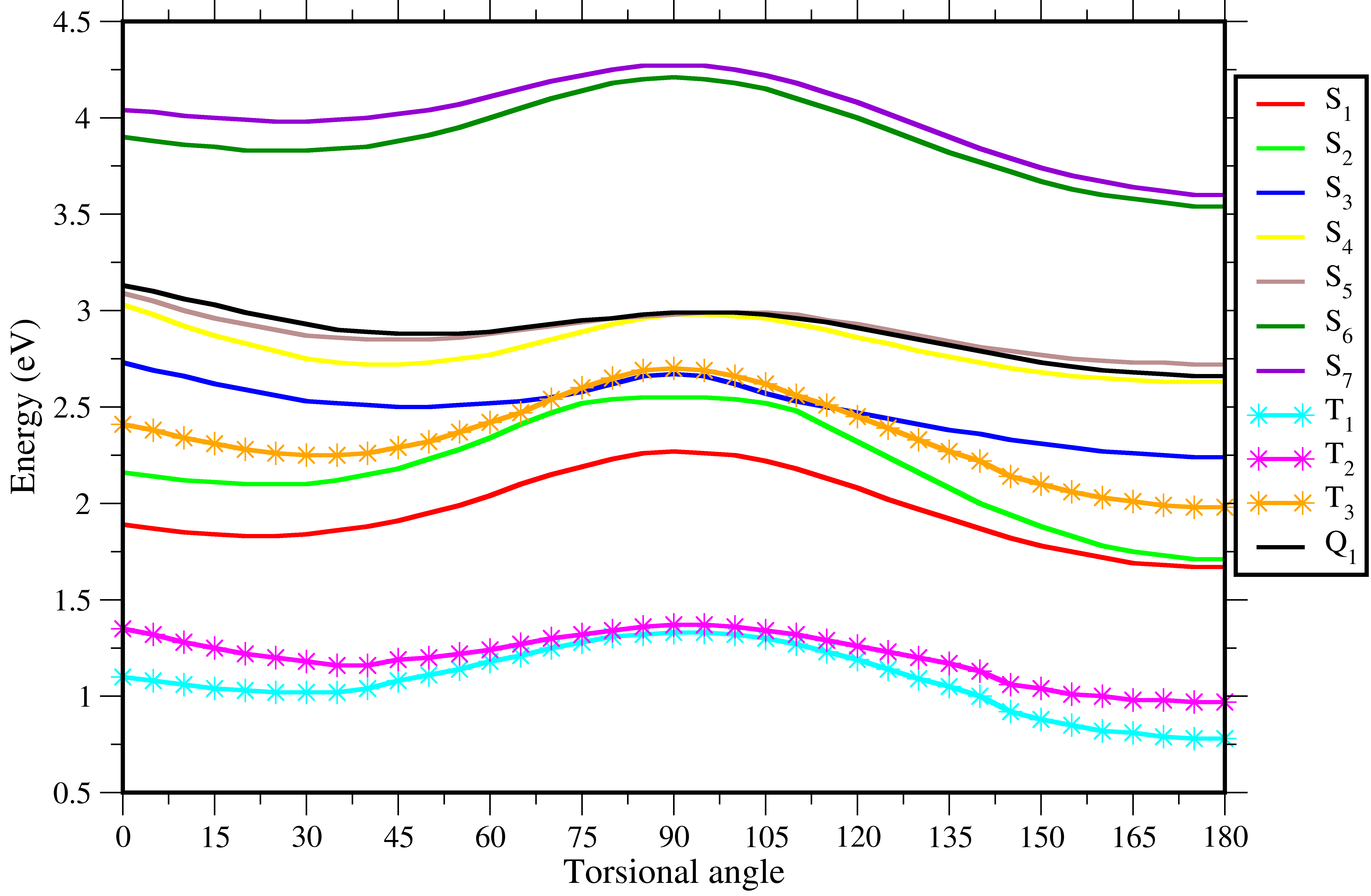}
		\caption{D[1,1] scan}
		\label{fig:aza11} 
	\end{subfigure}
	\begin{subfigure}{0.50\textwidth}
		\includegraphics[width=0.95\textwidth]{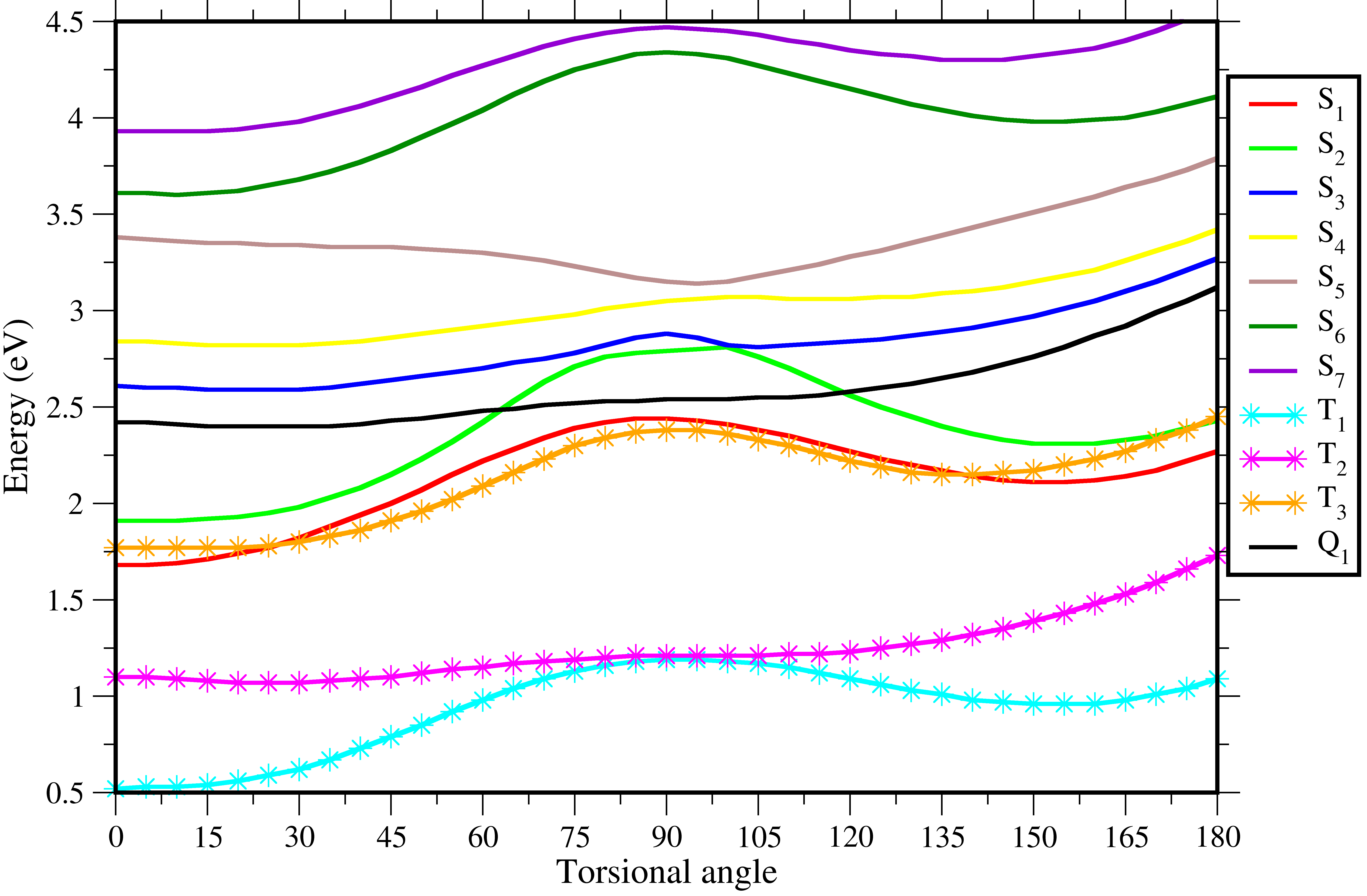}
		\caption{D[1,3] scan}
		\label{fig:aza13} 
	\end{subfigure}
\begin{subfigure}{0.50\textwidth}
	\includegraphics[width=0.95\textwidth]{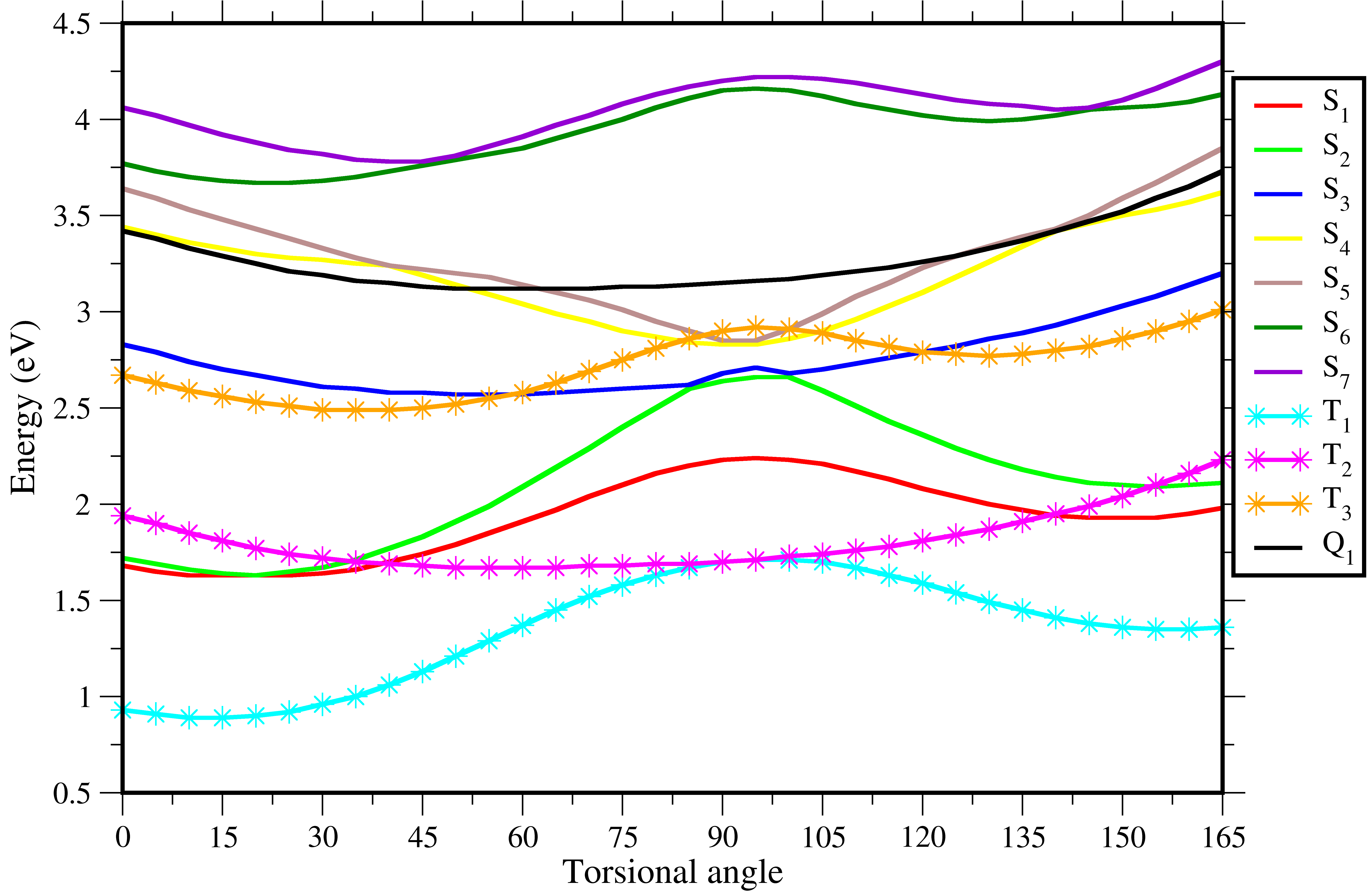}
	\caption{D[3,3] scan}
	\label{fig:aza33} 
\end{subfigure}
\begin{subfigure}{0.50\textwidth}
	\includegraphics[width=0.95\textwidth]{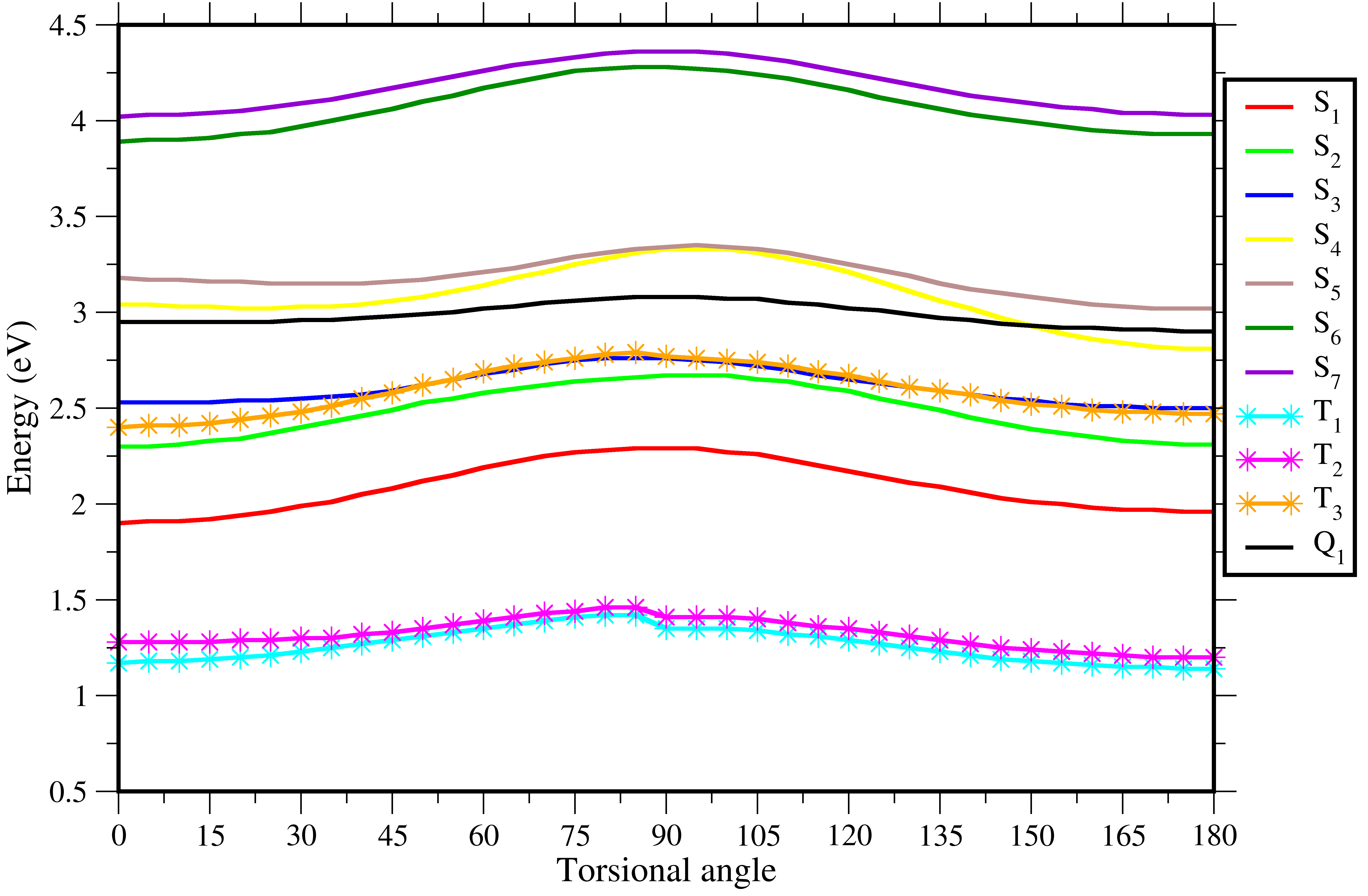}
	\caption{D[2,2] scan}
	\label{fig:aza22} 
\end{subfigure}
\label{fig:scan}
\end{figure}

Dimer-specific behavior underscores the role of regio-connectivity. D[1,3] displays an atypical trend where the axial orientation favors $\textrm{S}_1$–$\textrm{T}_3$ coupling more than the orthogonal one. The asymmetry introduced by its 1,3-linkage creates uneven charge distribution between the two monomer units, leading to distinct local electronic environments and altered SOC characteristics. The avoided crossing between $\textrm{S}_1$ and $\textrm{T}_3$ near the axial orientation (Figure \ref{fig:aza13}) supports this interpretation. Nevertheless, the difference in $k_\textrm{ISC}$ values between axial and orthogonal orientations is modest (Tables \ref{soc-13} and \ref{tab:rates_triplet13}). Conversely, D[2,2] maintains a consistently small $\textrm{S}_1$–$\textrm{T}_3$ gap across all torsional angles, indicating a unique electronic configuration that favors ISC in both axial and orthogonal forms.

Overall, the combined analysis of SOC, torsional scans, and singlet–triplet energetics demonstrates that orthogonal geometries generally favor SOCT-ISC in aza-BODIPY dimers, with the $\textrm{S}_1\to\textrm{T}_3$ transition emerging as the dominant pathway. However, molecular asymmetry and connectivity profoundly modulate this behavior, highlighting the delicate interplay between structure and electronic coupling in controlling triplet formation efficiency.

\section{Conclusions}
In this study, we aimed to understand the mechanism of triplet-state generation in aza-BODIPY dimers through either iSF or SOCT-ISC. Our results reveal that both mechanisms can contribute to triplet formation, showing a strong dependence on the torsional angle between the monomer units and a secondary influence of regio-connectivity. Among the four dimers investigated, D[2,2] is found to be highly endothermic toward SF, resulting in a low $k_\textrm{SF}$ value. In contrast, this dimer exhibits noticeable activity in the SOCT-ISC pathway for triplet formation in both its axial and orthogonal orientations. The other three dimers display higher $k_\textrm{SF}$ values than SOCT-ISC rates in their axial conformations but prefer SOCT-ISC when arranged orthogonally.

Dimers D[1,1] and D[1,3] emerge as favorable candidates for SF, whereas D[3,3] shows exothermic ME formation but a comparatively low iSF rate. This reduced iSF efficiency can be attributed to its coupling elements, which are nearly equal in magnitude and opposite in sign, effectively lowering the net electronic coupling and hence $k_\textrm{SF}$. The D[1,3] dimer provides particularly interesting insights: unlike the others, it favors SOCT-ISC in its axial conformation over the orthogonal one, likely due to asymmetric charge distribution within the molecular geometry. 

Further analysis of the transitions involved in the SOCT-ISC mechanism indicates that the dominant pathway arises from the $\textrm{S}_1$–$\textrm{T}_3$ transition, which features a small energy gap and relatively large spin–orbit coupling compared to other state pairs. These findings establish a clear relationship between molecular geometry, coupling interactions, and the operative triplet-generation mechanism in aza-BODIPY dimers. The insights gained here provide valuable design guidelines for tailoring torsional flexibility and electronic coupling in future aza-BODIPY-based triplet photosensitizers.

\begin{table}[h!] 
	\centering
	\caption{Spin-orbit couplings (SOC in cm$^{-1}$ units) and energy difference E (in cm$^{-1}$ units) between selected singlet and triplet states for non-orthogonal and orthogonal geometries of D[1,1] calculated at SA15-XMCQDPT(8,8)/cc-pVDZ level of theory.}
	\begin{center}
	\begin{tabular}{|c|cc|cc|}
		\hline
		\textbf{State} & \multicolumn{2}{c|}{\textbf{Non-orthogonal}} & \multicolumn{2}{c|}{\textbf{Orthogonal}} \\
		& SOC (cm$^{-1}$) & $\Delta E$ (cm$^{-1}$) & SOC (cm$^{-1}$) & $\Delta E$ (cm$^{-1}$) \\
		\hline
		$\textrm{S}_\textrm{1}$-$\textrm{T}_\textrm{1}$ & -0.007  & 6095.8464  &0.026  &6710.7446  \\
		$\textrm{S}_\textrm{1}$-$\textrm{T}_\textrm{2}$ & 0.112 & 4199.2319 &0.061  &6706.9035  \\
		$\textrm{S}_\textrm{1}$-$\textrm{T}_\textrm{3}$ & 0.002 & -1615.9400 &0.047  &-3592.6337  \\
		\hline
	\end{tabular}
\end{center}
\label{soc-11}
\end{table}

\begin{table}[h!]
	\centering
	\caption{Spin-orbit couplings (SOC in cm$^{-1}$ units) and energy difference E (in cm$^{-1}$ units) between selected singlet and triplet states for non-orthogonal and orthogonal geometries of D[1,3] calculated at SA15-XMCQDPT(8,8)/cc-pVDZ level of theory.}
	\begin{center}
	\begin{tabular}{|c|cc|cc|}
		\hline
		\textbf{State} & \multicolumn{2}{c|}{\textbf{Non-orthogonal}} & \multicolumn{2}{c|}{\textbf{Orthogonal}} \\
		& SOC (cm$^{-1}$) & $\Delta E$ (cm$^{-1}$) & SOC (cm$^{-1}$) & $\Delta E$ (cm$^{-1}$) \\
		\hline
		$\textrm{S}_\textrm{1}$-$\textrm{T}_\textrm{1}$ & 0.152 & 5528.29 & 0.180 & 5730.32 \\
		$\textrm{S}_\textrm{1}$-$\textrm{T}_\textrm{2}$ & 0.033 & 2131.15 & 0.053 & 5322.75 \\
		$\textrm{S}_\textrm{1}$-$\textrm{T}_\textrm{3}$ & 0.122 & -3576.68 & 0.051 & -4561.79 \\
		\hline
	\end{tabular}
\end{center}
\label{soc-13}
\end{table}

\begin{table}[h!]
	\centering
	\caption{Spin-orbit couplings (SOC in cm$^{-1}$ units) and energy difference E (in cm$^{-1}$ units) between selected singlet and triplet states for non-orthogonal and orthogonal geometries of D[3,3] calculated at SA15-XMCQDPT(8,8)/cc-pVDZ level of theory.}
	\begin{center}
	\begin{tabular}{|c|cc|cc|}
		\hline
		\textbf{State} & \multicolumn{2}{c|}{\textbf{Non-orthogonal}} & \multicolumn{2}{c|}{\textbf{Orthogonal}} \\
		& SOC (cm$^{-1}$) & $\Delta E$ (cm$^{-1}$) & SOC (cm$^{-1}$) & $\Delta E$ (cm$^{-1}$) \\
		\hline
		$\textrm{S}_\textrm{1}$-$\textrm{T}_\textrm{1}$ & 0.335  & 6816.4310 & 0.352 &5388.3722  \\
		$\textrm{S}_\textrm{1}$-$\textrm{T}_\textrm{2}$ & 0.001  & -854.8103 & 0.126 &4754.1086  \\
		$\textrm{S}_\textrm{1}$-$\textrm{T}_\textrm{3}$ & 0.074  & -6756.6204 & 0.270 &-4758.9813  \\
		\hline
	\end{tabular}
\end{center}
\label{soc-33}
\end{table}

\begin{table}[h!]
	\centering
	\caption{Spin-orbit couplings (SOC in cm$^{-1}$ units) and energy difference E (in cm$^{-1}$ units) between selected singlet and triplet states for non-orthogonal and orthogonal geometries of D[2,2] calculated at SA15-XMCQDPT(8,8)/cc-pVDZ level of theory.}
	\begin{center}
	\begin{tabular}{|c|cc|cc|}
		\hline
		\textbf{State} & \multicolumn{2}{c|}{\textbf{Non-orthogonal}} & \multicolumn{2}{c|}{\textbf{Orthogonal}} \\
		& SOC (cm$^{-1}$) & $\Delta E$ (cm$^{-1}$) & SOC (cm$^{-1}$) & $\Delta E$ (cm$^{-1}$) \\
		\hline
		$\textrm{S}_\textrm{1}$-$\textrm{T}_\textrm{1}$ &0.033  &4075.5900  &0.195  &5135.8681  \\
		$\textrm{S}_\textrm{1}$-$\textrm{T}_\textrm{2}$ &0.049  &3004.2567  &0.085  &5097.4526  \\
		$\textrm{S}_\textrm{1}$-$\textrm{T}_\textrm{3}$ &0.030  &-3638.0860  &0.086  &-5089.4591  \\
		\hline
	\end{tabular}
\end{center}
\label{soc-22}
\end{table}

\begin{table}[h!]
	\centering
	\caption{Computed SOCT-ISC rate constants (s$^{-1}$) for the $S_1 \rightarrow T_n$ ($n = 1$–3) transitions in the axial and orthogonal geometries of the investigated aza-BODIPY dimers}
	\begin{center}
		\setlength{\tabcolsep}{0.2pt}
		\begin{tabular}{|c|cc|cc|cc|cc|}
			\hline
  & \multicolumn{2}{c|}{\textbf{D[1,1]}} &\multicolumn{2}{c|}{\textbf{D[1,3]}} & \multicolumn{2}{c|}{\textbf{D[3,3]}} & \multicolumn{2}{c|}{\textbf{D[2,2]}} \\
  \toprule
&	\textbf{Axial} & \textbf{Orthogonal} & 	\textbf{Axial} & \textbf{Orthogonal} & 	\textbf{Axial} & \textbf{Orthogonal} & 	\textbf{Axial} & \textbf{Orthogonal}  \\
\midrule
  $k_{S_1 \rightarrow T_1}$ & ~1.37x$10^{-15}$           & ~1.24x$10^{-17}$ & ~3.30x$10^{-10}$ & ~5.41x$10^{-11}$ & ~5.53x$10^{-16}$ & ~7.92x$10^{-9}$ & ~1.66x$10^{-5}$ & ~3.21x$10^{-8}$\\
  $k_{S_1 \rightarrow T_2}$ & ~6.65x$10^{-5}$ & ~7.17x$10^{-17}$ &~1.14x$10^{1}$ & ~3.54x$10^{-10}$ & 2.85$10^{2}$ & ~5.54x$10^{-7}$ & ~1.34x$10^{-1}$ & ~8.96x$10^{-9}$  \\
  $k_{S_1 \rightarrow T_3}$ & ~1.88x$10^{3}$           & ~2.69x$10^{5}$   & ~1.79x$10^{6}$ & ~7.96x$10^{4}$ & ~2.94x$10^{3}$ & ~1.62x$10^{6}$ & ~1.03x$10^{5}$ & ~9.48x$10^{4}$  \\
  \bottomrule
	\end{tabular}
\end{center}
\label{kisc}
\end{table}

\begin{table}[ht]
	\caption{The rate constant($k_\textrm{ISC}$ in $s^\textrm{-1}$ units) of triplet state formation through the SOCT-ISC by varying the parameters reorganization energy ($\lambda$ in eV units), Haung-Rhys factor ($S$ in dimensionless units) and effective coupling ($\hbar \omega$ in cm$^\textrm{-1}$ units) for D[1,1].}
	\begin{tabular}{ccccccccc}
		\toprule
		&&&\multicolumn{3}{c}{Non-orthogonal} & \multicolumn{3}{c}{Orthogonal}\\
		$\lambda$ & $S$ & $\hbar\omega$ &  $k_{S_1 \rightarrow T_1}$ & $k_{S_1 \rightarrow T_2}$ & $k_{S_1 \rightarrow T_3}$ & $k_{S_1 \rightarrow T_1}$ & $k_{S_1 \rightarrow T_2}$ & $k_{S_1 \rightarrow T_3}$ \\
		\toprule
		\multirow{3}{*}{0.1} & \multirow{3}{*}{0.3} & ~700 & ~5.15x$10^{-27}$ & ~5.03x$10^{-10}$ & ~1.69x$10^{3}$ &~1.34x$10^{-31}$  & ~8.04x$10^{-31}$  & ~7.52x$10^{3}$   \\
		&                      & 1200 & ~5.15x$10^{-27}$ & ~5.03x$10^{-10}$ & ~1.49x$10^{3}$ & ~1.34x$10^{-31}$  &  ~8.04x$10^{-31}$  & ~5.98x$10^{4}$  \\
		&                      & 1600 & ~5.15x$10^{-27}$ & ~5.03x$10^{-10}$ & ~1.20x$10^{3}$ & ~1.34x$10^{-31}$  & ~8.04x$10^{-31}$ & ~9.59x$10^{4}$   \\
		\midrule                 
		\multirow{3}{*}{0.1} & \multirow{3}{*}{0.5} & ~700 & ~4.22x$10^{-27}$ & ~4.12x$10^{-10}$ & ~1.87x$10^{3}$ & ~1.10x$10^{-31}$ & ~6.59x$10^{-31}$  & ~2.30x$10^{4}$  \\
		&                      & 1200 & ~4.22x$10^{-27}$ & ~4.12x$10^{-10}$ & ~1.54x$10^{3}$ & ~1.10x$10^{-31}$ & ~6.59x$10^{-31}$ & ~1.31x$10^{5}$   \\
		&                      & 1600 & ~4.22x$10^{-27}$ & ~4.12x$10^{-10}$ & ~1.14x$10^{3}$ & ~1.10x$10^{-31}$  &~6.59x$10^{-31}$  & ~1.73x$10^{5}$   \\
		\midrule
		\multirow{3}{*}{0.2} & \multirow{3}{*}{0.3} & ~700 & ~1.37x$10^{-15}$ & ~6.65x$10^{-5}$ & ~2.08x$10^{3}$ & ~1.24x$10^{-17}$  & ~7.17x$10^{-17}$  & ~1.72x$10^{5}$  \\
		&                      & 1200 &  ~1.37x$10^{-15}$    & ~6.65x$10^{-5}$ & ~1.88x$10^{3}$ & ~1.24x$10^{-17}$  &~7.17x$10^{-17}$  & ~2.69x$10^{5}$   \\
		&                      & 1600 &  ~1.37x$10^{-15}$    & ~6.65x$10^{-5}$ & ~1.78x$10^{3}$ & ~1.24x$10^{-17}$  &~7.17x$10^{-17}$  & ~3.19x$10^{5}$  \\
		\midrule
		\multirow{3}{*}{0.2} & \multirow{3}{*}{0.5} & ~700 & ~1.12x$10^{-15}$  & ~5.45x$10^{-5}$ & ~1.92x$10^{3}$ &~1.02x$10^{-17}$  & ~5.87x$10^{-17}$  & ~2.48x$10^{5}$  \\
		&                      & 1200 & ~1.12x$10^{-15}$   & ~5.45x$10^{-5}$ & ~1.64x$10^{3}$ & ~1.02x$10^{-17}$  &~5.87x$10^{-17}$  & ~3.74x$10^{5}$   \\
		&                      & 1600 & ~1.12x$10^{-15}$    & ~5.45x$10^{-5}$ & ~1.50x$10^{3}$ & ~1.02x$10^{-17}$  &~5.87x$10^{-17}$  &~4.20x$10^{5}$   \\	
		\bottomrule		 
	\end{tabular}
	\label{tab:rates_triplet11}
\end{table}

\begin{table}[ht]
	\caption{The rate constant($k_\textrm{ISC}$ in $s^\textrm{-1}$ units) of triplet state formation through the SOCT-ISC by varying the parameters reorganization energy ($\lambda$ in eV units), Haung-Rhys factor ($S$ in dimensionless units) and effective coupling ($\hbar \omega$ in cm$^\textrm{-1}$ units) for D[1,3].}
	\begin{tabular}{ccccccccc}
		\toprule
		&&&\multicolumn{3}{c}{Non-orthogonal} & \multicolumn{3}{c}{Orthogonal}\\
		$\lambda$ & $S$ & $\hbar\omega$ &  $k_{S_1 \rightarrow T_1}$ & $k_{S_1 \rightarrow T_2}$ & $k_{S_1 \rightarrow T_3}$ &  $k_{S_1 \rightarrow T_1}$ & $k_{S_1 \rightarrow T_2}$ & $k_{S_1 \rightarrow T_3}$\\
		\toprule
		\multirow{3}{*}{0.1} & \multirow{3}{*}{0.3} & ~700 & ~1.72x$10^{-19}$ & 1.76 & ~5.33x$10^{4}$   & ~5.06x$10^{-21}$  & ~9.43x$10^{-19}$  & ~3.05x$10^{2}$ \\
		&                      & 1200 & ~1.72x$10^{-19}$ &  1.76 & ~4.13x$10^{5}$  & ~5.06x$10^{-21}$  & ~9.43x$10^{-19}$  & ~1.15x$10^{4}$  \\
		&                      & 1600 & ~1.72x$10^{-19}$ &  1.76 & ~6.59x$10^{5}$  & ~5.06x$10^{-21}$  & ~9.43x$10^{-19}$  & ~4.65x$10^{4}$  \\
		\midrule                 
		\multirow{3}{*}{0.1} & \multirow{3}{*}{0.5} & ~700 & ~1.41x$10^{-19}$ & 1.44 & ~1.61x$10^{5}$  & ~4.14x$10^{-21}$  & ~7.72x$10^{-19}$  & ~1.68x$10^{3}$   \\
		&                      & 1200 & ~1.41x$10^{-19}$ & 1.44 & ~8.99x$10^{5}$  & ~4.14x$10^{-21}$  & ~7.72x$10^{-19}$  & ~3.70x$10^{4}$   \\
		&                      & 1600 & ~1.41x$10^{-19}$ & 1.44 & ~1.18x$10^{6}$  & ~4.14x$10^{-21}$  & ~7.72x$10^{-19}$  & ~1.07x$10^{5}$   \\
		\midrule
		\multirow{3}{*}{0.2} & \multirow{3}{*}{0.3} & ~700 & ~3.39x$10^{-10}$    & ~1.39x$10^{1}$ & ~1.20x$10^{6}$ & ~5.41x$10^{-11}$  & ~3.54x$10^{-10}$ & ~2.12x$10^{4}$  \\
		&                      & 1200 & ~3.39x$10^{-10}$    & ~1.38x$10^{1}$ & ~1.85x$10^{6}$  & ~5.41x$10^{-11}$  & ~3.54x$10^{-10}$  & ~7.96x$10^{4}$  \\
		&                      & 1600 & ~3.39x$10^{-10}$    & ~1.38x$10^{1}$ & ~2.18x$10^{6}$ & ~5.41x$10^{-11}$ & ~3.54x$10^{-10}$  & ~1.36x$10^{5}$  \\
		\midrule
		\multirow{3}{*}{0.2} & \multirow{3}{*}{0.5} & ~700 & ~2.78x$10^{-10}$    & ~1.14x$10^{1}$ & ~1.71x$10^{6}$  & ~4.43x$10^{-11}$ & ~2.90x$10^{-10}$  & ~4.47x$10^{4}$  \\
		&                      & 1200 & ~2.78x$10^{-10}$    & ~1.13x$10^{1}$ & ~2.56x$10^{6}$  & ~4.43x$10^{-11}$  & ~2.90x$10^{-10}$  & ~1.53x$10^{5}$  \\
		&                      & 1600 & ~2.78x$10^{-10}$    & ~1.13x$10^{1}$ & ~2.86x$10^{6}$  & ~4.43x$10^{-11}$  & ~2.90x$10^{-10}$  & ~2.29x$10^{5}$  \\	
		\bottomrule		 
	\end{tabular}
	\label{tab:rates_triplet13}
\end{table}

\begin{table}[ht]
	\caption{The rate constant($k_\textrm{ISC}$ in $s^\textrm{-1}$ units) of triplet state formation through the SOCT-ISC by varying the parameters reorganization energy ($\lambda$ in eV units), Haung-Rhys factor ($S$ in dimensionless units) and effective coupling ($\hbar \omega$ in cm$^\textrm{-1}$ units) for D[3,3].}
	\begin{tabular}{ccccccccc}
		\toprule
		&&&\multicolumn{3}{c}{Non-orthogonal} & \multicolumn{3}{c}{Orthogonal}\\
		$\lambda$ & $S$ & $\hbar\omega$ &  $k_{S_1 \rightarrow T_1}$ & $k_{S_1 \rightarrow T_2}$ & $k_{S_1 \rightarrow T_3}$ &  $k_{S_1 \rightarrow T_1}$ & $k_{S_1 \rightarrow T_2}$ & $k_{S_1 \rightarrow T_3}$ \\
		\toprule
		\multirow{3}{*}{0.1} & \multirow{3}{*}{0.3} & ~700 & ~2.06x$10^{-30}$ &  ~6.99x$10^{2}$ & ~4.46x$10^{-2}$ & ~1.25x$10^{-17}$ & ~1.04x$10^{-13}$  & ~4.09x$10^{3}$ \\
		&                      & 1200 & ~2.06x$10^{-30}$ & ~6.62x$10^{2}$   & ~2.27x$10^{2}$ & ~1.25x$10^{-17}$  & ~1.04x$10^{-13}$ & ~2.25x$10^{5}$ \\
		&                      & 1600 & ~2.06x$10^{-30}$ & ~6.06x$10^{2}$  &  ~2.91x$10^{3}$ & ~1.25x$10^{-17}$ & ~1.04x$10^{-13}$  & ~9.24x$10^{5}$ \\
		\midrule                 
		\multirow{3}{*}{0.1} & \multirow{3}{*}{0.5} & ~700 & ~1.69x$10^{-30}$  & ~6.27x$10^{2}$ & 1.53 & ~1.02x$10^{-17}$ & ~8.54x$10^{-14}$ & ~2.55x$10^{4}$ \\
		&                      & 1200 & ~1.69x$10^{-30}$ & ~5.26x$10^{2}$ & ~1.71x$10^{3}$ & ~1.02x$10^{-17}$ & ~8.54x$10^{-14}$  & ~7.81x$10^{5}$ \\
		&                      & 1600 & ~1.69x$10^{-30}$ & ~4.99x$10^{2}$ &  ~1.31x$10^{4}$ & ~1.02x$10^{-17}$ & ~8.54x$10^{-14}$  & ~2.20x$10^{6}$ \\
		\midrule
		\multirow{3}{*}{0.2} & \multirow{3}{*}{0.3} & ~700 & ~5.53x$10^{-16}$ & ~3.05x$10^{2}$  &~3.51x$10^{1}$  & ~7.92x$10^{-9}$ &  ~5.54x$10^{-7}$    & ~3.47x$10^{5}$ \\
		&                      & 1200 &  ~5.53x$10^{-16}$    & ~2.85x$10^{2}$  & ~2.94x$10^{3}$ & ~7.92x$10^{-9}$ & ~5.54x$10^{-7}$  & ~1.62x$10^{6}$  \\
		&                      & 1600 &  ~5.53x$10^{-16}$    & ~2.80x$10^{2}$ &~1.62x$10^{4}$  & ~7.92x$10^{-9}$ & ~5.54x$10^{-7}$  & ~3.01x$10^{6}$ \\
		\midrule
		\multirow{3}{*}{0.2} & \multirow{3}{*}{0.5} & ~700 & ~4.53x$10^{-16}$  & ~2.65x$10^{2}$ & ~2.23x$10^{2}$ & ~6.49x$10^{-9}$ & ~4.54x$10^{-7}$  & ~7.98x$10^{5}$ \\
		&                      & 1200 & ~4.53x$10^{-16}$     & ~2.37x$10^{2}$  & ~1.25x$10^{4}$ & ~6.49x$10^{-9}$ & ~4.54x$10^{-7}$ & ~3.35x$10^{6}$ \\
		&                      & 1600 & ~4.53x$10^{-16}$    & ~2.30x$10^{2}$ &~5.25x$10^{4}$  & ~6.49x$10^{-9}$ & ~4.54x$10^{-7}$ & ~5.41x$10^{6}$ \\	
		\bottomrule		 
	\end{tabular}
	\label{tab:rates_triplet33}
\end{table}

\begin{table}[ht]
	\caption{The rate constant($k_\textrm{ISC}$ in $s^\textrm{-1}$ units) of triplet state formation through the SOCT-ISC by varying the parameters reorganization energy ($\lambda$ in eV units), Haung-Rhys factor ($S$ in dimensionless units) and effective coupling ($\hbar \omega$ in cm$^\textrm{-1}$ units) for D[2,2].}
	\begin{tabular}{ccccccccc}
		\toprule
		&&&\multicolumn{3}{c}{Non-orthogonal} & \multicolumn{3}{c}{Orthogonal}\\
		$\lambda$ & $S$ & $\hbar\omega$ &  $k_{S_1 \rightarrow T_1}$ & $k_{S_1 \rightarrow T_2}$ & $k_{S_1 \rightarrow T_3}$ &  $k_{S_1 \rightarrow T_1}$ & $k_{S_1 \rightarrow T_2}$ & $k_{S_1 \rightarrow T_3}$\\
		\toprule
		\multirow{3}{*}{0.1} & \multirow{3}{*}{0.3} & ~700 & ~2.69x$10^{-10}$  & ~6.09x$10^{-4}$  &  ~2.65x$10^{3}$ & ~3.65x$10^{-16}$ & ~1.36x$10^{-16}$  &  ~1.16x$10^{2}$ \\
		&                      & 1200 & ~2.69x$10^{-10}$ & ~6.09x$10^{-4}$ & ~2.26x$10^{4}$  & ~3.65x$10^{-16}$  & ~1.36x$10^{-16}$ & ~1.21x$10^{4}$ \\
		&                      & 1600 & ~2.69x$10^{-10}$ &  ~6.09x$10^{-4}$ &  ~3.71x$10^{4}$ & ~3.65x$10^{-16}$ & ~1.36x$10^{-16}$ & ~4.86x$10^{4}$ \\
		\midrule                 
		\multirow{3}{*}{0.1} & \multirow{3}{*}{0.5} & ~700 & ~2.19x$10^{-10}$ & ~4.99x$10^{-4}$  & ~8.30x$10^{3}$ & 2.99x$10^{-16}$ & ~1.12x$10^{-16}$ & ~8.92x$10^{2}$ \\
		&                      & 1200 & ~2.19x$10^{-10}$  &  ~4.99x$10^{-4}$  & ~5.02x$10^{4}$ & ~2.99x$10^{-16}$  & ~1.12x$10^{-16}$ & ~4.68x$10^{4}$ \\
		&                      & 1600 & ~2.19x$10^{-10}$ & ~4.99x$10^{-4}$ & ~6.87x$10^{4}$ & ~2.99x$10^{-16}$ & ~1.12x$10^{-16}$  & ~1.31x$10^{5}$ \\
		\midrule
		\multirow{3}{*}{0.2} & \multirow{3}{*}{0.3} & ~700 & ~1.66x$10^{-5}$  & ~1.34x$10^{-1}$ & ~6.42x$10^{4}$ & ~3.21x$10^{-8}$ & ~8.96x$10^{-9}$ & ~1.35x$10^{4}$ \\
		&                      & 1200 & ~1.66x$10^{-5}$  & ~1.34x$10^{-1}$ & ~1.03x$10^{5}$ & ~3.21x$10^{-8}$ & ~8.96x$10^{-9}$ & ~9.48x$10^{4}$ \\
		&                      & 1600 & ~1.66x$10^{-5}$ &  ~1.34x$10^{-1}$ & ~1.25x$10^{5}$ & ~3.21x$10^{-8}$ & ~8.96x$10^{-9}$ & ~2.07x$10^{5}$ \\
		\midrule
		\multirow{3}{*}{0.2} & \multirow{3}{*}{0.5} & ~700 & ~1.36x$10^{-5}$ & ~1.10x$10^{-1}$ & ~9.40x$10^{4}$  & ~2.63x$10^{-8}$ & ~7.34x$10^{-9}$ & ~3.63x$10^{4}$ \\
		&                      & 1200 & ~1.36x$10^{-5}$ & ~1.10x$10^{-1}$ &  ~1.46x$10^{5}$ & ~2.63x$10^{-8}$ & ~7.34x$10^{-9}$ & ~2.19x$10^{5}$ \\
		&                      & 1600 & ~1.36x$10^{-5}$ & ~1.10x$10^{-1}$  & ~1.66x$10^{5}$  & ~2.63x$10^{-8}$ & ~7.34x$10^{-9}$ & ~4.14x$10^{5}$\\	
		\bottomrule		 
	\end{tabular}
	\label{tab:rates_triplet22}
\end{table}

\section*{Conflicts of interest}
There are no conflicts to declare.

\section*{Acknowledgements}
S. G. expresses gratitude to the University Grants Commission (UGC), India, for providing the Senior Research Fellowship. SRR acknowledges financial support from SERB, India, for computational resources under the project (SRG/2021/001684), the UGC Startup Grant (F.30-547/2021(BSR)) and ANRF pair grant (ANRF/PAIR/2025/000006/PAIR-A).
\bibliography{ref}
\end{document}


\section{Excited states: Dimers}
	15 lowest lying singlet and triplet states are calculated by using state-averaged extended multiconfiguration quasi degenerate perturbation theory (SA-XMCQDPT)\cite{Granovsky2011} and cc-pVDZ\cite{Dunning2013} basis set. An intruder avoidance shift of 0.02 was applied in these calculations. In addition to the singlet and triplet states, we have also calculated 3 lowest lying quintet states at the same level of theory. A shift of 0.02 a.u. was applied to mitigate intruder state effects on energy calculations. Exploratory calculations in terms of active space (See Table \ref{tab:active-dimer}) with basis set cc-pVDZ are performed to find a compromise between accuracy and computational load. An active space (8,8) found to be optimal for the current problem.
	
	\begin{table}[ht]
		\caption{The VEEs for aza-BODIPY Dimers with Different Active Spaces }
		\begin{center}
			\begin{tabular}{cccccccccccc}
				\toprule
				& \multicolumn{3}{c}{4,4} && \multicolumn{3}{c}{8,8}&& \multicolumn{3}{c}{12,10} \\ 
				\cmidrule{2-4} \cmidrule{6-8} \cmidrule{10-12}
				Dimer & $\textrm{S}_\textrm{1}$ & $\textrm{T}_\textrm{1}$ & $\Delta$SF && $\textrm{S}_\textrm{1}$  & $\textrm{T}_\textrm{1}$ & $\Delta$SF && $\textrm{S}_\textrm{1}$  & $\textrm{T}_\textrm{1}$ & $\Delta$SF \\ 
				\toprule
				D[1,1] & 1.88 & 1.06 & ~0.24 && 1.85 & 1.09 & ~0.33 && 1.68 & 0.95 & ~0.22 \\
				D[1,3] & 1.66 & 0.93 & ~0.21 && 1.63 & 0.94 & ~0.25 && 1.52 & 0.95 & ~0.37 \\
				D[3,3] & 1.64 & 0.75 & -0.13 && 1.54 & 0.69 & -0.15 && 1.49 & 0.53 & -0.43 \\
				D[2,2] & 1.78 & 1.13 & ~0.49 && 1.76 & 1.26 & ~0.75 && 1.77 & 1.30 & ~0.83 \\
				\bottomrule
			\end{tabular}
			\label{tab:active-dimer}
		\end{center}
	\end{table}
	
	\begin{table}[!ht]
		\caption{The distance $r$ between the two $\textrm{BF}_\textrm{2}$ units and the torsional angle $\Phi$ between the two planes of aza-BODIPY monomer units (See Fig. \ref{fig:aza} in main text).}
		\begin{tabular}{c c c}
			\toprule
			Dimer & $r$ (in \AA) & $\Phi$ (in $\circ$) \\
			\toprule
			D[1,1] & 8.99 & ~0.0\\
			D[1,3] & 6.83 & 14.9\\
			D[3,3] & 5.62 & 23.0\\
			D[2,2] & 8.19 & ~0.0\\
			\bottomrule
		\end{tabular}
		\label{tab:bond-angle-distance}
	\end{table}
	
 \begin{figure}[!ht]
	 \caption{The relaxed scan of the ground singlet state energy of the aza-BODIPY dimers D[1,1], D[1,3], D[3,3] and D[2,2].}
	 \begin{subfigure}{0.45\textwidth}
                \includegraphics[width=0.95\textwidth]{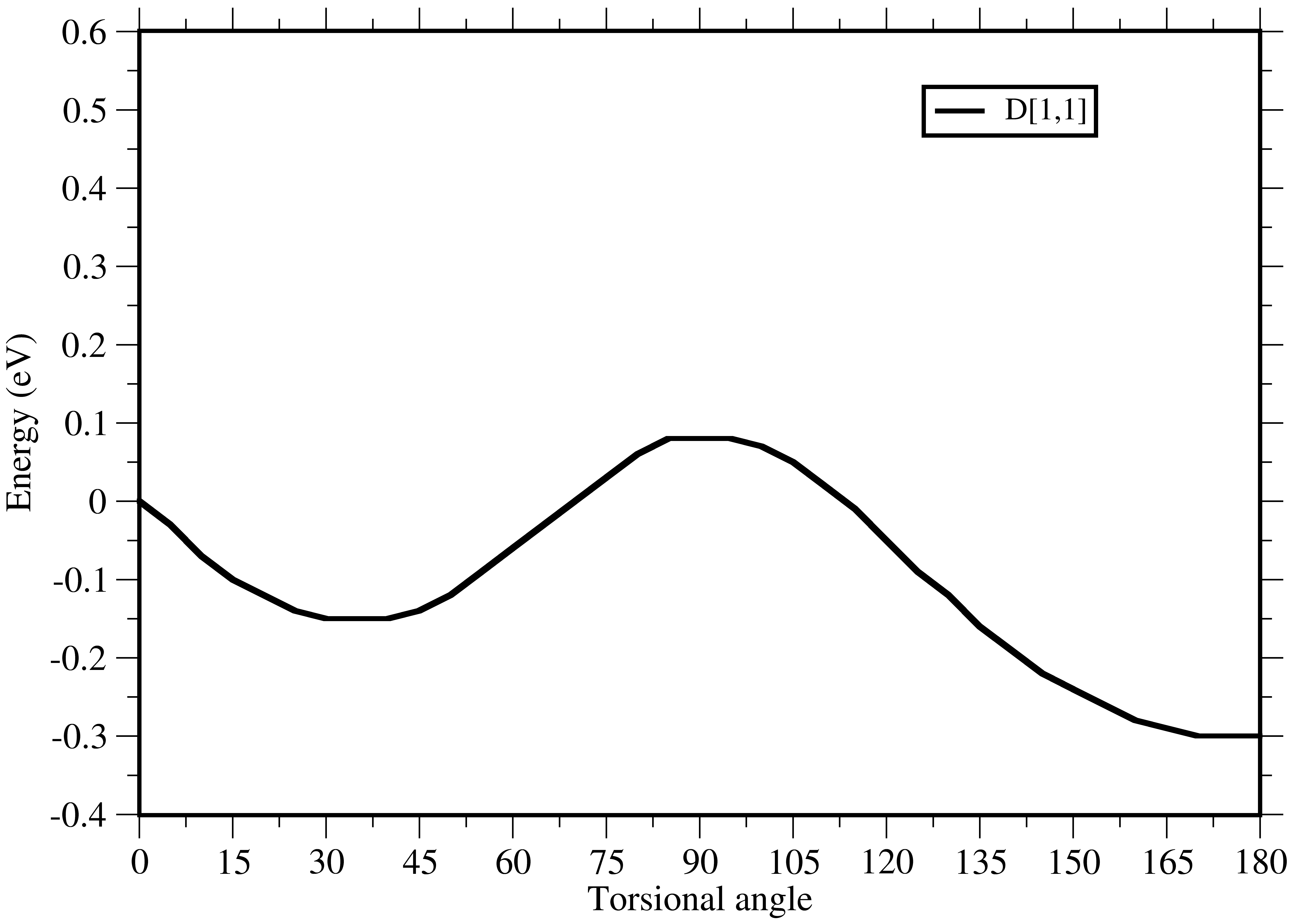}
                \label{fig:aza11}
        \end{subfigure}
        \begin{subfigure}{0.45\textwidth}
                \includegraphics[width=0.95\textwidth]{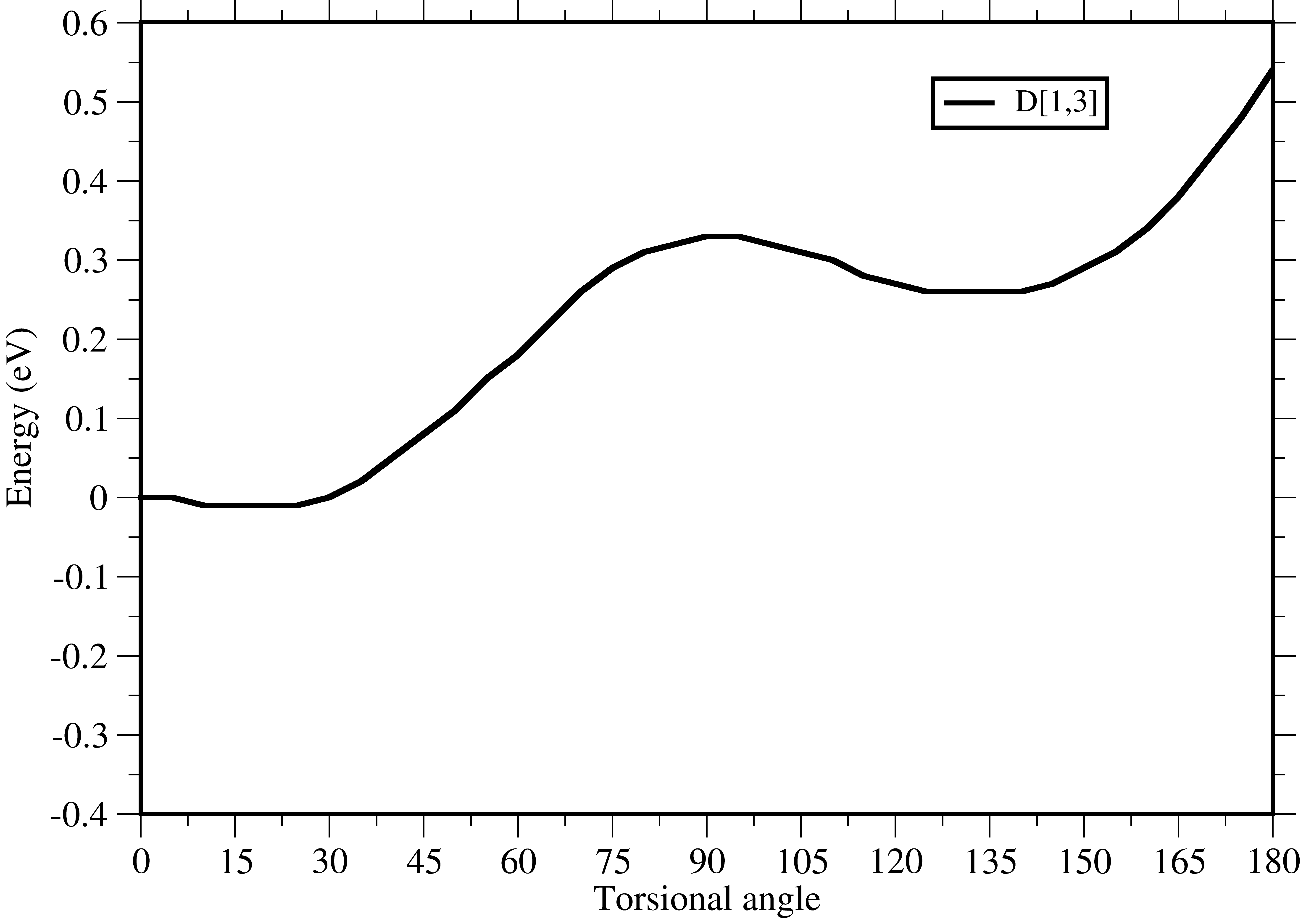}
                \label{fig:aza13}
        \end{subfigure}
        \begin{subfigure}{0.45\textwidth}
                \includegraphics[width=0.95\textwidth]{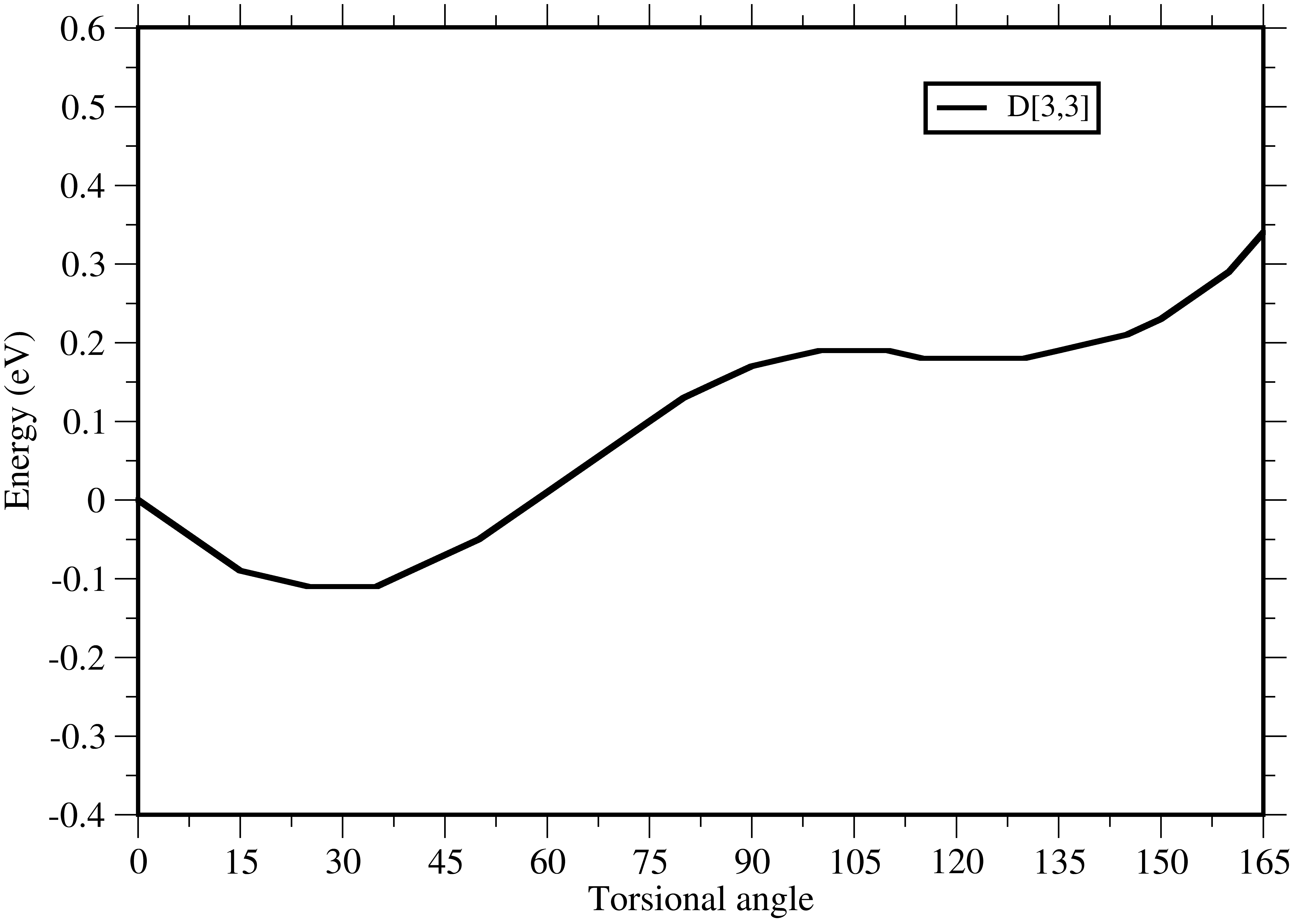}
                \label{fig:aza33}
        \end{subfigure}
        \begin{subfigure}{0.45\textwidth}
                \includegraphics[width=0.95\textwidth]{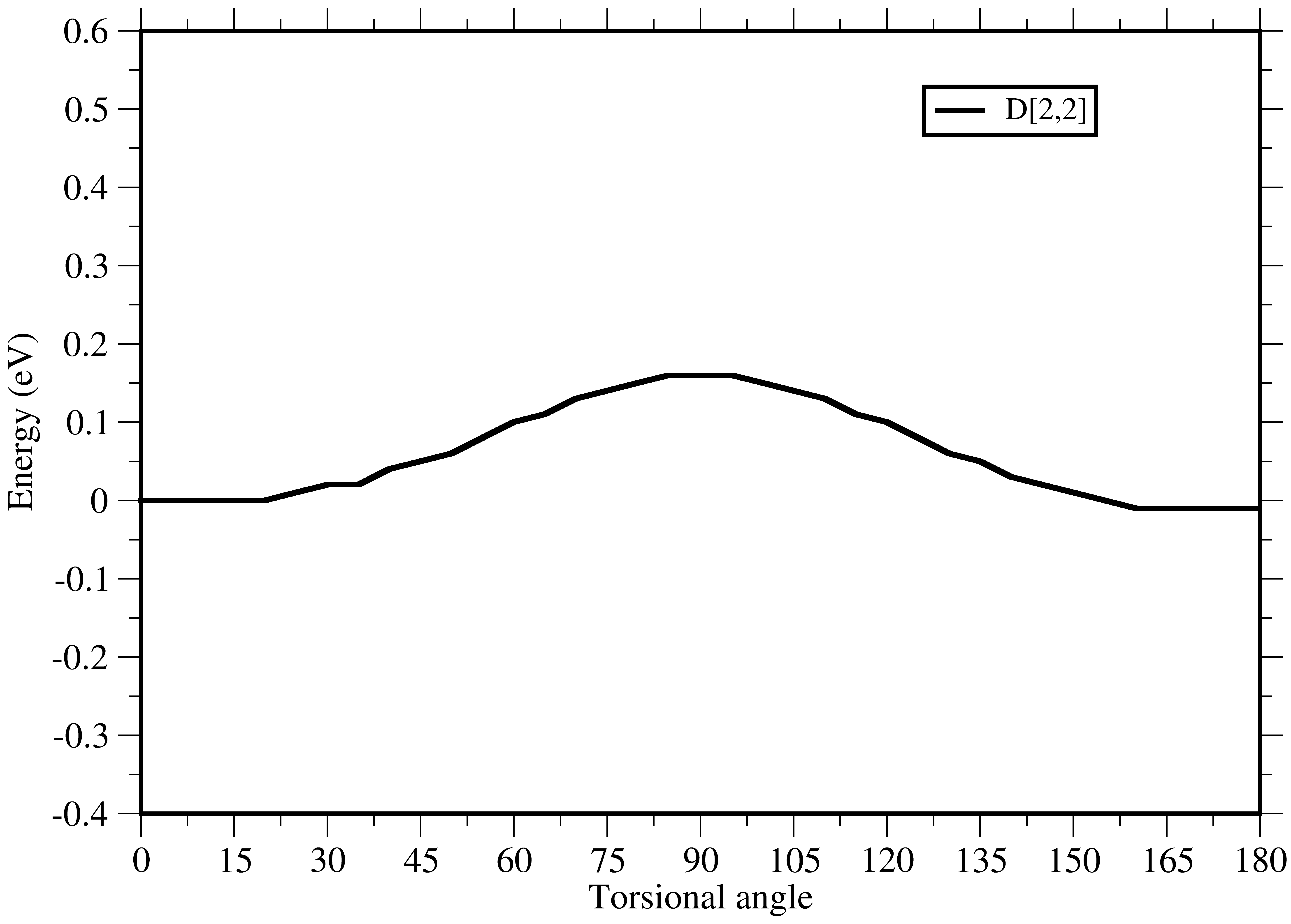}
		\label{fig:aza22}
        \end{subfigure}
        \end{figure}

\begin{figure}[H]
		\caption{The characterization of the excited states of dimer D[1,1] by using the localised HOMO-1 (a), HOMO (b), LUMO (c) and LUMO+1 (d) orbitals. Single and double arrows represents single and double excitations. The value on top of the arrow represent the \% of the particular excitation to the total wavefunction. The black and blue boxes represent the constructive and distructive combinations.}
		\includegraphics[width=0.6\textwidth]{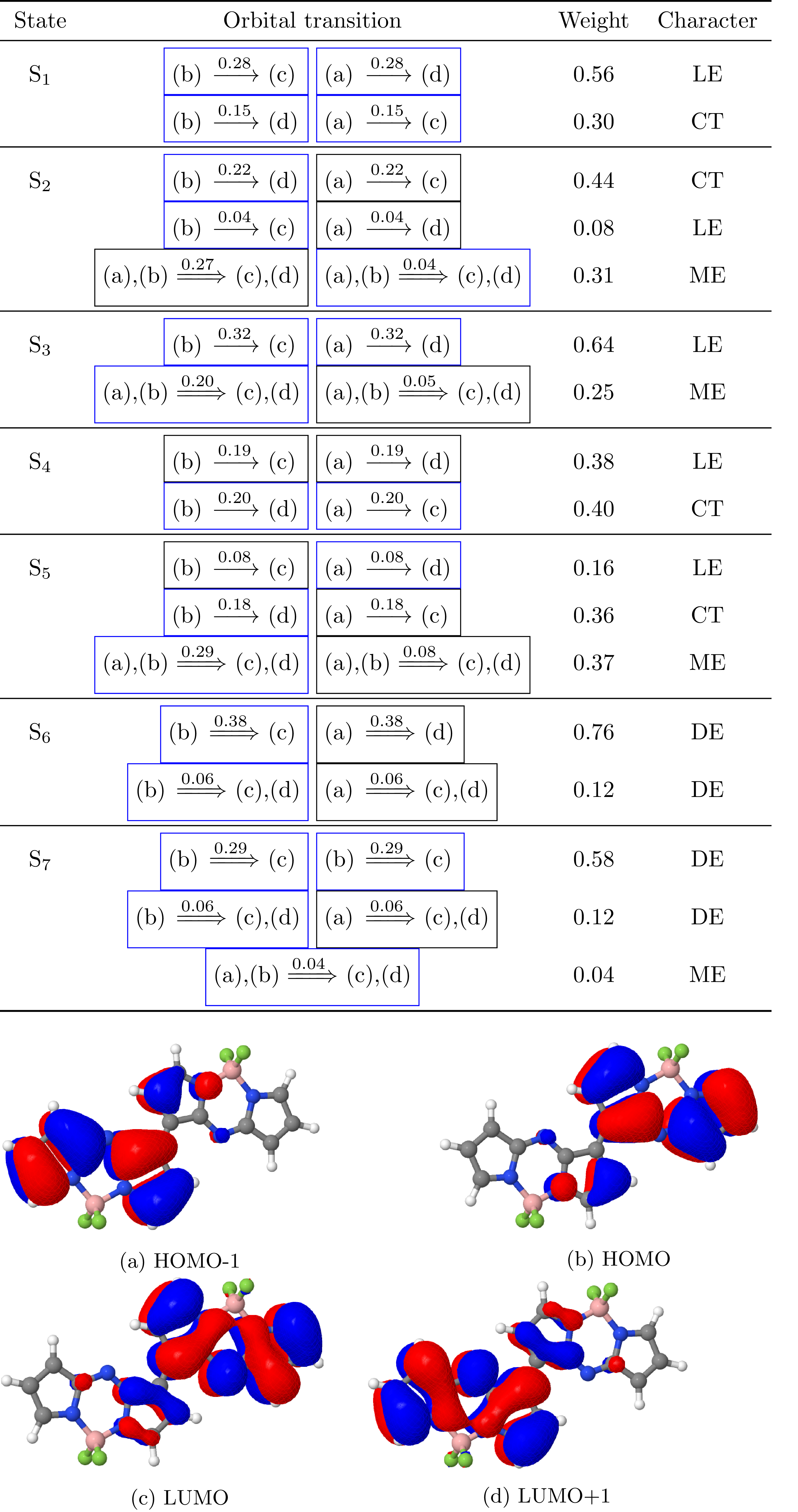}
		\label{fig:dimer11-localized}
	\end{figure}

\begin{figure}[H]
		\caption{The characterization of the excited states of dimer D[1,3] by using the localised HOMO-1 (a), HOMO (b), LUMO (c) and LUMO+1 (d) orbitals. Single and double arrows represents single and double excitations. The value on top of the arrow represent the \% of the particular excitation to the total wavefunction. The black and blue boxes represent the constructive and distructive combinations.}
		\includegraphics[width=0.6\textwidth]{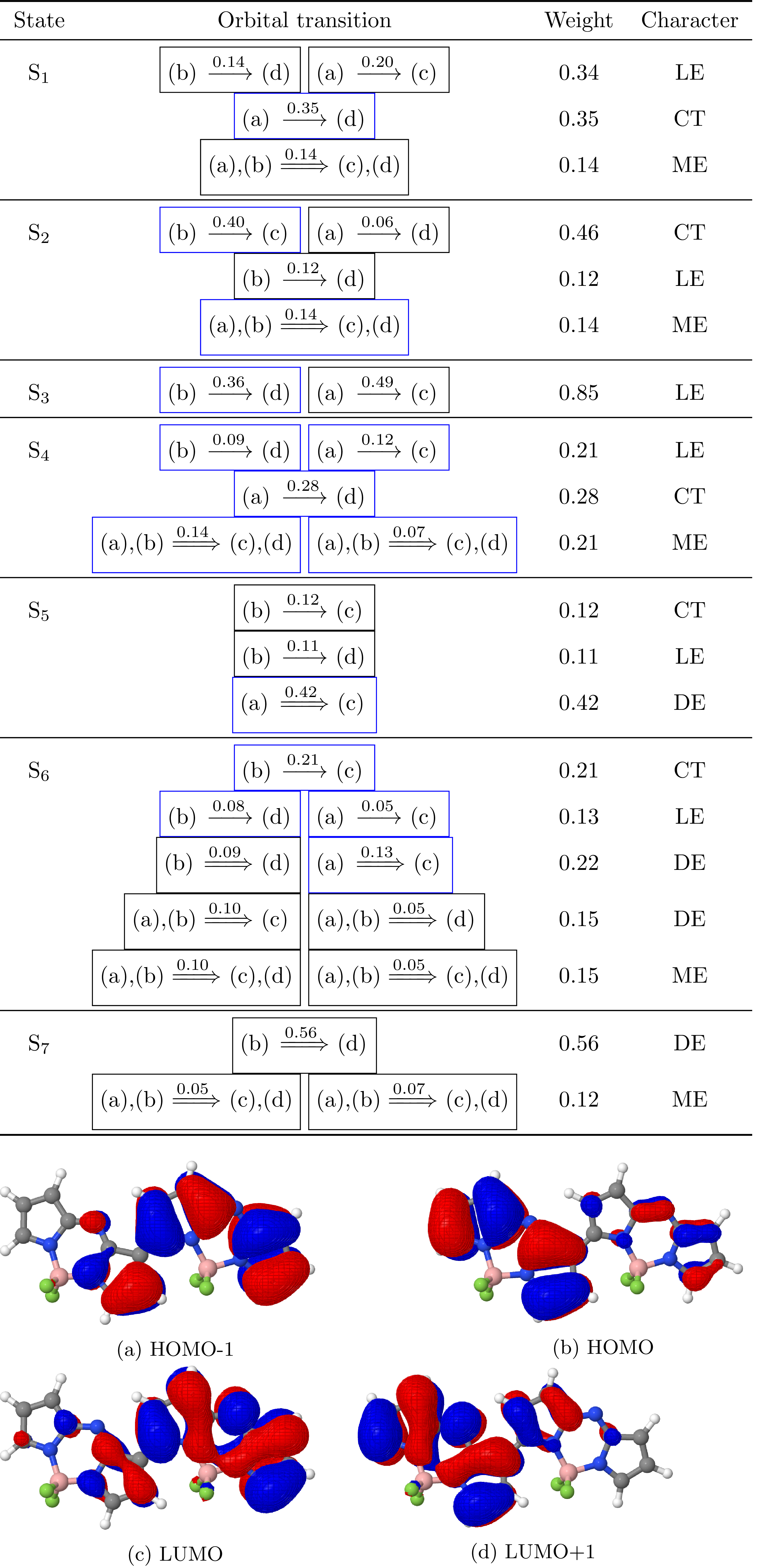}
		\label{fig:dimer13-localized}	
	\end{figure}

\begin{figure}[H]
		\caption{The characterization of the excited states of dimer D[3,3] by using the localised HOMO-1 (a), HOMO (b), LUMO (c) and LUMO+1 (d) orbitals. Single and double arrows represents single and double excitations. The value on top of the arrow represent the \% of the particular excitation to the total wavefunction. The black and blue boxes represent the constructive and distructive combinations.}
		\includegraphics[width=0.60\textwidth]{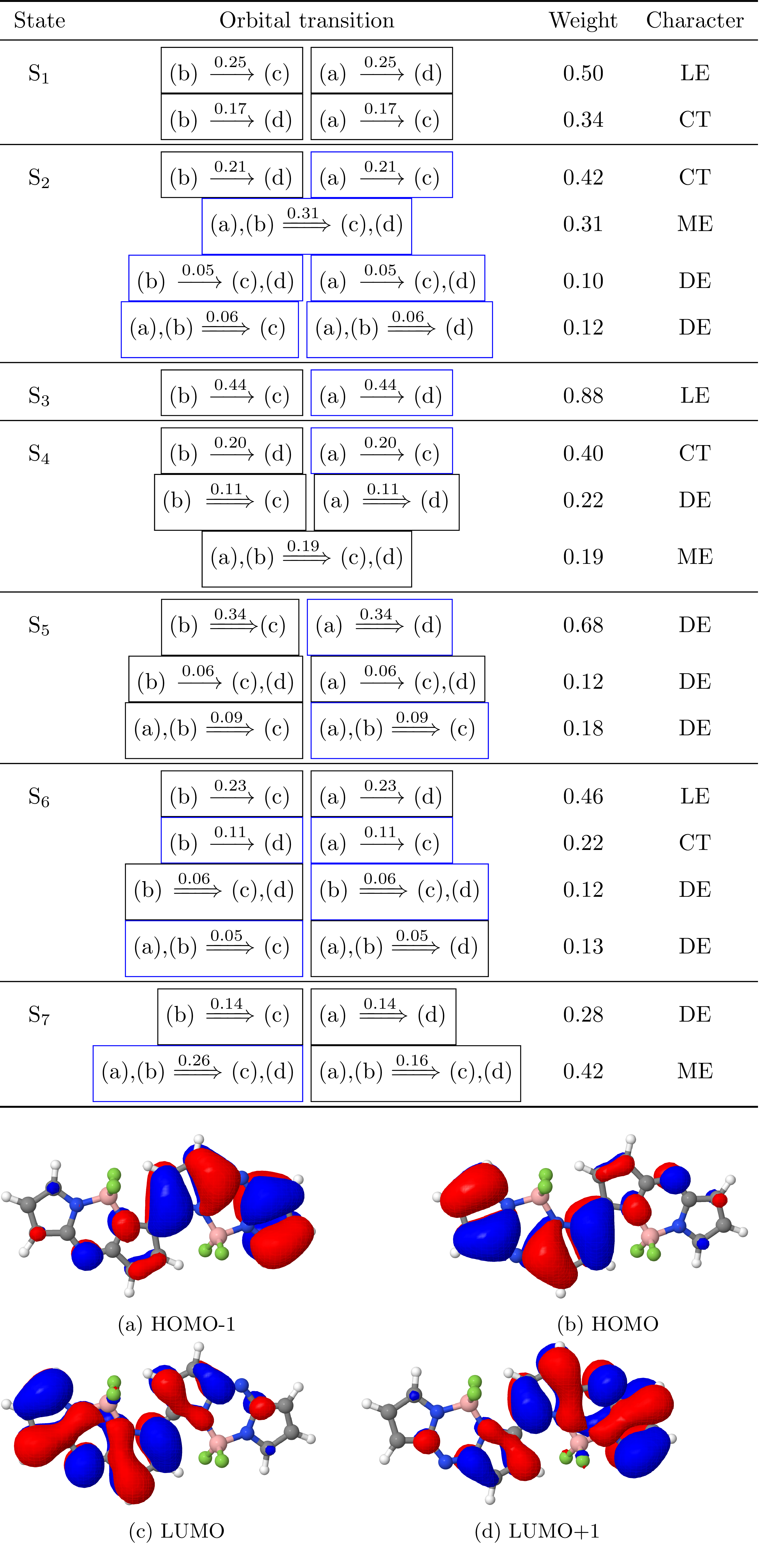}
		\label{fig:dimer33-localized}	
	\end{figure}

\begin{figure}[H]
		\caption{The characterization of the excited states of dimer D[2,2] by using the localised HOMO-1 (a), HOMO (b), LUMO (c) and LUMO+1 (d) orbitals. Single and double arrows represents single and double excitations. The value on top of the arrow represent the \% of the particular excitation to the total wavefunction. The black and blue boxes represent the constructive and distructive combinations.}
		\includegraphics[width=0.7\textwidth]{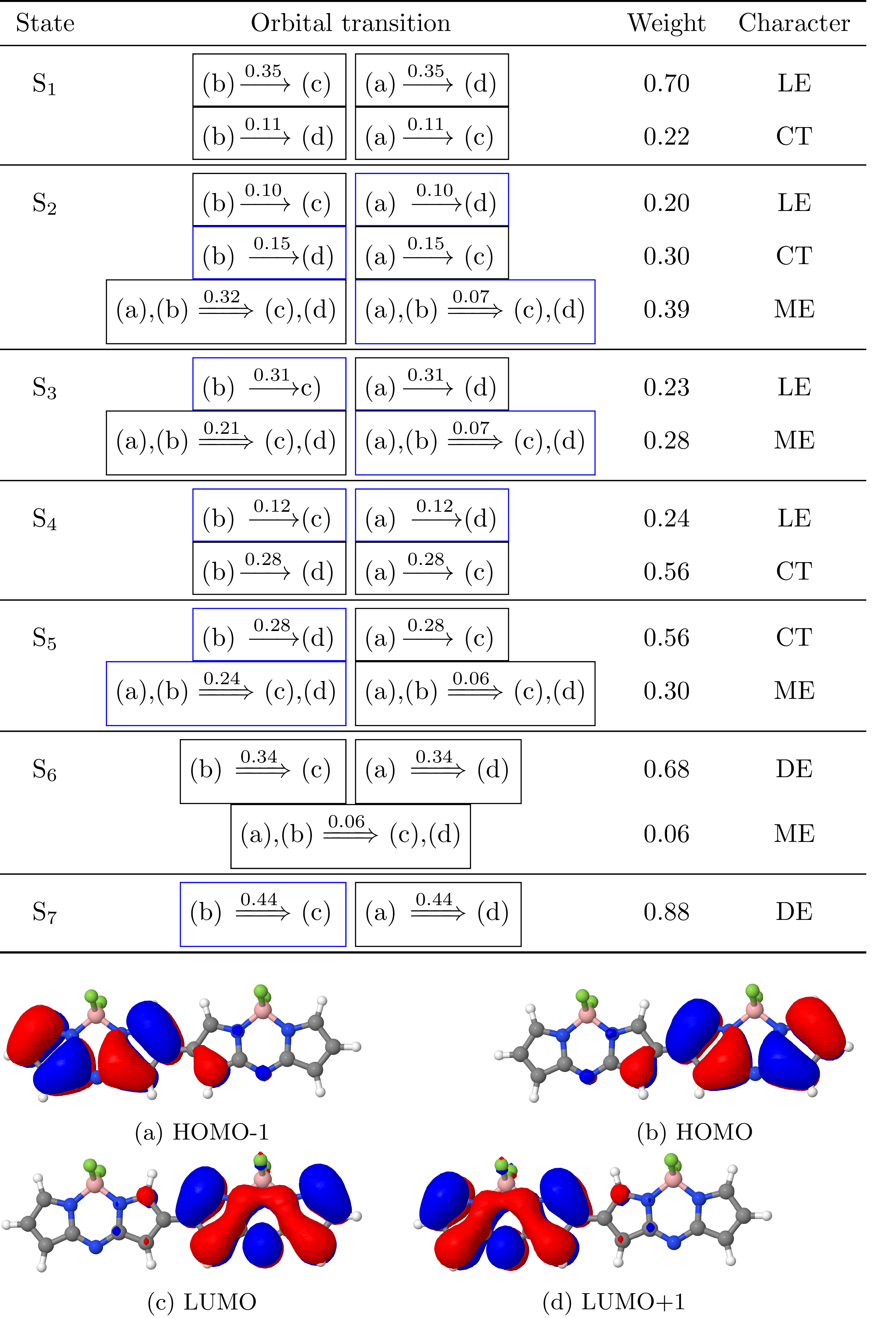}
		\label{fig:dimer22-localized}	
	\end{figure}

The excitation energies ($\Delta$E, in eV), oscillator strength ($f$, dimensionless) and dipolemoment ($\mu$, in Debye) for dimers D[1,1], D[1,3], D[3,3] and D[2,2] at SA15-XMCQDPT(8,8)/cc-pVDZ level of theory are provided in Table \ref{tab:vees-regio}. The optimized orbitals are delocalized. To further characterize the electronic states, we performed a localization calculation for the (4,4) active space (see Figs. \ref{fig:dimer11-localized}- \ref{fig:dimer22-localized}). The characterization of the first 7 singlet states for these dimers at 
	SA15-XMCQDPT(8,8)/cc-pVDZ level of theory are shown in Tables \ref{tab:11-excited}-\ref{tab:22-excited}. 
	
	\clearpage
	\begin{table}[H]
		\caption{The excitation energy (in eV units), oscillator strength (in dimensionless) and dipole moment (in Debye units) and electronic configuration of the dimer D[1,1] calculated by using SA15-XMCQDPT(8,8) and SA8-XMCQDPT(4,4) level of theory.}
		\small
		\begin{tabular}{cccccccc}
			\toprule
			\multirow{2}{*}{S. No.} & \multicolumn{4}{c}{SA15-XPT(8,8)}  & \multicolumn{2}{c}{SA8-XPT(4,4)} & \multirow{2}{*}{Char.$^a$}  \\			
			& $\Delta E$ &  $f$   & $\mu$ & config$^b$  & $\Delta E$ & config$^b$ & \\
			\toprule
			\multirow{3}{*}{S$_\text{0}$} & \multirow{3}{*}{0.00} & \multirow{3}{*}{0.000} & \multirow{3}{*}{0.00} & 22220000 & \multirow{3}{*}{0.00} & 2200 & \multirow{3}{*}{GS} \\
			& & & & 22202000  & & 2020 & \\
			& & & & 22+-+-00 & & +-+- & \\
			\midrule
			\multirow{4}{*}{S$_\text{1}$} & \multirow{4}{*}{1.85} & \multirow{4}{*}{$<$0.001} & \multirow{4}{*}{0.88} & 22+2-000 &  \multirow{4}{*}{1.98} & 2+0- & \multirow{4}{*}{$\textrm{LE}_\textrm{1}$} \\
			&  &    &  & 22202000    &  & 2020  & \\
			&  &    &  & 222+0-00    &  & +2-0  & \\
			&  &    &  & 22200200    &  & 0220  & \\
			\midrule
			\multirow{2}{*}{S$_\text{2}$}    & \multirow{2}{*}{1.90} & \multirow{2}{*}{1.021} & \multirow{2}{*}{5.26} & 222+-000 &   \multirow{2}{*}{1.88} & 2+-0 & \multirow{2}{*}{$\textrm{CT}_\textrm{1}$$^*$} \\
			&  &    &  & 22+-2000   &   & +-20 & \\ 
			\midrule
			\multirow{3}{*}{S$_\text{3}$}    & \multirow{3}{*}{2.41} & \multirow{3}{*}{0.224} & \multirow{3}{*}{2.70} & 22+2-000 &  \multirow{3}{*}{2.41} & +2-0  &  \multirow{3}{*}{ $\textrm{LE}_\textrm{1}$} \\
			&   &    &  & 222+0-00    &  & 2+0- & \\
			&   &    &  & 22++--00    &  &  & \\
			\midrule
			\multirow{3}{*}{S$_\text{4}$} & \multirow{3}{*}{2.47} & \multirow{3}{*}{0.164} & \multirow{3}{*}{3.68} & 2+22-000  &  &  & \multirow{3}{*}{$\textrm{LE}_\textrm{2}$} \\
			&  &   &  &  2+220-00  &  &  & \\
			&  &   &  &  2+220-00  &  &      & \\
			\midrule
			\multirow{2}{*}{S$_\text{5}$} & \multirow{2}{*}{2.81} & \multirow{2}{*}{0.018} & \multirow{2}{*}{4.01} & 22+20-00 &  \multirow{2}{*}{2.78} & +20- & \multirow{2}{*}{ $\textrm{CT}_\textrm{1}$$^*$} \\
			&   &   &   & 22+02-00 &  & +-20  &  \\
			\midrule
			\multirow{4}{*}{S$_\text{6}$} & \multirow{4}{*}{2.87} & \multirow{4}{*}{0.169} & \multirow{4}{*}{2.68} & 22022000  & \multirow{4}{*}{2.85} & 2+0- & \multirow{4}{*}{ME}\\
			&  &  &  & 22020200 &  & 0220 & \\
			&  &  &  & 222+0-00 &  & 0202 & \\
			&  &  &  & 22++--00 &  & ++-- & \\		
			\midrule
			\multirow{2}{*}{S$_\text{7}$} & \multirow{2}{*}{3.04} & \multirow{2}{*}{0.424} & \multirow{2}{*}{2.82} & +222-000 &  & & \multirow{2}{*}{ $\textrm{LE}_2$}\\
			&  &  &  & +22-2000   &  &  & \\
			\bottomrule
		\end{tabular}\\
		\small $^a$ In the localised orbitals. Please see Fig. \ref{fig:dimer11-localized} $^b$ In the delocalised orbitals
		\label{tab:11-excited}
	\end{table}
	
	\begin{table}[H]
		\caption{The excitation energy (in eV units), oscillator strength (in dimensionless) and dipole moment (in Debye units) and electronic configuration of the dimer D[1,3] calculated by using SA15-XMCQDPT(8,8) and SA8-XMCQDPT(4,4) level of theory.}
		\small
		\begin{tabular}{cccccccc}
			\toprule
			\multirow{2}{*}{S. No.} & \multicolumn{4}{c}{SA15-XPT(8,8)} &  \multicolumn{2}{c}{SA8-XPT(4,4)} & \multirow{2}{*}{Char.$^a$}   \\
			& $\Delta E$ & $f$ & $\mu$ & config$^b$ & $\Delta E$ & config$^b$  &  \\
			\toprule	
			\multirow{3}{*}{S$_\text{0}$} & \multirow{3}{*}{0.00} & \multirow{3}{*}{0.00} & \multirow{3}{*}{4.46} & 22220000 &  \multirow{3}{*}{0.00} & \multirow{3}{*}{2200} &\multirow{3}{*}{GS}\\
			& & & & 222+-000 & & &\\
			\midrule
			\multirow{4}{*}{S$_\text{1}$} & \multirow{4}{*}{1.63} & \multirow{4}{*}{0.010} & \multirow{4}{*}{4.88} & 222+-000 &  \multirow{4}{*}{1.66} & 2+-0 &\multirow{4}{*}{ $\textrm{CT}_\textrm{1}$$^*$}\\
			& & & & 22+2-000 & & +2-0 &\\
			& & & & 222+0-00  & & 2+0- &\\
			& & & & 22202000  & & 2020 &\\
			\midrule
			\multirow{5}{*}{S$_\text{2}$} & \multirow{5}{*}{1.81} & \multirow{5}{*}{0.573} & \multirow{5}{*}{5.57} & 222+-000 &  \multirow{5}{*}{1.87} & 2+-0 &\multirow{5}{*}{ $\textrm{CT}_\textrm{1}$}\\		
			& & & & 22+2-000 & & 2+0- &\\
			& & & & 222+0-00 & & 2020 &\\
			& & & & 22202000 & & +2-0 &\\
			& & & & 22+-2000 & &      &\\
			\midrule
			\multirow{4}{*}{S$_\text{3}$} & \multirow{4}{*}{2.40} & \multirow{4}{*}{0.054} & \multirow{4}{*}{4.17} & 222+0-00 &  \multirow{4}{*}{2.45} & 2+0- & \multirow{4}{*}{ $\textrm{LE}_1$} \\		
			& & & & 22+2-000  & & +2-0 & \\		
			& & & & 22202000  & & 0+2- & \\		
			& & & & 220+2-00  & &      & \\			
			\midrule
			\multirow{3}{*}{S$_\text{4}$} & \multirow{3}{*}{2.72} & \multirow{3}{*}{0.267} & \multirow{3}{*}{4.93} & 22+20-00 &  \multirow{3}{*}{2.66} & +20- & \multirow{3}{*}{ $\textrm{LE}_\textrm{1}$$^*$} \\
			& & & & 22+-2000 & & +-20 & \\
			& & & & 22++--00 & & ++-- & \\
			\midrule		
			\multirow{5}{*}{S$_\text{5}$} & \multirow{5}{*}{2.89} & \multirow{5}{*}{0.052} & \multirow{5}{*}{6.55} & 2+22-000 &  & &\multirow{5}{*}{ $\textrm{LE}_2$} \\		
			& & & & 22+-2000  & & & \\
			& & & & 2+2-2000  & & & \\
			\midrule
			\multirow{3}{*}{S$_\text{6}$} & \multirow{3}{*}{3.11} & \multirow{3}{*}{0.090} & \multirow{3}{*}{5.36} & +222-000 &  & &\multirow{3}{*}{ $\textrm{LE}_\textrm{2}$} \\		
			& & & & 222+0-00 &  & & \\
			& & & & 2+2-2000 &  & & \\
			\midrule
			\multirow{8}{*}{S$_\text{7}$} & \multirow{8}{*}{3.32} & \multirow{8}{*}{0.032} & \multirow{8}{*}{1.56} & 22+-+-00 &  \multirow{8}{*}{3.19} & 2+0- & \multirow{8}{*}{ME$^*$} \\		
			& & & & 22+20-00 & & 2020 & \\
			& & & & 222+0-00 & & +2-0 & \\
			& & & & 22202000 & & +20- & \\
			& & & & 2220+-00 & & +-+- & \\
			& & & & 22+2-000 & & 0220 & \\
			& & & & 22022000 & &      & \\
			& & & & 22020200 & &      & \\
			\bottomrule		
		\end{tabular}\\
		\small $^a$ In the localised orbitals. Please see Fig. \ref{fig:dimer13-localized}  $^b$ In the delocalised orbitals $^*$Mixed-state character(LE, CT and ME) consistent with super-exchange mechanism
		\label{tab:13-excited}
	\end{table}

	\begin{table}[H]
		\caption{The excitation energy (in eV units), oscillator strength (in dimensionless) and dipole moment (in Debye units) and electronic configuration of the dimer D[3,3] calculated by using SA15-XMCQDPT(8,8) and SA8-XMCQDPT(4,4) level of theory.}
		\small
		\begin{tabular}{cccccccc}
			\toprule
			\multirow{2}{*}{S. No.} & \multicolumn{4}{c}{SA15-XPT(8,8)}  & \multicolumn{2}{c}{SA8-XPT(4,4)} & \multirow{2}{*}{Char.} \\
			& $\Delta E$ &  $f$   & $\mu$ & config$^b$ & $\Delta E$ & config$^b$ & \\
			\toprule
			\multirow{2}{*}{S$_\text{0}$}&\multirow{2}{*}{0.00} &\multirow{2}{*}{0.000}& \multirow{2}{*}{1.25} & 22220000 &  \multirow{2}{*}{0.00} & 2200&\multirow{2}{*}{ GS} \\
			&&&&22202000&&2020&\\
			\midrule
			\multirow{4}{*}{S$_\text{1}$} & \multirow{4}{*}{1.54} & \multirow{4}{*}{$<$0.001} & \multirow{4}{*}{1.32}  & 22202000 &  \multirow{4}{*}{1.69} & 2020 & \multirow{4}{*}{ $\textrm{LE}_1$} \\
			&&&&22+2-000&&+2-0 &\\
			&  & &  & 222+0-00&  & 2+0-& \\
			&&&&22022000&&&\\
			\midrule
			\multirow{3}{*}{S$_\text{2}$}& \multirow{3}{*}{1.73} & \multirow{3}{*}{1.066}&\multirow{3}{*}{0.07}  & 222+-000&\multirow{3}{*}{1.64} &2+-0&\multirow{3}{*}{ $\textrm{CT}_1$$^*$}\\
			&&&&22+-2000&&+-20& \\
			&&&&&&20+-& \\
			\midrule
			\multirow{6}{*}{S$_\text{3}$} & \multirow{6}{*}{2.71} & \multirow{6}{*}{$<$0.001} & \multirow{6}{*}{0.66} & 222+0-00 &  \multirow{6}{*}{2.63} & 2+0- & \multirow{6}{*}{ $\textrm{LE}_1$}\\
			&&&&+222-000&&+2-0&\\
			&&&&22+2-000&&0+2-&\\
			&&&&&&+0-2&\\
		
			\midrule
			\multirow{5}{*}{S$_\text{4}$} & \multirow{5}{*}{3.03} & \multirow{5}{*}{0.466} &\multirow{5}{*}{0.75} & 2+22-000 & &  & \multirow{5}{*}{ $\textrm{LE}_2$}\\
			&&&&2+2-+-00&&&\\
			&&&&+2220-00&&&\\
			&&&&22+-2000&&&\\
		
			\midrule
			\multirow{5}{*}{S$_\text{5}$} & \multirow{5}{*}{3.06} & \multirow{5}{*}{0.065} & \multirow{5}{*}{3.24} & +222-000 &  & & \multirow{5}{*}{ $\textrm{LE}_2$} \\
			&&&&222+0-00&&&\\
			&&&&+22-+-00&&&\\
			&&&&22+2-000&&&\\
		
			\midrule
			\multirow{7}{*}{S$_\text{6}$} & \multirow{7}{*}{3.27} & \multirow{7}{*}{0.038} & \multirow{7}{*}{1.03} & 22202000 &  \multirow{7}{*}{3.23} & 2020 & \multirow{7}{*}{ $\textrm{CT}_1$}\\
			&&&&22+-+-00&&+-+-&\\
			&&&&22+2-000&&+2-0&\\
			&&&&22220000&&2200&\\
			&&&&222+0-00&&2+0-&\\
			&&&&&&0220&\\
			&&&&&&0202&\\
			\midrule
			\multirow{4}{*}{S$_\text{7}$}&\multirow{4}{*}{3.60}&\multirow{4}{*}{0.170}&\multirow{4}{*}{0.86}&2220+-00&\multirow{4}{*}{3.37}&20+-&\multirow{4}{*}{ ME}\\
			&&&&22+20-00&&+20-&\\
			&&&&22+02-00&&+02-&\\
			&&&&&&+-20&\\
			\bottomrule
		\end{tabular}\\
		\small $^a$ In the localised orbitals. Please see Fig. \ref{fig:dimer33-localized} $^b$ In the delocalised orbitals
		\label{tab:33-excited}
	\end{table}
	
	\begin{table}[H]
		\caption{The excitation energy (in eV units), oscillator strength (in dimensionless) and dipole moment (in Debye units) and electronic configuration of the dimer D[2,2] calculated by using SA15-XMCQDPT(8,8) and SA8-XMCQDPT(4,4) level of theory.}
		\small
		\begin{tabular}{cccccccc}
			\toprule
			\multirow{2}{*}{S. No.} & \multicolumn{4}{c}{SA15-XPT(8,8)} & \multicolumn{2}{c}{SA8-XPT(4,4)}& \multirow{2}{*}{Char.}  \\			
			& $\Delta E$ &  $f$   & $\mu$ & config$^b$ &  $\Delta E$ & config$^b$& \\
			\toprule
			S$_\text{0}$&0.000 & 0.00&4.97  & 22220000  & 0.00 & 2200&GS\\
			\midrule
			\multirow{3}{*}{S$_\text{1}$} & \multirow{3}{*}{1.76} &\multirow{3}{*}{0.756}  &\multirow{3}{*}{5.91}  & 222+-000&  \multirow{3}{*}{1.78}&2+-0&\multirow{3}{*}{ $\textrm{LE}_1$}\\
			&&&&222+0-00  &  & +20- &\\
			\midrule
			\multirow{5}{*}{S$_\text{2}$}&\multirow{5}{*}{2.10} &\multirow{5}{*}{0.617} &\multirow{5}{*}{9.09}  & 222+0-00& \multirow{5}{*}{2.24}&2+0-&\multirow{5}{*}{ $\textrm{ME}$$^*$}\\
			&&&&22202000&&2020 &\\
			&&&&2+220-00&&2002&\\
			&&&&&&2002&\\
			&&&&&&++--&\\
			\midrule
			\multirow{5}{*}{S$_\text{3}$} & \multirow{5}{*}{2.40} &\multirow{5}{*}{0.001} &\multirow{5}{*}{3.75}  & 22+2-000&\multirow{5}{*}{2.42}  & +2-0 &\multirow{5}{*}{$\textrm{LE}_{1}$$^*$}\\
			&&&&2220+-00&&20+-&\\
			&&&&22++--00&&++--&\\
			&&&&22200200&&2002&\\
			&&&&222+0-00&&2+0-&\\
			\midrule
			\multirow{5}{*}{S$_\text{4}$} & \multirow{5}{*}{2.72} & \multirow{5}{*}{0.826} & \multirow{5}{*}{5.61} & 2+22-000 &  \multirow{5}{*}{2.87} &  +20- & \multirow{5}{*}{ $\textrm{CT}_1$}\\
			&&&&22+20-00&&20+-&\\
			&&&&22+2-000&&2+-0&\\
			&&&&&&+-02&\\
			
			\midrule
			\multirow{4}{*}{S$_\text{5}$}&\multirow{4}{*}{2.82}&\multirow{4}{*}{0.092}&\multirow{4}{*}{8.35}&22+2-000&\multirow{4}{*}{3.00}&+2-0&\multirow{4}{*}{ $\textrm{CT}_1$}\\
			&&&&222+0-00&&2+0-&\\
			&&&&2+220-00&&++--&\\
			&&&&22020200&&0202&\\
			\midrule
			\multirow{6}{*}{S$_\text{6}$}&\multirow{6}{*}{2.97}&\multirow{6}{*}{0.070}&\multirow{6}{*}{5.08}&2+22-000&&&\multirow{6}{*}{ $\textrm{LE}_2$}\\
			&&&&222+-000&&&\\
			&&&&22+-2000&&&\\
			&&&&222+0-00&&&\\
			&&&&22+2-000&&&\\
			&&&&2+-22000&&&\\
			\midrule
			\multirow{6}{*}{S$_\text{7}$}&\multirow{6}{*}{3.12}&\multirow{6}{*}{0.045}&\multirow{6}{*}{6.78}&2+220-00&&&\multirow{6}{*}{ $\textrm{LE}_2$}\\
			&&&&22+20-00&&&\\
			&&&&22+-+-00&&&\\
			&&&&222+0-00&&&\\
			&&&&2+2-+-00&&&\\
			&&&&2+2-2000&&&\\
			\bottomrule
		\end{tabular}\\
		\small $^a$ In the localised orbitals. Please see Fig. \ref{fig:dimer22-localized} $^b$In the delocalised orbitals
		\label{tab:22-excited}
	\end{table}
	
	\clearpage
	\begin{table}[!ht]
		\caption{The excitation energy (in eV units), oscillator strength (in dimensionless) and dipole moment (in Debye units) and electronic configuration of the orthogonal dimer D[1,1] calculated by using SA15-XMCQDPT(8,8) and SA8-XMCQDPT(4,4) level of theory.}
		\small
		\begin{tabular}{cccccccc}
			\toprule
			\multirow{2}{*}{S. No.} & \multicolumn{4}{c}{SA15-XPT(8,8)}  & \multicolumn{2}{c}{SA8-XPT(4,4)} & \multirow{2}{*}{Char.$^a$}  \\			
			& $\Delta E$ &  $f$   & $\mu$ & config$^b$  & $\Delta E$ & config$^b$ & \\
			\toprule
			S$_\text{0}$ & 0.00 & 0.000 & 1.77 & 22220000 & 0.00 & 2200 & GS \\
			\midrule
			\multirow{2}{*}{S$_\text{1}$} & \multirow{2}{*}{2.069} & \multirow{2}{*}{0.742} & \multirow{2}{*}{1.07} & 222+-000 & \multirow{2}{*}{2.058} & 2+-0 & \multirow{2}{*}{LE$_1$} \\
			& & & & 22+20-00 & & +20- & \\
			\midrule
			\multirow{2}{*}{S$_\text{2}$} & \multirow{2}{*}{2.298} & \multirow{2}{*}{1.035} & \multirow{2}{*}{0.42} & 222+0-00 & \multirow{2}{*}{2.328} & 2+0- & \multirow{2}{*}{LE$_1$} \\
			& & & & 22+2-000 & & +2-0 & \\
			\midrule
			
			\multirow{3}{*}{S$_\text{3}$} & \multirow{3}{*}{2.481} & \multirow{3}{*}{0.000} & \multirow{3}{*}{1.88} & 22202000 & \multirow{3}{*}{2.566} & 2020 & \multirow{3}{*}{ME} \\
			& & & & 22200200 & & 2002 & \\
			& & & & 22++--00 & & ++-- & \\
			& & & & 22022000 & & 0220 & \\
			& & & & 22020200 & & 0202 & \\
			\midrule
			
			\multirow{2}{*}{S$_\text{4}$} & \multirow{2}{*}{2.883} & \multirow{2}{*}{0.352} & \multirow{2}{*}{1.86} & 22+20-00 & \multirow{2}{*}{2.793} & +20- & \multirow{2}{*}{CT$_1$} \\
			& & & & 222+-000 & & 2+-0 & \\
			\midrule
			
			\multirow{2}{*}{S$_\text{5}$} & \multirow{2}{*}{2.885} & \multirow{2}{*}{0.004} & \multirow{2}{*}{2.12} & 22+2-000 & \multirow{2}{*}{2.799} & +2-0 & \multirow{2}{*}{CT$_1$} \\
			& & & & 222+0-00 & & 2+0- & \\
			\midrule
			
			\multirow{2}{*}{S$_\text{6}$} & \multirow{2}{*}{3.012} & \multirow{2}{*}{0.121} & \multirow{2}{*}{0.31} & 2+220-00 & \multirow{2}{*}{--} & +222-000 & \multirow{2}{*}{LE$_2$} \\
			& & & & +222-000 & &  & \\
			\midrule
			\multirow{2}{*}{S$_\text{7}$} & \multirow{2}{*}{3.034} & \multirow{2}{*}{0.114} & \multirow{2}{*}{0.22} & 2+22-000 & \multirow{2}{*}{--} & +2220-00 & \multirow{2}{*}{LE$_2$}\\
			& & & & +2220-00 & &  & \\
			\bottomrule
		\end{tabular}\\
		\small $^a$ In the localised orbitals. Please see Fig. \ref{fig:dimer11-ortho-localized} $^b$ In the delocalised orbitals
		\label{tab:11-ortho}
	\end{table}
	
		\begin{figure}[!ht]
		\caption{The characterization of the excited states of orthogonal dimer D[1,1] by using the localised HOMO-1 (a), HOMO (b), LUMO (c) and LUMO+1 (d) orbitals. Single and double arrows represents single and double excitations. The value on top of the arrow represent the \% of the particular excitation to the total wavefunction. The black and blue boxes represent the constructive and distructive combinations.}
		\includegraphics[width=0.7\textwidth]{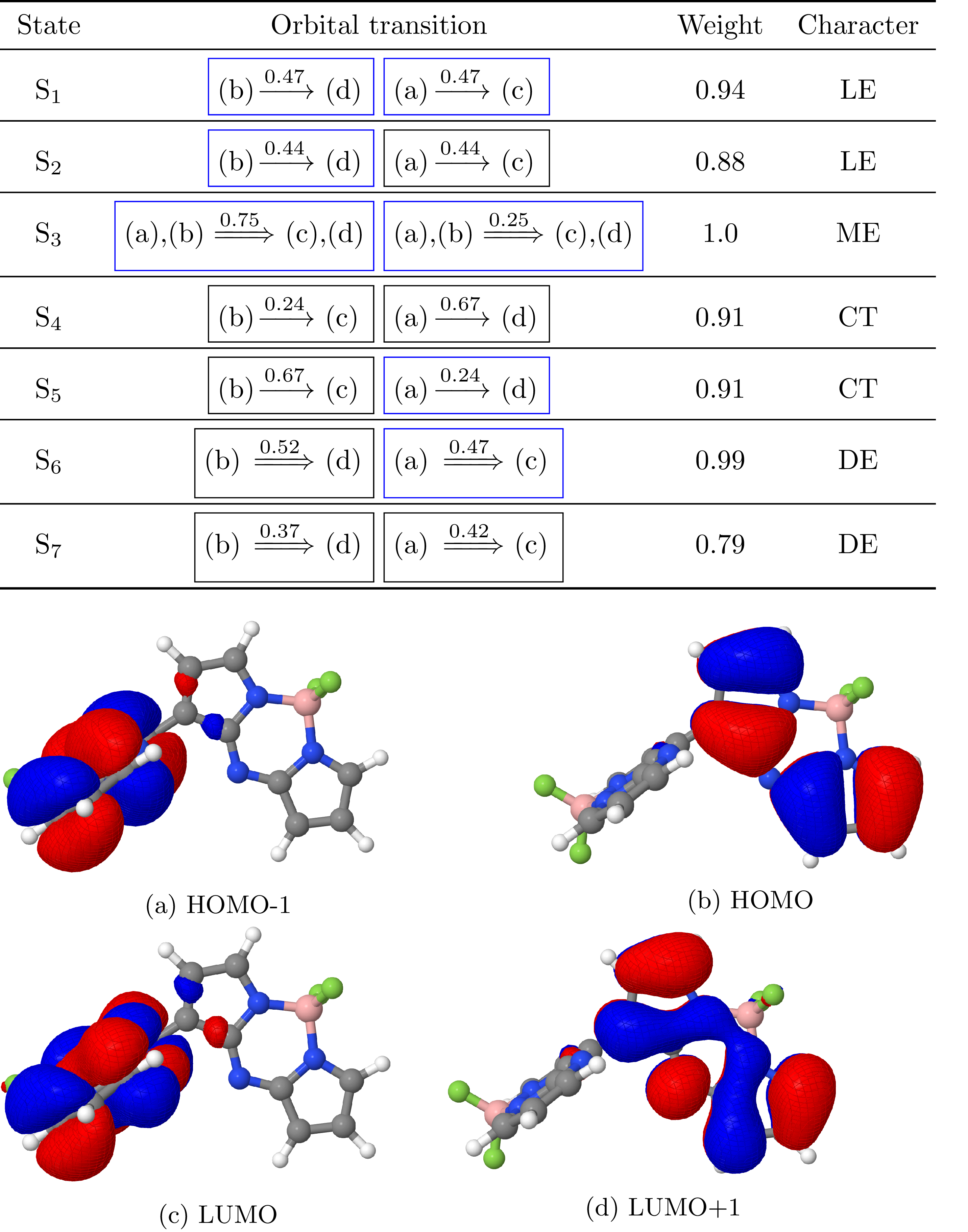}
		\label{fig:dimer11-ortho-localized}	
	\end{figure}
	
	\begin{table}[!ht]
		\caption{The excitation energy (in eV units), oscillator strength (in dimensionless) and dipole moment (in Debye units) and electronic configuration of the orthogonal dimer D[1,3] calculated by using SA15-XMCQDPT(8,8) and SA8-XMCQDPT(4,4) level of theory.}
		\small
		\begin{tabular}{cccccccc}
			\toprule
			\multirow{2}{*}{S. No.} & \multicolumn{4}{c}{SA15-XPT(8,8)}  & \multicolumn{2}{c}{SA8-XPT(4,4)} & \multirow{2}{*}{Char.$^a$}  \\			
			& $\Delta E$ &  $f$   & $\mu$ & config$^b$  & $\Delta E$ & config$^b$ & \\
			\toprule
			S$_\text{0}$ & 0.00 & 0.000 & 4.48 & 22220000 & 0.00 & 2200 & GS \\
			\midrule
			\multirow{2}{*}{S$_\text{1}$} & \multirow{2}{*}{1.97} & \multirow{2}{*}{1.087} & \multirow{2}{*}{5.01} & 222+0-00 & \multirow{2}{*}{1.98} & 2+0- & \multirow{2}{*}{LE$_1$} \\
			& & & & 22+2-000 & & +2-0 & \\
			\midrule
			
			\multirow{4}{*}{S$_\text{2}$} & \multirow{4}{*}{2.28} & \multirow{4}{*}{1.031} & \multirow{4}{*}{6.15} & 222+-000 & \multirow{4}{*}{2.32} & 2+-0 & \multirow{4}{*}{LE$_1$} \\
			& & & & 22+20-00 & & +20- & \\
			& & & & 222+0-00 & & 2+0- & \\
			& & & &  & & +2-0 & \\
			\midrule
			
			\multirow{5}{*}{S$_\text{3}$} & \multirow{5}{*}{2.44} & \multirow{5}{*}{0.000} & \multirow{5}{*}{4.04} & 22+-+-00 & \multirow{5}{*}{2.51} & +-+- & \multirow{5}{*}{ME} \\
			& & & & 22++--00 & & ++-- & \\
			& & & & 2220+-00 & & +-02 & \\
			& & & & 2202+-00 & & +-20 & \\
			& & & & & & 2+-0 & \\
			\midrule
			
			\multirow{2}{*}{S$_\text{4}$} & \multirow{2}{*}{2.63} & \multirow{2}{*}{0.003} & \multirow{2}{*}{22.12} & 222+0-00 & \multirow{2}{*}{2.60} & 2+0- & \multirow{2}{*}{CT$_1$} \\
			& & & & 22+20-00 & & 2+-0  & \\
			& & & &  & &  +-20 & \\
			\midrule
			
			\multirow{2}{*}{S$_\text{5}$} & \multirow{2}{*}{2.73} & \multirow{2}{*}{0.002} & \multirow{2}{*}{16.29} & 22+2-000 & \multirow{2}{*}{2.71} & +2-0 & \multirow{2}{*}{CT$_1$} \\
			& & & & 222+-000 & & +20-   & \\
			& & & &  & &  +-02 & \\
			\midrule
			
			\multirow{2}{*}{S$_\text{6}$} & 	\multirow{2}{*}{3.07} & 	\multirow{2}{*}{0.121} & 	\multirow{2}{*}{5.24} & 2+220-00 &  & & 	\multirow{2}{*}{LE$_2$} \\
			& & & & 2+-20200 & &  & \\
			\midrule
			\multirow{2}{*}{S$_\text{7}$} & 	\multirow{2}{*}{3.09} & 	\multirow{2}{*}{0.111} & 	\multirow{2}{*}{5.30} & +222-000 &  &  & \multirow{2}{*}{LE$_2$} \\
			& & & & +22-2000 & &  & \\
			\bottomrule
		\end{tabular}\\
		\small $^a$ In the localised orbitals. Please see Fig. \ref{fig:dimer13-ortho-localized} $^b$ In the delocalised orbitals
		\label{tab:13-ortho}
	\end{table}

		\begin{figure}[!ht]
		\caption{The characterization of the excited states of orthogonal dimer D[3,3] by using the localised HOMO-1 (a), HOMO (b), LUMO (c) and LUMO+1 (d) orbitals. Single and double arrows represents single and double excitations. The value on top of the arrow represent the \% of the particular excitation to the total wavefunction. The black and blue boxes represent the constructive and distructive combinations.}
		\includegraphics[width=0.7\textwidth]{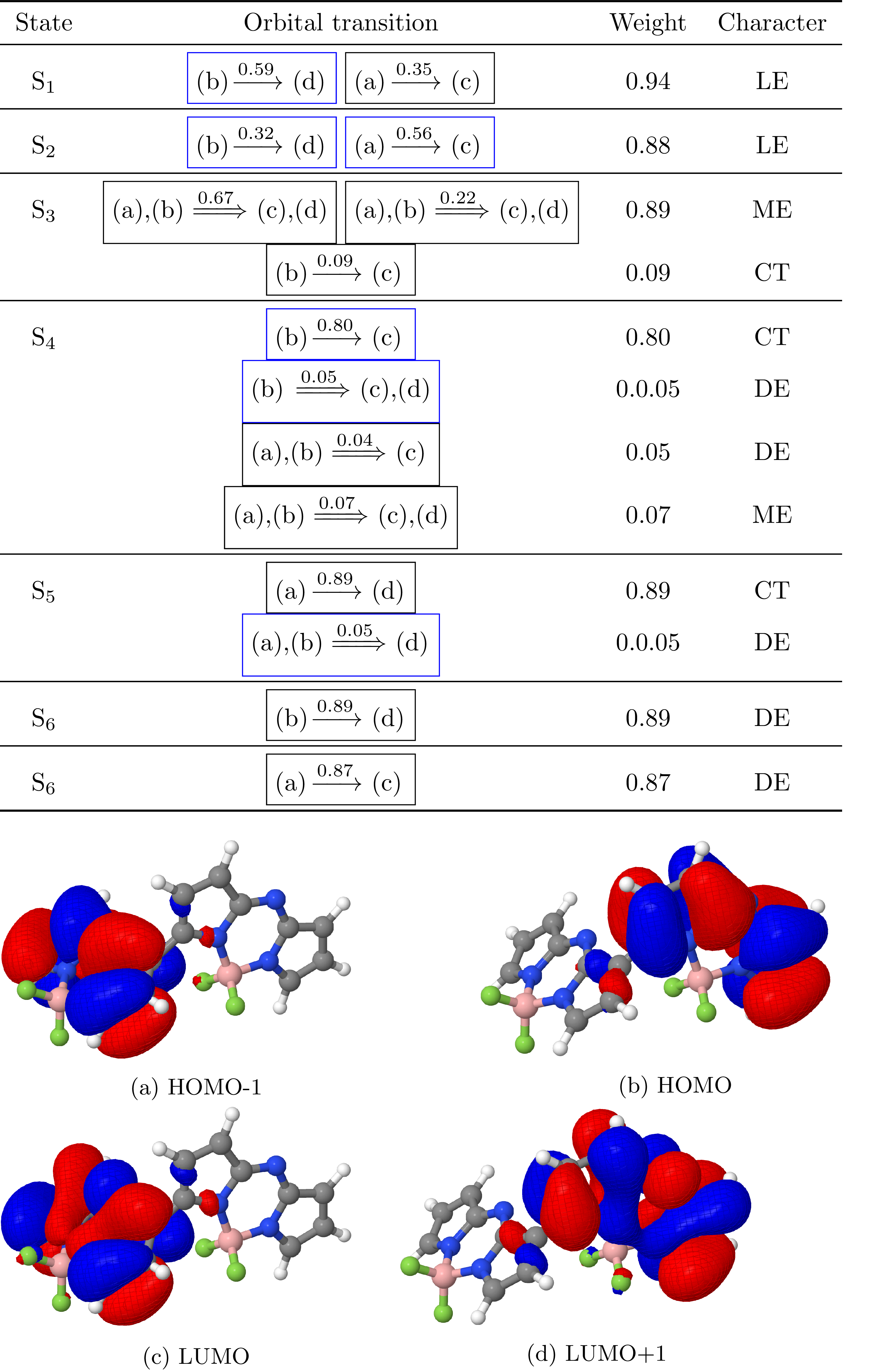}
		\label{fig:dimer13-ortho-localized}	
	\end{figure}

\begin{table}[!ht]
	\caption{The excitation energy (in eV units), oscillator strength (in dimensionless) and dipole moment (in Debye units) and electronic configuration of the orthogonal dimer D[3,3] calculated by using SA15-XMCQDPT(8,8) and SA8-XMCQDPT(4,4) level of theory.}
	\small
	\begin{tabular}{cccccccc}
		\toprule
		\multirow{2}{*}{S. No.} & \multicolumn{4}{c}{SA15-XPT(8,8)}  & \multicolumn{2}{c}{SA8-XPT(4,4)} & \multirow{2}{*}{Char.$^a$}  \\			
		& $\Delta E$ &  $f$   & $\mu$ & config$^b$  & $\Delta E$ & config$^b$ & \\
		\toprule
		S$_\text{0}$ & 0.00 & 0.000 & 1.23 & 22220000 & 0.00 & 2200 & GS \\
		\midrule
		\multirow{2}{*}{S$_\text{1}$} & \multirow{2}{*}{1.95} & \multirow{2}{*}{0.956} & \multirow{2}{*}{1.40} & 222+-000 & \multirow{2}{*}{1.92} & 2+-0 & \multirow{2}{*}{LE$_1$} \\
	& & & & 22+20-00 & & +20- & \\
	\midrule
		\multirow{2}{*}{S$_\text{2}$} & \multirow{2}{*}{2.35} & \multirow{2}{*}{1.027} & \multirow{2}{*}{1.80} & 222+0-00 & \multirow{2}{*}{2.32} & 2+0- & \multirow{2}{*}{LE$_1$} \\
	& & & & 22+2-000 & & +2-0 & \\
	\midrule
		\multirow{5}{*}{S$_\text{3}$} & \multirow{5}{*}{2.41} & \multirow{5}{*}{0.000} & \multirow{5}{*}{1.39} & 22202000 & \multirow{5}{*}{2.48} & 2020 & \multirow{5}{*}{ME} \\
	& & & & 22++--00 & & ++-- & \\
	& & & & 22022000 & & 0220 & \\
	& & & & 22+2-000 & & +2-0 & \\
	& & & & 222+0-00 & & 2+0- & \\
	\midrule
		\multirow{2}{*}{S$_\text{4}$} & \multirow{2}{*}{2.51} & \multirow{2}{*}{0.051} & \multirow{2}{*}{4.20} & 22+20-00 & \multirow{2}{*}{2.54} & +20- & \multirow{2}{*}{CT$_1$} \\
	& & & & 222+-000 & & 2+-0 & \\
	\midrule
		\multirow{3}{*}{S$_\text{5}$} & \multirow{3}{*}{2.53} & \multirow{3}{*}{$<$0.001} & \multirow{3}{*}{3.48} & 22+2-000 & \multirow{3}{*}{2.57} & +2-0 & \multirow{3}{*}{CT$_1$} \\
	& & & & 222+0-00 & & 2+0- & \\
	& & & & 22020200 & & 0202 & \\
	\midrule
		\multirow{3}{*}{S$_\text{6}$} & \multirow{3}{*}{3.13} & \multirow{3}{*}{0.220} & \multirow{3}{*}{0.47} & 2+22-000 &  &  & \multirow{3}{*}{{LE$_2$}} \\
	& & & & +2220-00 & &  & \\
	& & & & 2+220-00 & &  & \\
	\midrule
		\multirow{3}{*}{S$_\text{7}$} & \multirow{3}{*}{3.14} & \multirow{3}{*}{0.012} & \multirow{3}{*}{0.24} & +222-000 &  &  & \multirow{3}{*}{{LE$_2$}} \\
	& & & & 2+220-00 & &  & \\
	& & & & +2220-00 & &  & \\
	\midrule
		\bottomrule
	\end{tabular}\\
	\small $^a$ In the localised orbitals. Please see Fig. \ref{fig:dimer33-ortho-localized} $^b$ In the delocalised orbitals
	\label{tab:33-ortho}
\end{table}

	\begin{figure}[!ht]
	\caption{The characterization of the excited states of orthogonal dimer D[3,3] by using the localised HOMO-1 (a), HOMO (b), LUMO (c) and LUMO+1 (d) orbitals. Single and double arrows represents single and double excitations. The value on top of the arrow represent the \% of the particular excitation to the total wavefunction. The black and blue boxes represent the constructive and distructive combinations.}
	\includegraphics[width=0.7\textwidth]{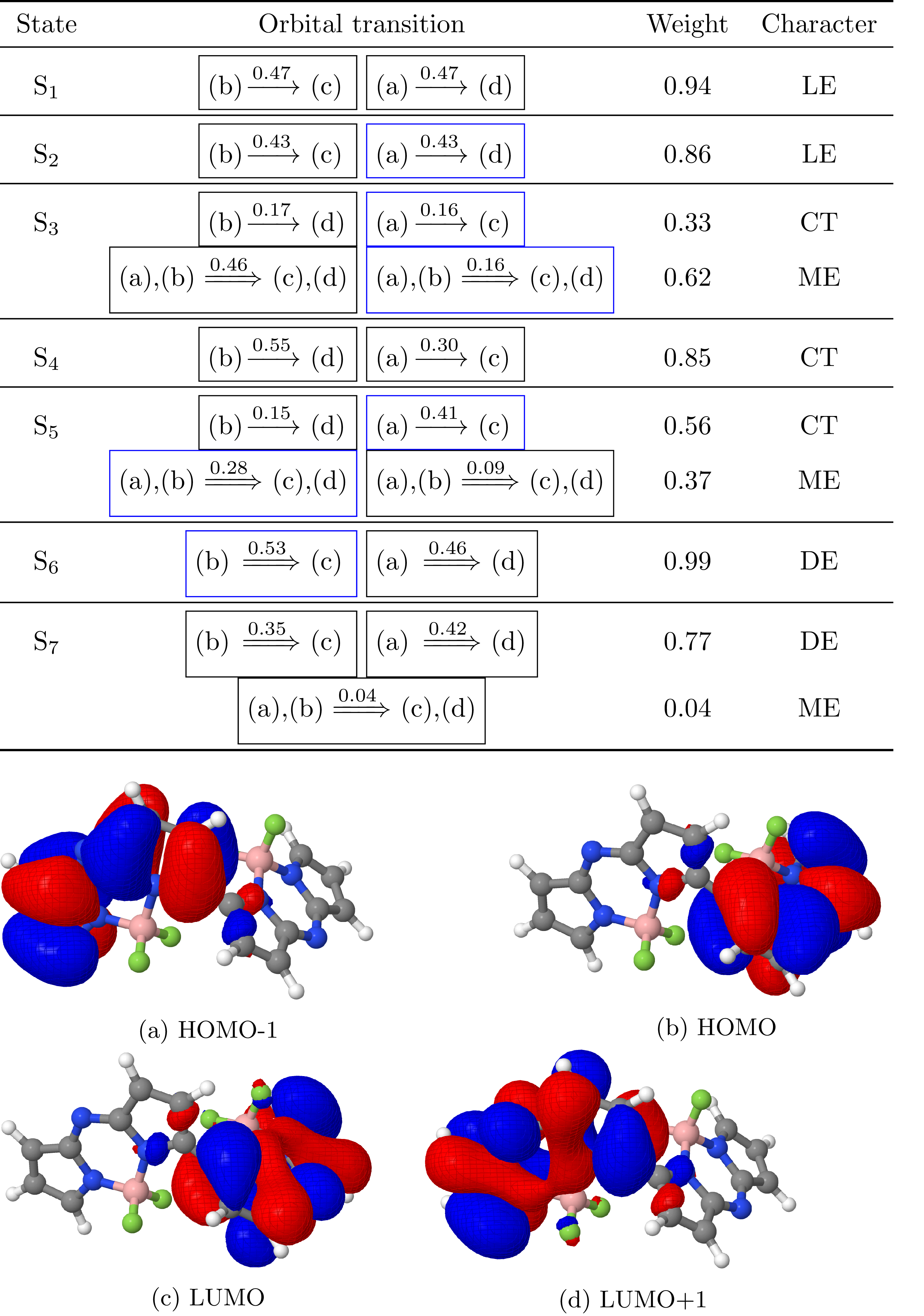}
	\label{fig:dimer33-ortho-localized}	
\end{figure}

\begin{table}[!ht]
	\caption{The excitation energy (in eV units), oscillator strength (in dimensionless) and dipole moment (in Debye units) and electronic configuration of the orthogonal dimer D[2,2] calculated by using SA15-XMCQDPT(8,8) and SA8-XMCQDPT(4,4) level of theory.}
	\small
	\begin{tabular}{cccccccc}
		\toprule
		\multirow{2}{*}{S. No.} & \multicolumn{4}{c}{SA15-XPT(8,8)}  & \multicolumn{2}{c}{SA8-XPT(4,4)} & \multirow{2}{*}{Char.$^a$}  \\			
		& $\Delta E$ &  $f$   & $\mu$ & config$^b$  & $\Delta E$ & config$^b$ & \\
		\toprule
		S$_\text{0}$ & 0.00 & 0.000 & 3.33 & 22220000 & 0.00 & 2200 & GS \\
		\midrule
		\multirow{2}{*}{S$_\text{1}$} & \multirow{2}{*}{2.00} & \multirow{2}{*}{1.884} & \multirow{2}{*}{2.88} & 222+-000 & \multirow{2}{*}{2.01} & 2+-0 & \multirow{2}{*}{LE$_1$} \\
		& & & & 22+20-00 & & +20- & \\
		\midrule
		\multirow{2}{*}{S$_\text{2}$} & \multirow{2}{*}{2.35} & \multirow{2}{*}{$<$0.001} & \multirow{2}{*}{2.77} & 22+2-000 & \multirow{2}{*}{2.38} & +2-0 & \multirow{2}{*}{LE$_1$} \\
	& & & & 222+0-00 & & 2+0- & \\
	\midrule
		\multirow{5}{*}{S$_\text{3}$} & \multirow{5}{*}{2.51} & \multirow{5}{*}{$<$0.001} & \multirow{5}{*}{3.09} & 22++--00 & \multirow{5}{*}{2.58} & ++-- & \multirow{5}{*}{ME} \\
	& & & & 22202000 & & 2020 & \\
	& & & & 22022000 & & 0220 & \\
	& & & & 22200200 & & 2002 & \\
	& & & & 22020200 & & 0202 & \\
	\midrule
		\multirow{2}{*}{S$_\text{4}$} & \multirow{2}{*}{3.05} & \multirow{2}{*}{0.224} & \multirow{2}{*}{2.40} & 22+20-00 & \multirow{2}{*}{3.04} & +20- & \multirow{2}{*}{CT$_1$} \\
	& & & & 222+-000 & & 2+-0 & \\
	\midrule
		\multirow{2}{*}{S$_\text{5}$} & \multirow{2}{*}{3.08} & \multirow{2}{*}{$<$0.001} & \multirow{2}{*}{2.49} & 222+0-00 & \multirow{2}{*}{3.06} & 2+0- & \multirow{2}{*}{CT$_1$} \\
	& & & & 22+2-000 & & +2-0 & \\
	\midrule
	
		\multirow{2}{*}{S$_\text{6}$} & \multirow{2}{*}{3.13} & \multirow{2}{*}{0.035} & \multirow{2}{*}{3.47} & +222-000 &  &  & \multirow{2}{*}{LE$_2$} \\
	& & & & 2+220-00 & &  & \\
	\midrule
		\multirow{3}{*}{S$_\text{7}$} & \multirow{3}{*}{3.14} & \multirow{3}{*}{0.324} & \multirow{3}{*}{3.22} & 2+22-000 &  &  & \multirow{3}{*}{LE$_2$} \\
	& & & & +2220-00 & &  & \\
	& & & & 22+20-00 & &  & \\
		\bottomrule
	\end{tabular}\\
	\small $^a$ In the localised orbitals. Please see Fig. \ref{fig:dimer22-ortho-localized} $^b$ In the delocalised orbitals
	\label{tab:22-ortho}
\end{table}
	
	\begin{figure}[!ht]
	\caption{The characterization of the excited states of orthogonal dimer D[2,2] by using the localised HOMO-1 (a), HOMO (b), LUMO (c) and LUMO+1 (d) orbitals. Single and double arrows represents single and double excitations. The value on top of the arrow represent the \% of the particular excitation to the total wavefunction. The black and blue boxes represent the constructive and distructive combinations.}
	\includegraphics[width=0.7\textwidth]{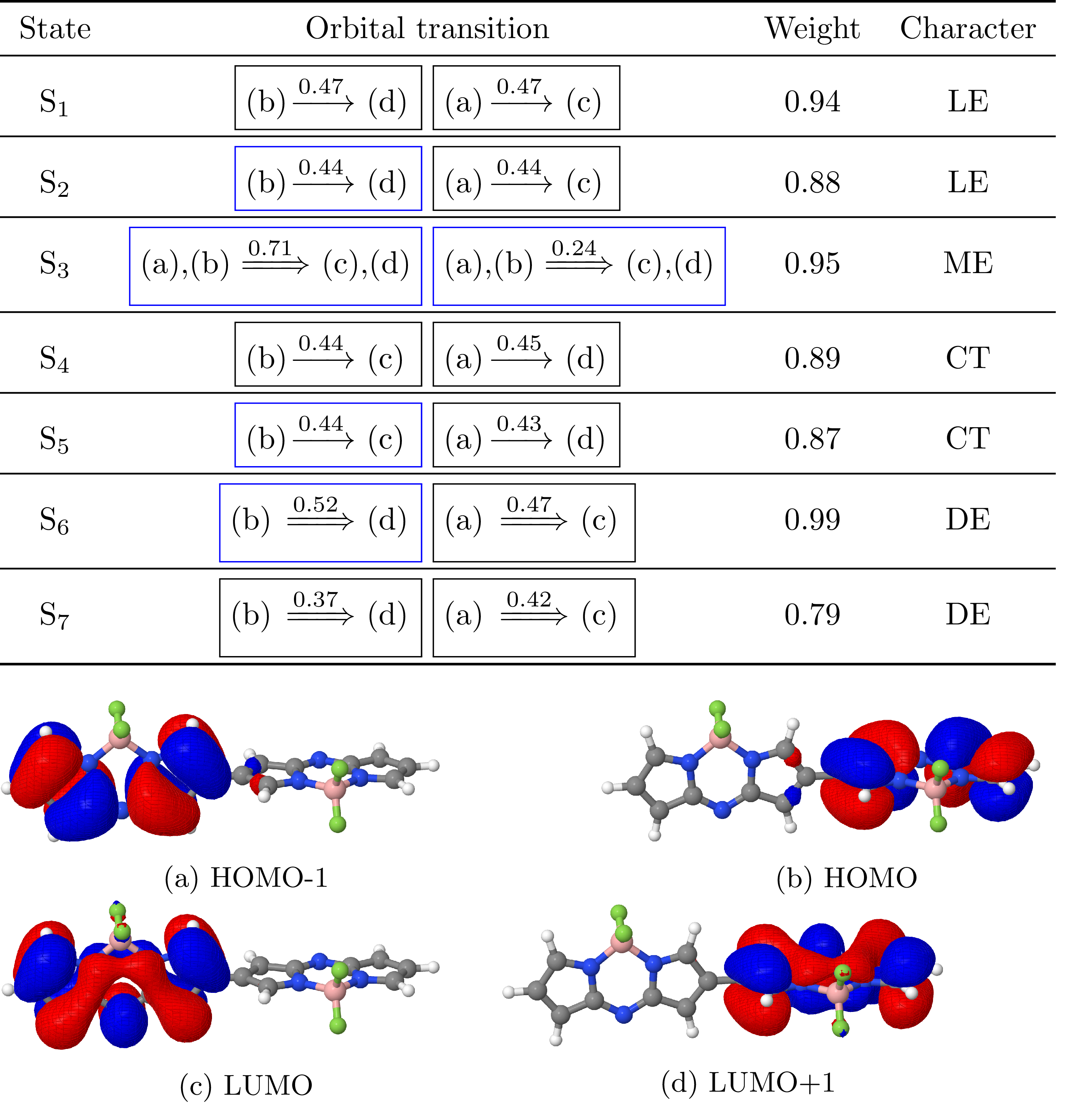}
	\label{fig:dimer22-ortho-localized}	
\end{figure}

\begin{figure}
\caption{The contribution of the diabatic states $\ket{^{1}(\textrm{S}_{0}\textrm{S}_{0})}$, $\ket{^{1}(\textrm{S}_{0}\textrm{S}_{1})}$, $\ket{^{1}(\textrm{S}_{0}\textrm{S}_{1})}$, $\ket{^{1}(\textrm{T}_{1}\textrm{T}_{1})}$, $\ket{^{1}(\textrm{C}\textrm{A})}$, $\ket{^{1}(\textrm{A}\textrm{C})}$, $\ket{^{1}(\textrm{DE})_{1}}$ and $\ket{^{1}(\textrm{DE})_{2}}$ to the adiabatic states (S$_\textrm{1}$ - S$_\textrm{7}$) for dimer D[1,3].}
        \begin{subfigure}{0.45\textwidth}
                \includegraphics[width=0.9\textwidth]{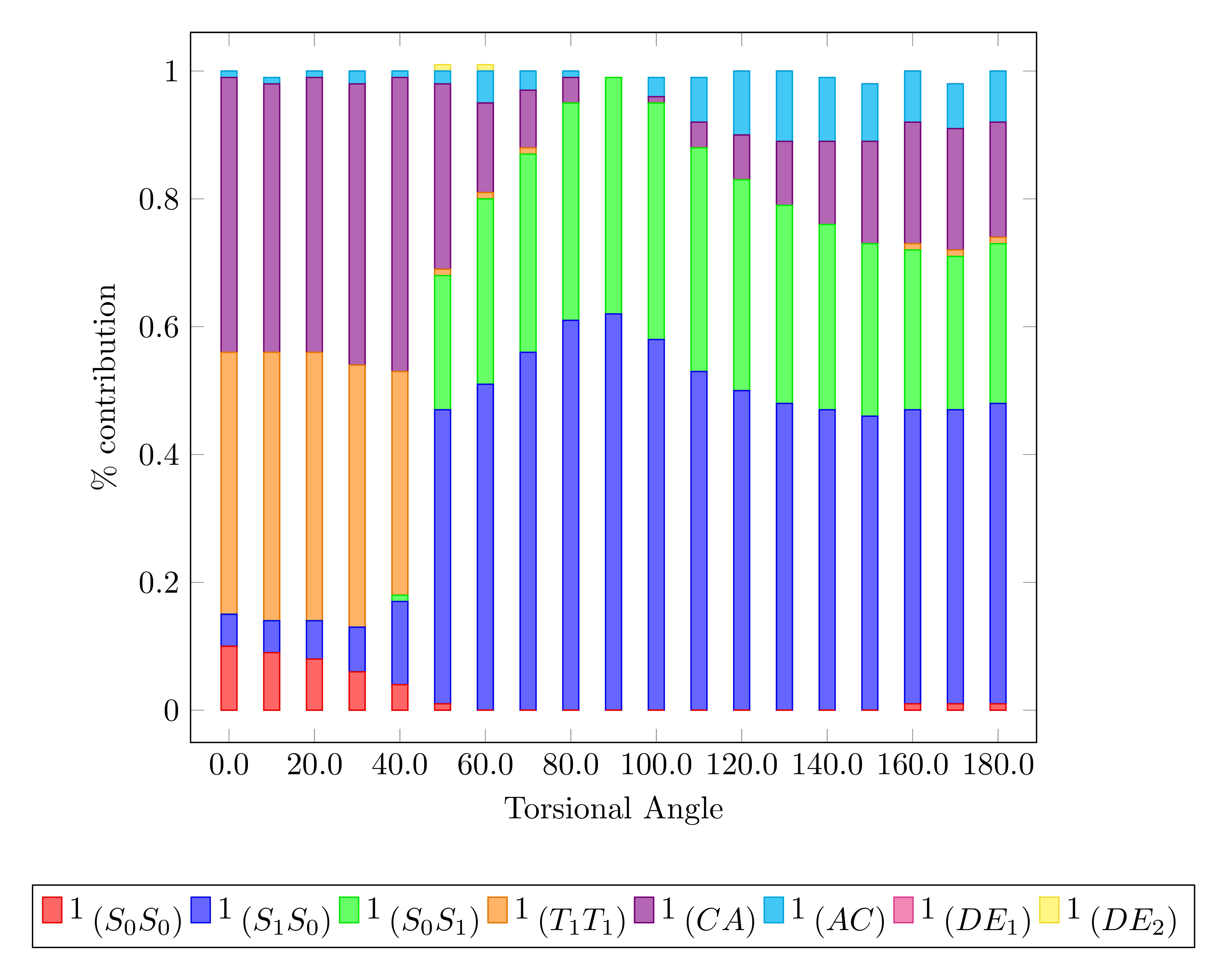}
                \caption{S$_\textrm{1}$}
        \end{subfigure}
        \begin{subfigure}{0.45\textwidth}
                \includegraphics[width=0.9\textwidth]{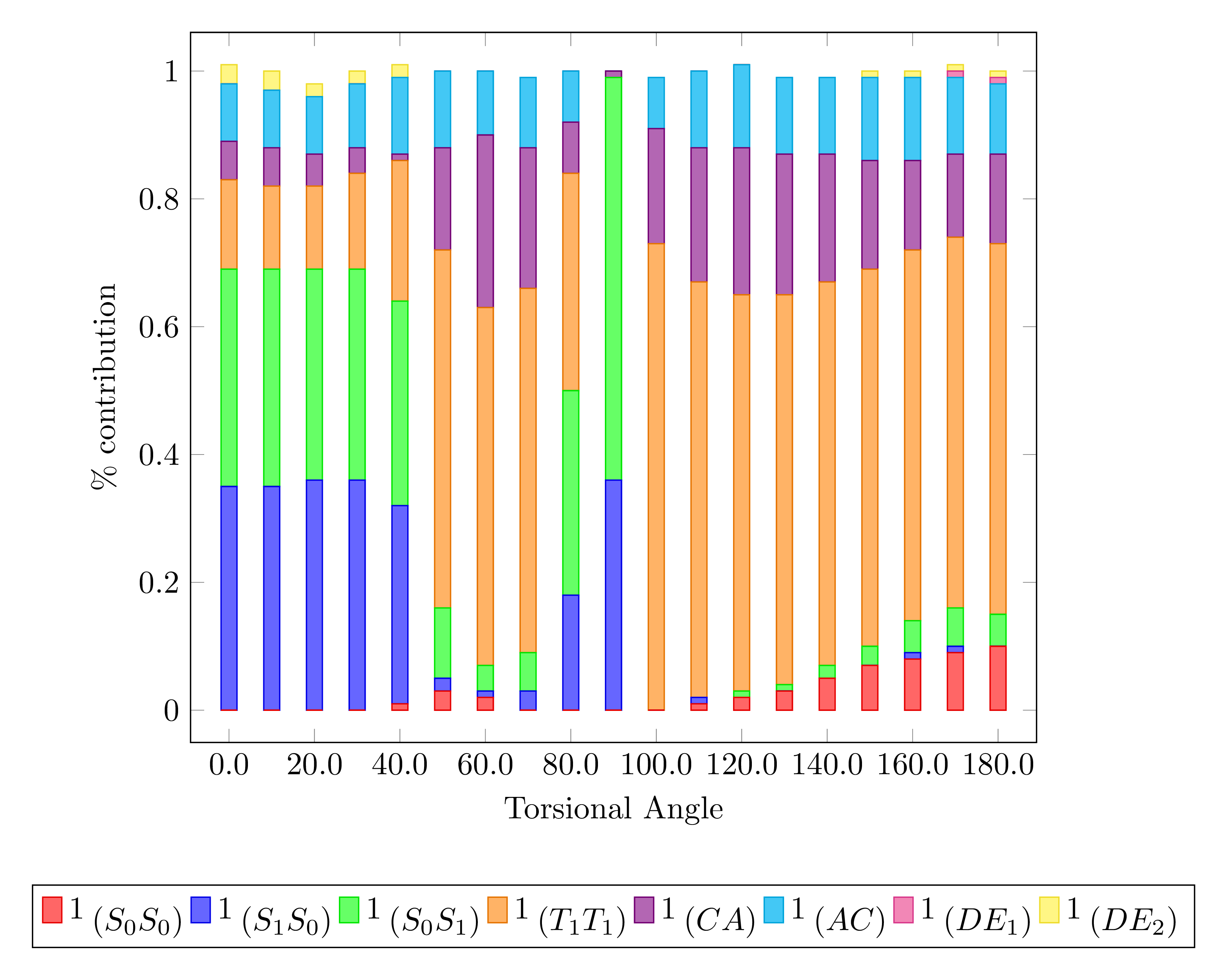}
                \caption{S$_\textrm{2}$}
        \end{subfigure}
        \begin{subfigure}{0.45\textwidth}
                \includegraphics[width=0.9\textwidth]{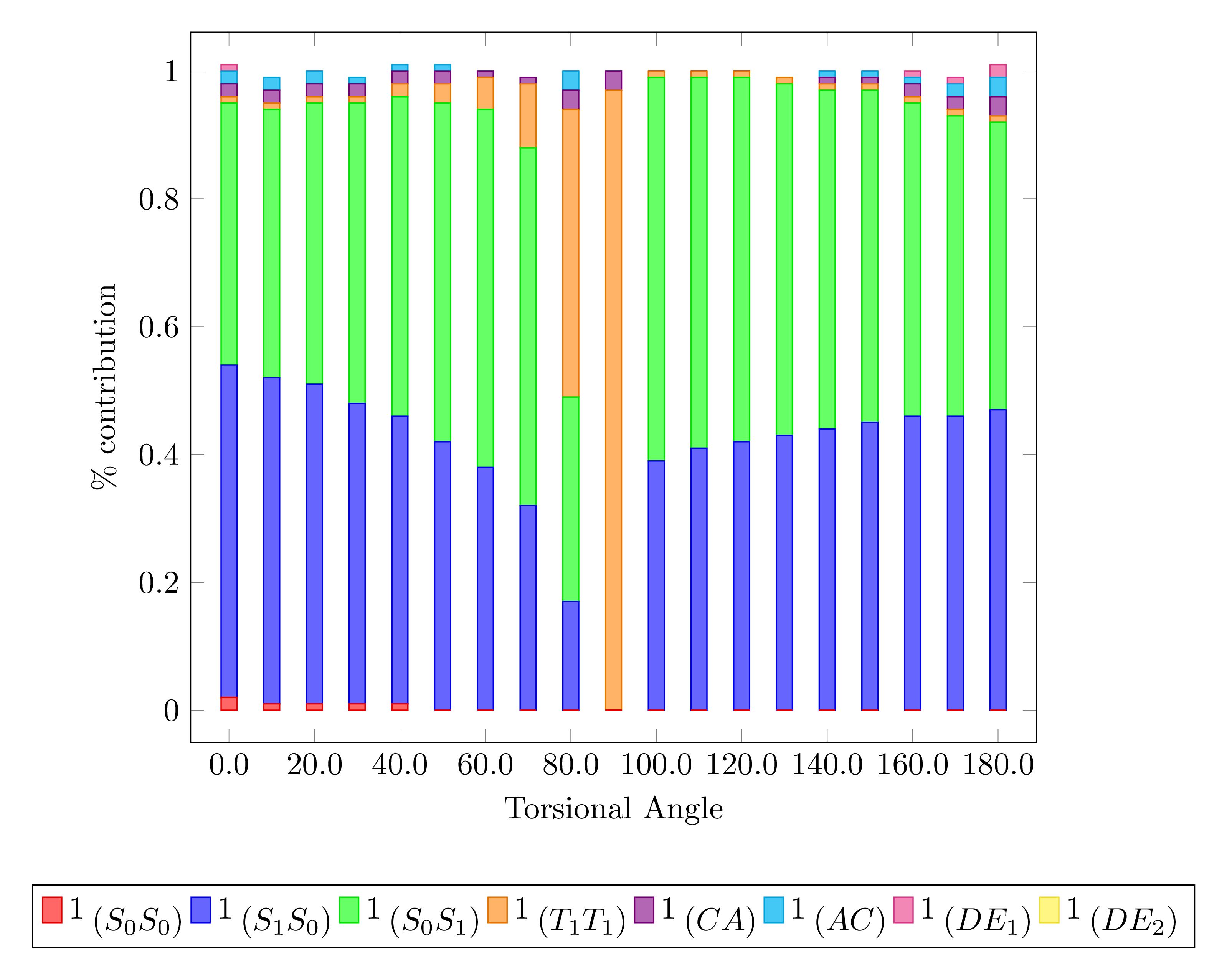}
                \caption{S$_\textrm{3}$}
        \end{subfigure}
        \begin{subfigure}{0.45\textwidth}
                \includegraphics[width=0.9\textwidth]{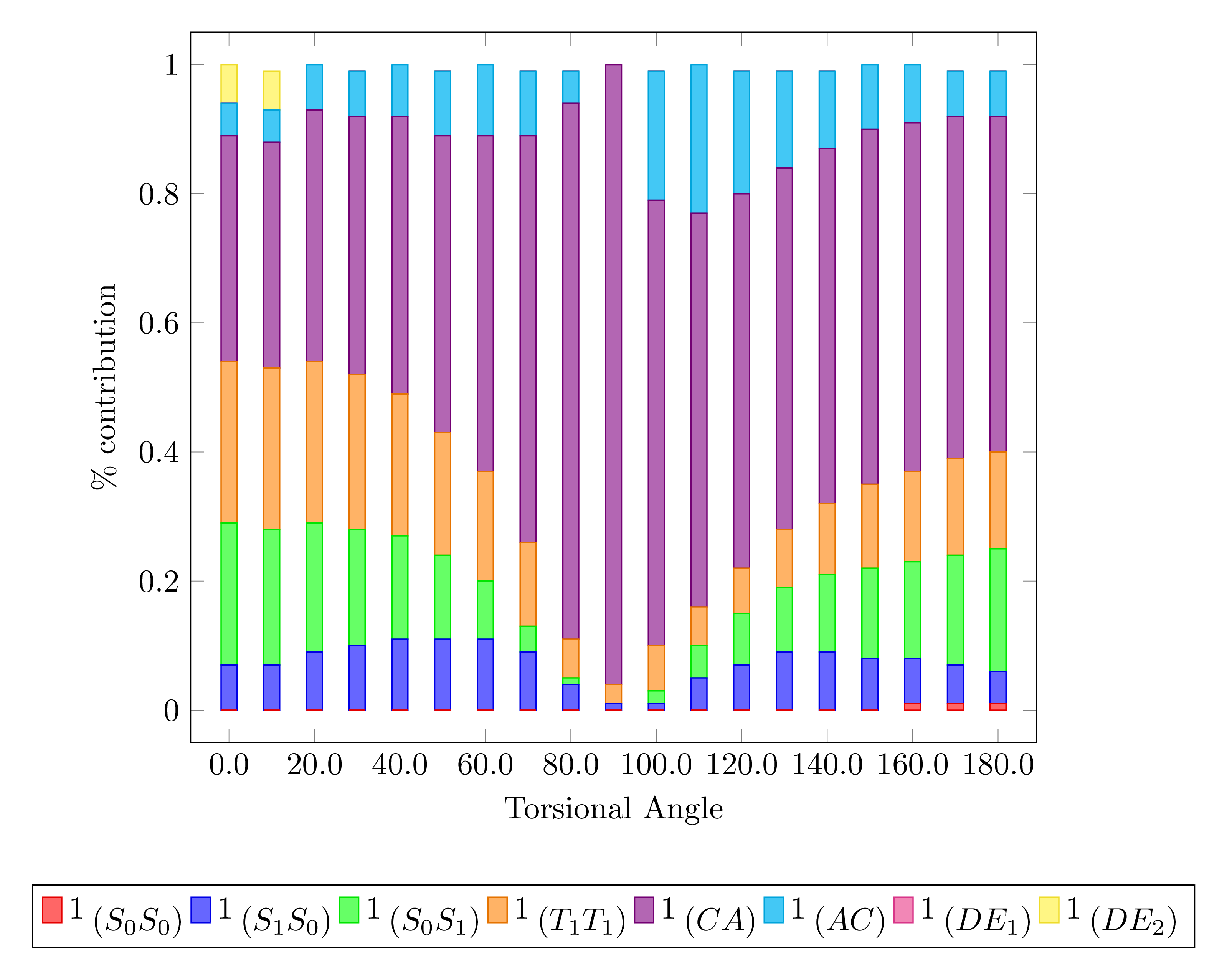}
                \caption{S$_\textrm{4}$}
        \end{subfigure}
	\begin{subfigure}{0.45\textwidth}
                \includegraphics[width=0.9\textwidth]{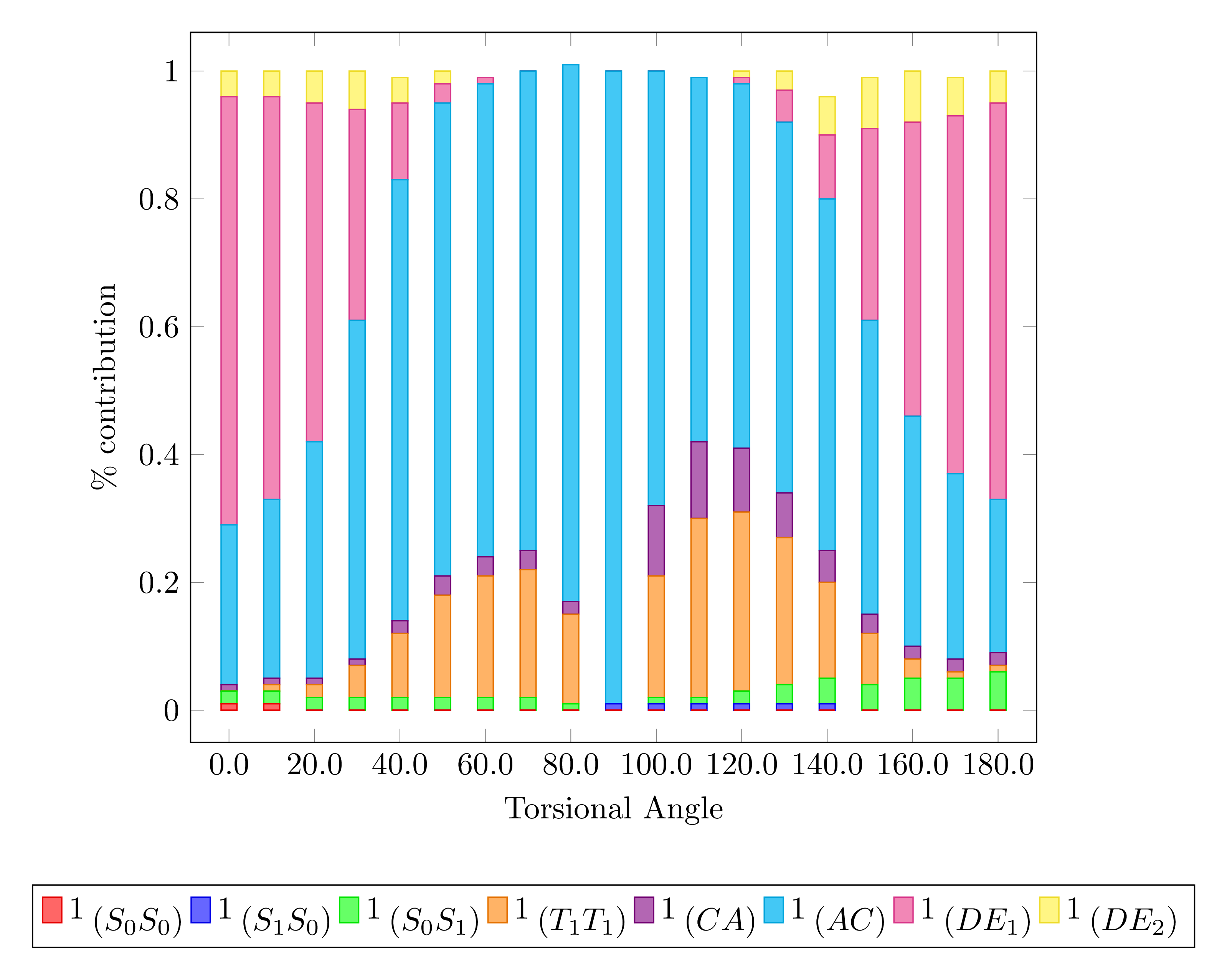}
                \caption{S$_\textrm{5}$}
        \end{subfigure}
        \begin{subfigure}{0.45\textwidth}
                \includegraphics[width=0.9\textwidth]{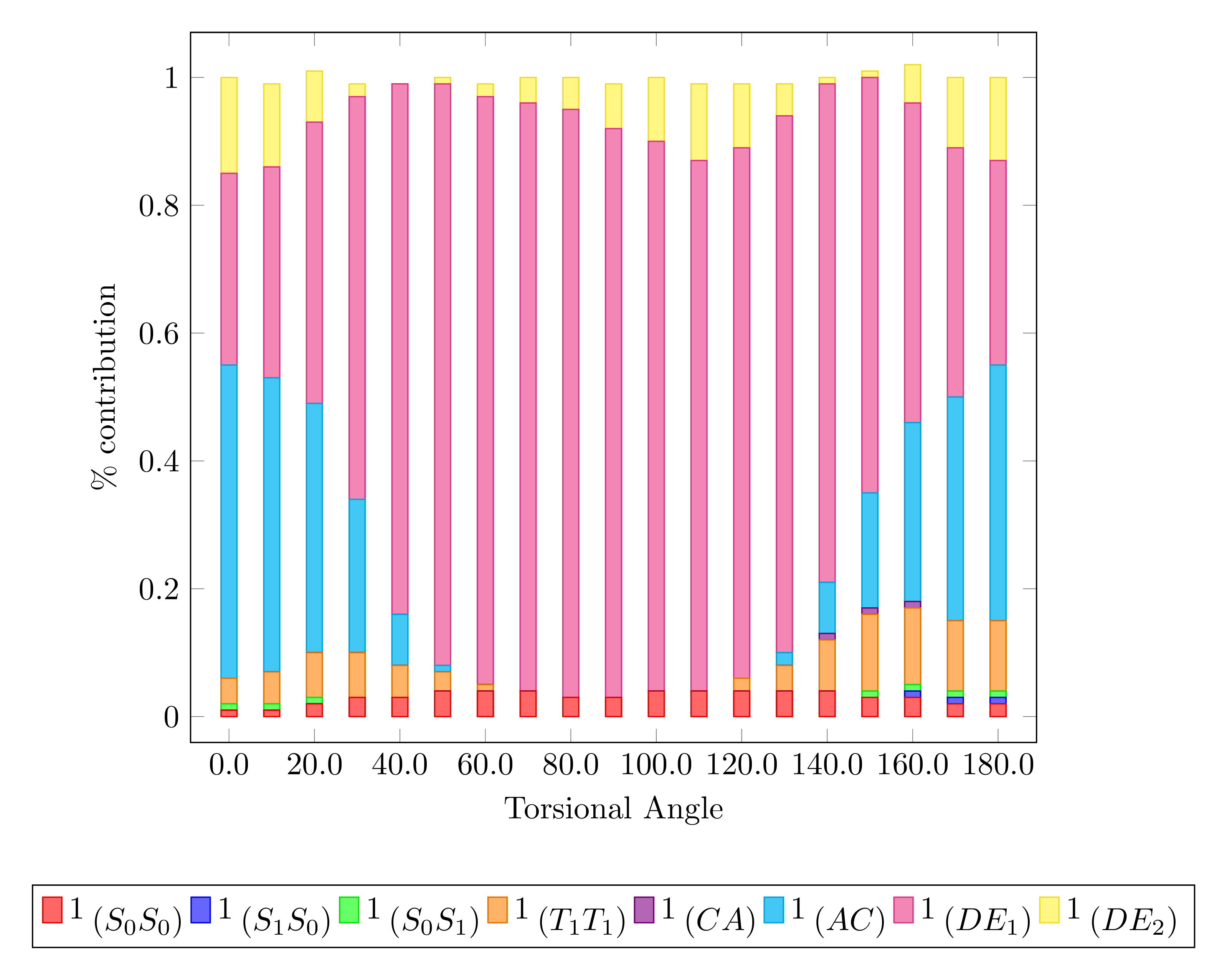}
                \caption{S$_\textrm{6}$}
		\begin{subfigure}{0.45\textwidth}
                \includegraphics[width=0.9\textwidth]{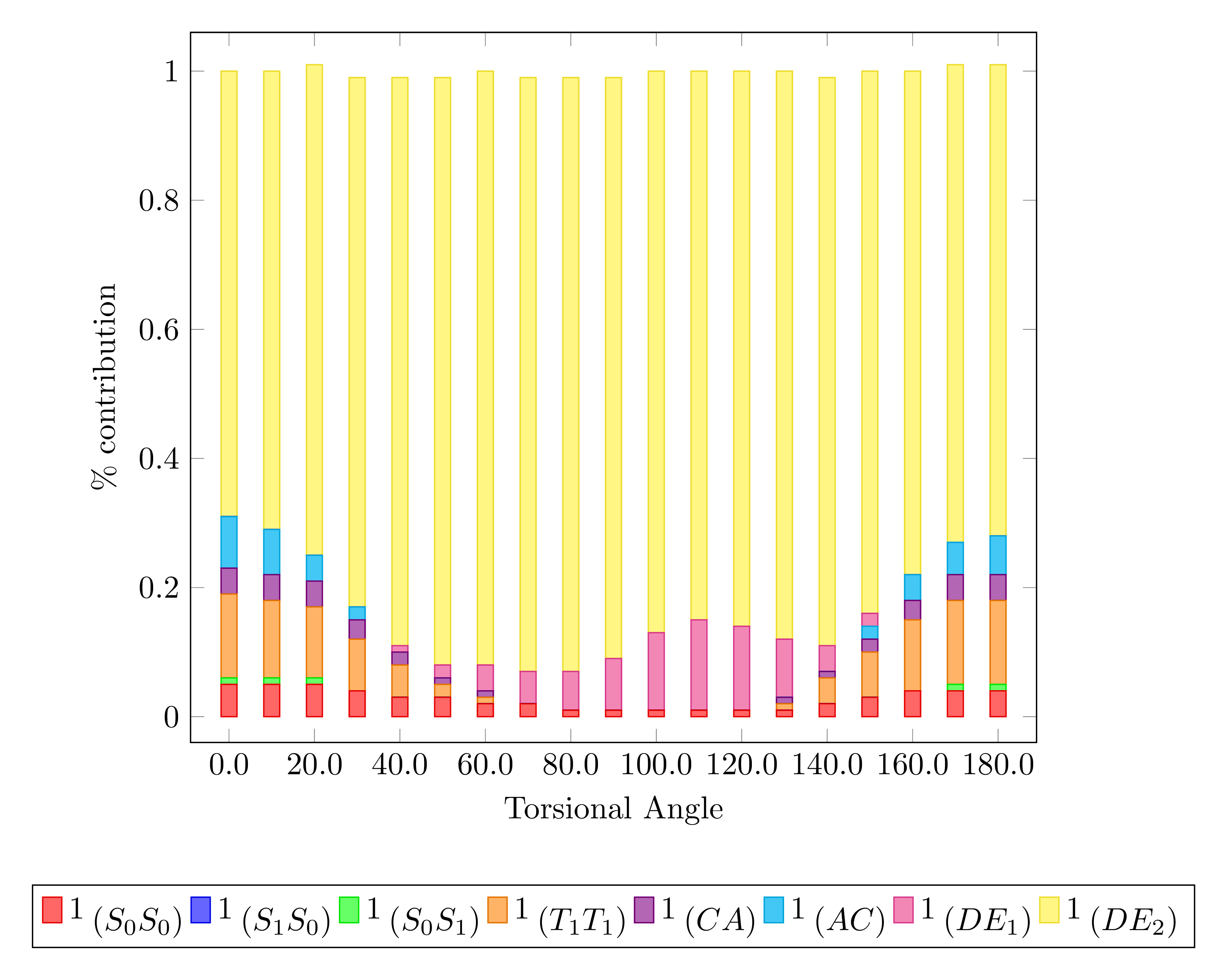}
                \caption{S$_\textrm{7}$}
        \end{subfigure}

        \end{subfigure}
  
\label{fig:b33-cont}
\end{figure}

	\clearpage
	
\section{Diabatic states and couplings}

For investigating the intramolecular singlet fission (iSF) mechanism in aza-BODIPY dimers D[1,1], D[1,3], D[3,3], and D[2,2], we computed the eight lowest-lying adiabatic states using the SA8-XMCQDPT(4,4)/cc-pVDZ method. The eight relevant adiabatic states were then used to derive the diabatic electronic states via Nakamura and Truhlar's threefold diabatization method \cite{Nakamura2001, Nakamura2002}. The diabatization procedure retained only the configuration state functions (CSFs) with weights greater than 0.20 in any adiabatic state. After applying the three-fold density criterion and MORMO conditions, diabatic molecular orbitals (DMOs) were derived from the adiabatic (canonical) orbitals. Finally, a unitary transformation was performed to convert the adiabatic states, expressed in the DMO basis, into diabatic states.

For each of these four conformers, we selected CSFs with coefficients greater than 0.20. The weights of the dominant CSFs obtained from the DMOs in the adiabatic states S$_0$, S$_1$, S$_2$, S$_3$, S$_4$, S$_5$, S$_6$ and S$_7$ for D[1,1] are 100.0\%, 100.0\%, 100.0\%, 100.0\%, 100.0\%, 100.0\%, 100.0\% and 100.0\% respectively. For D[1,3], these weights are 99.8\%, 99.1\%, 100.0\%, 99.9\%, 99.2\%, 97.2\%, 98.9\% and 96.4\%. Similarly, for D[3,3] these weights are 99.7\%, 96.6\%, 96.8\%, 100.0\%, 94.5\%, 100.0\%, 97.3\% and 98.6\%. For D[2,2] these weights are 100.0\%, 100.0\%, 100.0\%, 100.0\%, 100.0\%, 100.0\%, 100.0\% and 100.0\%. The results indicate that for each of the four regioisomers, the contributions from the dominating CSFs are $\sim$94\% or higher.

The diabatic electronic states include the electronic ground state $\ket{^{1}(\textrm{S}_{0}\textrm{S}_{0})}$, one correlated triplet pair state or the multiexcitonic (ME) state $\ket{^{1}(\textrm{T}_{1}\textrm{T}_{1})}$, the locally excited (LE) states $\ket{^{1}(\textrm{S}_{1}\textrm{S}_{0})}$ and $\ket{^{1}(\textrm{S}_{0}\textrm{S}_{1})}$ where the excitation ($\textrm{S}_{1}$) is localized in one of the aza-BODIPY moieties, and the charge transfer (CT) states $\ket{^{1}(\textrm{C}\textrm{A})}$ and $\ket{^{1}(\textrm{A}\textrm{C})}$, where C and A denote the radical cation and radical anion forms of the aza-BODIPY moiety, respectively. Additionally, two doubly excited (DE) diabatic states $\ket{^{1}(\textrm{DE})_{1}}$ and $\ket{^{1}(\textrm{DE})_{2}}$ were included due to energy considerations.  The electronic Hamiltonian ($H_{\textrm{el}}$) in the diabatic basis defined by these states is reported in Tables \ref{tab:diab11}–\ref{tab:diab22} for the non-orthogonal dimers and in Tables \ref{tab:diab11-90}–\ref{tab:diab22-90} for the orthogonal dimers. The corresponding diabatic-to-adiabatic transformation matrices are provided in Eqs. \ref{eq:rot11}–\ref{eq:rot22} and Eqs. \ref{eq:rot11-90}–\ref{eq:rot22-90}, respectively, for the dimers D[1,1], D[1,3], D[3,3], and D[2,2].

\begin{equation}
	\begin{pmatrix}
		\textrm{S}_\textrm{0} \\
		\textrm{S}_\textrm{1} \\
		\textrm{S}_\textrm{2} \\
		\textrm{S}_\textrm{3} \\
		\textrm{S}_\textrm{4} \\
		\textrm{S}_\textrm{5} \\
		\textrm{S}_\textrm{6} \\
		\textrm{S}_\textrm{7} 
	\end{pmatrix}
	=
	\begin{pmatrix}
		~0.95 & ~0.10 & ~0.11 & -0.04 & ~0.04 & ~0.04 & ~0.19 & -0.18 \\
		~0.00 & -0.47 & ~0.46 & -0.54 & -0.52 & ~0.00 & -0.05 & -0.04 \\
		-0.07 & ~0.52 & ~0.52 & -0.22 & ~0.22 & -0.58 & -0.07 & ~0.10 \\
		-0.08 & ~0.02 & ~0.03 & -0.59 & ~0.61 & ~0.52 & ~0.00 & -0.01 \\
		~0.00 & -0.53 & ~0.53 & ~0.47 & ~0.46 & ~0.00 & -0.03 & ~0.00 \\	
		-0.12 & ~0.47 & ~0.47 & ~0.29 & -0.30 & ~0.62 & ~0.02 & -0.03 \\
		~0.00 & -0.05 & ~0.03 & -0.02 & -0.03 & ~0.02 & ~0.67 & ~0.74 \\
		-0.26 & ~0.00 & ~0.04 & -0.02 & ~0.03 & -0.11 & ~0.71 & -0.64 \\
	\end{pmatrix}
	\begin{pmatrix}
		\ket{^{1}(\textrm{S}_{0}\textrm{S}_{0})} \\
		\ket{^{1}(\textrm{C}\textrm{A})} \\
		\ket{^{1}(\textrm{A}\textrm{C})}\\
		\ket{^{1}(\textrm{S}_{1}\textrm{S}_{0})}\\
		\ket{^{1}(\textrm{S}_{0}\textrm{S}_{1})} \\
		\ket{^{1}(\textrm{T}_{1}\textrm{T}_{1})} \\
		\ket{^{1}(\textrm{DE})_{1}} \\
		\ket{^{1}(\textrm{DE})_{2}} 
	\end{pmatrix}
	\label{eq:rot11}
\end{equation}

\begin{equation}
	\begin{pmatrix}
		\textrm{S}_\textrm{0} \\
		\textrm{S}_\textrm{1} \\
		\textrm{S}_\textrm{2} \\
		\textrm{S}_\textrm{3} \\
		\textrm{S}_\textrm{4} \\
		\textrm{S}_\textrm{5} \\
		\textrm{S}_\textrm{6} \\
		\textrm{S}_\textrm{7} 
	\end{pmatrix}
	=
	\begin{pmatrix}
		~0.95 &	-0.11 &	-0.03 &	~0.06 &	-0.03 &	~0.08 &	-0.21 &	~0.18 \\ 
		~0.01 &	~0.69 &	~0.46 &	~0.14 &	-0.36 &	~0.39 &	-0.09 &	~0.07 \\
		-0.02 &	-0.20 &	~0.20 &	~0.76 &	-0.35 &	-0.47 &	-0.02 &	-0.08 \\
		~0.07 &	-0.04 &	~0.73 &	~0.00 &	~0.65 &	-0.17 &	~0.10 &	~0.02 \\
		~0.11 &	~0.68 &	-0.37 &	~0.12 &	~0.33 &	-0.50 &	-0.07 &	-0.09 \\
		~0.23 &	~0.09 &	~0.11 &	-0.33 &	-0.33 &	-0.30 &	~0.79 &	-0.02 \\
		-0.01 &	~0.03 &	-0.25 &	~0.51 &	~0.30 &	~0.40 &	~0.54 &	~0.36 \\
		-0.18 &	~0.01 &	~0.04 &	-0.16 &	-0.11 &	-0.31 &	-0.16 &	~0.90 \\	
	\end{pmatrix}
	\begin{pmatrix}
		\ket{^{1}(\textrm{S}_{0}\textrm{S}_{0})} \\
		\ket{^{1}(\textrm{C}\textrm{A})} \\
		\ket{^{1}(\textrm{S}_{1}\textrm{S}_{0})} \\
		\ket{^{1}(\textrm{A}\textrm{C})}\\
		\ket{^{1}(\textrm{S}_{0}\textrm{S}_{1})}	 \\
		\ket{^{1}(\textrm{T}_{1}\textrm{T}_{1})} \\
		\ket{^{1}(\textrm{DE})_{1}} \\
		\ket{^{1}(\textrm{DE})_{2}} 
	\end{pmatrix}
	\label{eq:rot13}
\end{equation}

\begin{equation}
	\begin{pmatrix}
		\textrm{S}_\textrm{0} \\
		\textrm{S}_\textrm{1} \\
		\textrm{S}_\textrm{2} \\
		\textrm{S}_\textrm{3} \\
		\textrm{S}_\textrm{4} \\
		\textrm{S}_\textrm{5} \\
		\textrm{S}_\textrm{6} \\
		\textrm{S}_\textrm{7} 
	\end{pmatrix}
	=
	\begin{pmatrix}
		~0.93 & -0.04 & -0.04 & -0.01 & ~0.02 & -0.13 & ~0.23 & -0.23 \\
		~0.00 & ~0.50 & -0.50 & ~0.51 & ~0.49 & ~0.00 & ~0.04 & ~0.04 \\
		~0.09 & -0.58 & -0.58 & -0.02 & ~0.01 & ~0.53 & -0.13 & ~0.13 \\
		-0.01 & ~0.02 & ~0.01 & -0.69 & ~0.72 & -0.02 & -0.03 & ~0.03 \\
		-0.33 & -0.34 & -0.34 & -0.02 & ~0.02 & -0.45 & ~0.48 & -0.48 \\
		~0.00 & ~0.08 & -0.08 & -0.02 & -0.02 & ~0.00 & -0.70 & -0.70 \\
		~0.00 & -0.49 & ~0.49 & ~0.51 & ~0.49 & ~0.00 & -0.07 & -0.07 \\
		-0.10 & ~0.22 & ~0.22 & -0.02 & ~0.02 & ~0.70 & ~0.45 & -0.45 \\
	\end{pmatrix}
	\begin{pmatrix}
		\ket{^{1}(\textrm{S}_{0}\textrm{S}_{0})} \\
		\ket{^{1}(\textrm{C}\textrm{A})} \\
		\ket{^{1}(\textrm{A}\textrm{C})} \\
		\ket{^{1}(\textrm{S}_{1}\textrm{S}_{0})} \\
		\ket{^{1}(\textrm{S}_{0}\textrm{S}_{1})} \\
		\ket{^{1}(\textrm{T}_{1}\textrm{T}_{1})} \\
		\ket{^{1}(\textrm{DE})_{1}} \\
		\ket{^{1}(\textrm{DE})_{2}} 
	\end{pmatrix}
	\label{eq:rot33}
\end{equation}

\begin{equation}
	\begin{pmatrix}
		\textrm{S}_\textrm{0} \\
		\textrm{S}_\textrm{1} \\
		\textrm{S}_\textrm{2} \\
		\textrm{S}_\textrm{3} \\
		\textrm{S}_\textrm{4} \\
		\textrm{S}_\textrm{5} \\
		\textrm{S}_\textrm{6} \\
		\textrm{S}_\textrm{7} 
	\end{pmatrix}
	=
	\begin{pmatrix}
		~0.97 & ~0.01 & -0.01 & -0.01 & -0.06 & ~0.06 & ~0.15 & ~0.15 \\
		~0.00 & -0.62 & -0.59 & ~0.00 & ~0.37 & ~0.37 & ~0.02 & -0.02 \\
		~0.01 & ~0.34 & -0.35 & -0.63 & ~0.42 & -0.42 & ~0.08 & ~0.08 \\
		~0.00 & -0.58 & ~0.60 & -0.53 & ~0.07 & -0.07 & ~0.04 & ~0.04 \\
		~0.00 & ~0.37 & ~0.36 & ~0.00 & ~0.60 & ~0.60 & ~0.06 & -0.06 \\
		~0.07 & -0.18 & ~0.18 & ~0.55 & ~0.56 & -0.56 & ~0.03 & ~0.02 \\
		-0.22 & ~0.00 & ~0.00 & ~0.09 & -0.06 & ~0.06 & ~0.71 & ~0.66 \\
		-0.01 & ~0.01 & ~0.01 & ~0.00 & ~0.06 & ~0.06 & -0.68 & ~0.73 \\
	\end{pmatrix}
	\begin{pmatrix}
		\ket{^{1}(\textrm{S}_{0}\textrm{S}_{0})} \\
		\ket{^{1}(\textrm{S}_{1}\textrm{S}_{0})} \\
		\ket{^{1}(\textrm{S}_{0}\textrm{S}_{1})} \\
		\ket{^{1}(\textrm{T}_{1}\textrm{T}_{1})} \\
		\ket{^{1}(\textrm{C}\textrm{A})} \\
		\ket{^{1}(\textrm{A}\textrm{C})} \\
		\ket{^{1}(\textrm{DE})_{1}} \\
		\ket{^{1}(\textrm{DE})_{2}} 
	\end{pmatrix}
	\label{eq:rot22}
\end{equation}

\begin{equation}
	\begin{pmatrix}
		\textrm{S}_\textrm{0} \\
		\textrm{S}_\textrm{1} \\
		\textrm{S}_\textrm{2} \\
		\textrm{S}_\textrm{3} \\
		\textrm{S}_\textrm{4} \\
		\textrm{S}_\textrm{5} \\
		\textrm{S}_\textrm{6} \\
		\textrm{S}_\textrm{7} 
	\end{pmatrix}
	=
	\begin{pmatrix}
		~0.98	& ~0.00	&-0.01	&~0.00	&~0.00	&~0.00	&-0.15	&-0.15 \\
		~0.00	&~0.72	&~0.69	&~0.00	&-0.04	&-0.06	&-0.02	&~0.02 \\
		~0.00	&-0.69	&~0.72	&~0.01	&-0.05	&~0.05	&-0.01	&-0.01 \\
		~0.00	&~0.00	&-0.01	&~1.00	&~0.03	&-0.04	&~0.00	&~0.00 \\
		~0.00	&~0.03	&~0.06	&~0.00	&~0.85	&~0.51	&-0.02	&~0.01 \\ 
		~0.00	&~0.07	&-0.03	&~0.04	&-0.51	&~0.85	&~0.01	&~0.02 \\
		~0.01	&~0.02	&~0.02	&~0.00	&~0.01	&~0.01	&~0.75	&-0.66 \\
		~0.21	&-0.01	&~0.01	&~0.00	&~0.02	&-0.02	&~0.65	&~0.73 \\	
	\end{pmatrix}
	\begin{pmatrix}
		\ket{^{1}(\textrm{S}_{0}\textrm{S}_{0})} \\
		\ket{^{1}(\textrm{C}\textrm{A})} \\
		\ket{^{1}(\textrm{A}\textrm{C})}\\
		\ket{^{1}(\textrm{T}_{1}\textrm{T}_{1})} \\
		\ket{^{1}(\textrm{S}_{1}\textrm{S}_{0})}\\
		\ket{^{1}(\textrm{S}_{0}\textrm{S}_{1})} \\
		\ket{^{1}(\textrm{DE})_{1}} \\
		\ket{^{1}(\textrm{DE})_{2}} 
	\end{pmatrix}
	\label{eq:rot11-90}
\end{equation}

\begin{equation}
	\begin{pmatrix}
		\textrm{S}_\textrm{0} \\
		\textrm{S}_\textrm{1} \\
		\textrm{S}_\textrm{2} \\
		\textrm{S}_\textrm{3} \\
		\textrm{S}_\textrm{4} \\
		\textrm{S}_\textrm{5} \\
		\textrm{S}_\textrm{6} \\
		\textrm{S}_\textrm{7} 
	\end{pmatrix}
	=
	\begin{pmatrix}
		~0.98	&~0.00	&~0.02	&~0.00	&~0.00	&~0.00	&-0.14	&~0.15 \\
		~0.01	&~0.80	&-0.59	&-0.01	&~0.04	&-0.05	&~0.00	&~0.03 \\
		-0.01	&~0.59	&~0.80	&~0.01	&~0.07	&-0.05	&~0.01	&-0.02 \\
		~0.00	&~0.01	&~0.00	&~0.95	&-0.31	&-0.09	&~0.00	&-0.01 \\
		~0.00	&-0.07	&-0.03	&~0.31	&~0.95	&-0.01	&-0.01	&~0.02 \\
		~0.00	&~0.07	&~0.01	&~0.09	&-0.02	&~0.99	&-0.02	&~0.02 \\
		~0.14	&~0.00	&~0.00	&~0.00	&~0.00	&~0.02	&~0.99	&~0.06 \\
		-0.16	&-0.01	&~0.03	&~0.00	&-0.02	&-0.02	&-0.03	&~0.99 \\
	\end{pmatrix}
	\begin{pmatrix}
		\ket{^{1}(\textrm{S}_{0}\textrm{S}_{0})} \\
		\ket{^{1}(\textrm{S}_{1}\textrm{S}_{0})}\\
		\ket{^{1}(\textrm{S}_{0}\textrm{S}_{1})} \\
		\ket{^{1}(\textrm{T}_{1}\textrm{T}_{1})} \\
		\ket{^{1}(\textrm{C}\textrm{A})} \\
		\ket{^{1}(\textrm{A}\textrm{C})}\\
		\ket{^{1}(\textrm{DE})_{1}} \\
		\ket{^{1}(\textrm{DE})_{2}} 
	\end{pmatrix}
	\label{eq:rot13-90}
\end{equation}

\begin{equation}
	\begin{pmatrix}
		\textrm{S}_\textrm{0} \\
		\textrm{S}_\textrm{1} \\
		\textrm{S}_\textrm{2} \\
		\textrm{S}_\textrm{3} \\
		\textrm{S}_\textrm{4} \\
		\textrm{S}_\textrm{5} \\
		\textrm{S}_\textrm{6} \\
		\textrm{S}_\textrm{7} 
	\end{pmatrix}
	=
	\begin{pmatrix}
	~0.98	&~0.00	&~0.00	&~0.00	&~0.02	&~0.01	&-0.15	&~0.15 \\
	~0.00	&~0.72	&-0.68	&~0.00	&~0.03	&-0.11	&~0.00	&~0.00 \\
	~0.00	&~0.69	&~0.72	&~0.01	&-0.01	&~0.02	&~0.00	&~0.00 \\
	-0.02	&~0.02	&-0.03	&~0.79	&~0.43	&~0.43	&-0.01	&~0.00 \\
	-0.01	&-0.06	&~0.08	&-0.11	&~0.80	&-0.58	&-0.01	&~0.00 \\
	-0.02	&~0.04	&-0.04	&-0.60	&~0.42	&~0.68	&~0.00	&~0.00 \\
	~0.02	&~0.00	&~0.00	&~0.00	&~0.01	&~0.00	&~0.76	&~0.65 \\
	-0.21	&~0.00	&~0.00	&~0.00	&-0.02	&~0.00	&-0.63	&~0.74 \\	
	\end{pmatrix}
	\begin{pmatrix}
		\ket{^{1}(\textrm{S}_{0}\textrm{S}_{0})} \\
		\ket{^{1}(\textrm{S}_{1}\textrm{S}_{0})}\\
		\ket{^{1}(\textrm{S}_{0}\textrm{S}_{1})} \\
		\ket{^{1}(\textrm{T}_{1}\textrm{T}_{1})} \\
		\ket{^{1}(\textrm{C}\textrm{A})} \\
		\ket{^{1}(\textrm{A}\textrm{C})}\\
		\ket{^{1}(\textrm{DE})_{1}} \\
		\ket{^{1}(\textrm{DE})_{2}} 
	\end{pmatrix}
	\label{eq:rot33-90}
\end{equation}

\begin{equation}
	\begin{pmatrix}
		\textrm{S}_\textrm{0} \\
		\textrm{S}_\textrm{1} \\
		\textrm{S}_\textrm{2} \\
		\textrm{S}_\textrm{3} \\
		\textrm{S}_\textrm{4} \\
		\textrm{S}_\textrm{5} \\
		\textrm{S}_\textrm{6} \\
		\textrm{S}_\textrm{7} 
	\end{pmatrix}
	=
	\begin{pmatrix}
		~0.98	&-0.01	&~0.01	&~0.00	&-0.01	&-0.01	&-0.14	&~0.14 \\
 		~0.00	&~0.72	&~0.69	&~0.00	&-0.07	&~0.05	&~0.01	&~0.00 \\
		-0.01	&-0.69	&~0.72	&-0.04	&~0.04	&~0.05	&~0.02	&-0.02 \\
		-0.01	&-0.04	&~0.04	&~0.98	&-0.15	&-0.15	&~0.00	&~0.01 \\
		~0.00	&-0.05	&-0.07	&~0.00	&-0.70	&~0.71	&~0.03	&~0.03 \\
		~0.02	&~0.05	&-0.03	&~0.21	&~0.69	&~0.68	&~0.02	&-0.02 \\
 		-0.01	&~0.00	&~0.00	&~0.00	&~0.03	&-0.03	&~0.66	&~0.75 \\
		~0.20	&~0.01	&-0.01	&~0.00	&-0.02	&-0.03	&~0.74	&-0.64 \\
	\end{pmatrix}
	\begin{pmatrix}
		\ket{^{1}(\textrm{S}_{0}\textrm{S}_{0})} \\
		\ket{^{1}(\textrm{S}_{1}\textrm{S}_{0})}\\
		\ket{^{1}(\textrm{S}_{0}\textrm{S}_{1})} \\
		\ket{^{1}(\textrm{T}_{1}\textrm{T}_{1})} \\
		\ket{^{1}(\textrm{C}\textrm{A})} \\
		\ket{^{1}(\textrm{A}\textrm{C})}\\
		\ket{^{1}(\textrm{DE})_{1}} \\
		\ket{^{1}(\textrm{DE})_{2}} 
	\end{pmatrix}
	\label{eq:rot22-90}
\end{equation}

\begin{table}[!ht]
	\begin{center}
		\caption{The diabatic electronic energies and couplings (in eVs) between the diabatic states $\ket{^{1}(\textrm{S}_{0}\textrm{S}_{0})}$, $\ket{^{1}(\textrm{S}_{0}\textrm{S}_{1})}$, $\ket{^{1}(\textrm{S}_{0}\textrm{S}_{1})}$, $\ket{^{1}(\textrm{T}_{1}\textrm{T}_{1})}$, $\ket{^{1}(\textrm{C}\textrm{A})}$, $\ket{^{1}(\textrm{A}\textrm{C})}$, $\ket{^{1}(\textrm{DE})_{1}}$ and $\ket{^{1}(\textrm{DE})_{2}}$ for the system D[1,1] calculated at SA8-XMCQDPT(4,4)/cc-pVDZ level of theory.}
		\begin{tabular}{ccccccccc}
			\toprule
			$\mathcal{H}_{el}$ & $\ket{^{1}(\textrm{S}_\textrm{0}\textrm{S}_\textrm{0})}$ & $\ket{^{1}(\textrm{C}\textrm{A})}$   & $\ket{^{1}(\textrm{A}\textrm{C})}$ & $\ket{^{1}(\textrm{S}_{1}\textrm{S}_{0})}$ & $\ket{^{1}(\textrm{S}_{0}\textrm{S}_{1})}$ & $\ket{^{1}(\textrm{T}_{1}\textrm{T}_{1})}$ & $\ket{^{1}(\textrm{DE})_{1}}$ & $\ket{^{1}(\textrm{DE})_{2}}$ \\
			\toprule
			$\bra{^{1}(\textrm{S}_{0}\textrm{S}_{0})}$ &~0.0000 &-0.2253 &-0.2582 &~0.0636 &-0.0701 &-0.1271 &-0.6395 &~0.6027 \\
			$\bra{^{1}(\textrm{C}\textrm{A})}$         &-0.2253 &~1.9977 &-0.0452 &-0.1438 &-0.4319 &~0.3200 &-0.0626 &-0.0478 \\
			$\bra{^{1}(\textrm{A}\textrm{C})}$         &-0.2581 &-0.0452 &~1.9983 &~0.4377 &~0.1325 &~0.3138 &~0.0431 &~0.0171 \\
			$\bra{^{1}(\textrm{S}_{1}\textrm{S}_{0})}$ &~0.0636	&-0.1438 &~0.4377 &~1.9991 &-0.1172 &~0.0217 &-0.0501 &-0.0222\\
			$\bra{^{1}(\textrm{S}_{0}\textrm{S}_{1})}$ &-0.0701 &-0.4319 &~0.1325 &-0.1172 &~2.0070 &-0.0268 &-0.0351 &-0.0468 \\
			$\bra{^{1}(\textrm{T}_{1}\textrm{T}_{1})}$ &-0.1271 &~0.3200 &~0.3138 &~0.0217 &-0.0268 &~2.1049 &-0.1105 &~0.1449 \\
			$\bra{^{1}(\textrm{DE})_{1}}$              &-0.6395 &-0.0626 &~0.0431 &-0.0501 &-0.0350 &-0.1105 &~3.0682 &~0.1042\\
			$\bra{^{1}(\textrm{DE})_{2}}$              &~0.6027 &-0.0479 &~0.0171 &-0.0222 &-0.0468 &~0.1448 &~0.1042 &~3.0739 \\
			\bottomrule
		\end{tabular}
		\label{tab:diab11}
	\end{center}
\end{table}

\begin{table}[!ht]
	\begin{center}
		\caption{The diabatic electronic energies and couplings (in eVs) between the diabatic states $^{1}(\textrm{S}_{0}\textrm{S}_{0})$, $^{1}(\textrm{S}_{0}\textrm{S}_{1})$, $^{1}(\textrm{S}_{0}\textrm{S}_{1})$, $^{1}(\textrm{T}_{1}\textrm{T}_{1})$, $^{1}(\textrm{A}\textrm{C})$, $^{1}(\textrm{C}\textrm{A})$, $^{1}(\textrm{DE})_{1}$ and $^{1}(\textrm{DE})_{2}$ for the system D[1,3] calculated at SA8-XMCQDPT(4,4)/cc-pVDZ level of theory.}
		\begin{tabular}{ccccccccc}
			\toprule
			$\mathcal{H}_{el}$ &$\ket{^{1}(\textrm{S}_{0}\textrm{S}_{0})}$ &$\ket{^{1}(\textrm{C}\textrm{A})}$   &$\ket{^{1}(\textrm{S}_{1}\textrm{S}_{0})}$ &$\ket{^{1}(\textrm{A}\textrm{C})}$ &$\ket{^{1}(\textrm{S}_{0}\textrm{S}_{1})}$& $\ket{^{1}(\textrm{T}_{1}\textrm{T}_{1})}$ &$\ket{^{1}(\textrm{DE})_{1}}$             &$\ket{^{1}(\textrm{DE})_{2}}$ \\
			\toprule
			$\bra{^{1}(\textrm{S}_{0}\textrm{S}_{0})}$  & 0.0000 & 0.2668 & 0.0747 &-0.1421 & 0.0359 &-0.1949 & 0.6383 &-0.6198 \\
			$\bra{^{1}(\textrm{C}\textrm{A})}$          & 0.2668 & 1.7034 &-0.3807 & 0.0421 & 0.2435 &-0.4401 & 0.0546 &-0.0153 \\
			$\bra{^{1}(\textrm{S}_{1}\textrm{S}_{0})}$  & 0.0747 &-0.3807 & 1.9696 &-0.3353 & 0.1014 &-0.1886 &-0.0296 &-0.0370 \\
			$\bra{^{1}(\textrm{A}\textrm{C})}$          &-0.1421 & 0.0421 &-0.3353 & 1.9886 & 0.5023 & 0.4911 & 0.1281 & 0.0084 \\
			$\bra{^{1}(\textrm{S}_{0}\textrm{S}_{1})}$  & 0.0359 & 0.2435 & 0.1014 & 0.5023 & 2.0946 & 0.1961 & -0.0871 & -0.0018  \\
			$\bra{^{1}(\textrm{T}_{1}\textrm{T}_{1})}$  &-0.1949 &-0.4401 &-0.1886 & 0.4911 & 0.1961 & 2.1628 & 0.1461 &-0.2668 \\
			$\bra{^{1}(\textrm{DE})_{1}}$               & 0.6383 & 0.0546 &-0.0296 & 0.1281 &-0.0871 & 0.1461 & 2.7364 & 0.0920 \\
			$\bra{^{1}(\textrm{DE})_{2}}$               &-0.6198 &-0.0153 &-0.0370 & 0.0084 &-0.0018 &-0.2668 & 0.0920 & 3.0873 \\
			\bottomrule
		\end{tabular}
		\label{tab:diab13}
	\end{center}
\end{table}

\begin{table}[!ht]
	\begin{center}
		\caption{The diabatic electronic energies and couplings (in eVs) between the diabatic states $\ket{^{1}(\textrm{S}_{0}\textrm{S}_{0})}$, $\ket{^{1}(\textrm{S}_{0}\textrm{S}_{1})}$, $\ket{^{1}(\textrm{S}_{0}\textrm{S}_{1})}$, $\ket{^{1}(\textrm{T}_{1}\textrm{T}_{1})}$, $\ket{^{1}(\textrm{C}\textrm{A})}$, $\ket{^{1}(\textrm{A}\textrm{C})}$, $\ket{^{1}(\textrm{DE})_{1}}$ and $\ket{^{1}(\textrm{DE})_{2}}$ for the system D[3,3] calculated at SA8-XMCQDPT(4,4)/cc-pVDZ level of theory.}
		\begin{tabular}{ccccccccc}
			\toprule
			$\mathcal{H}_{el}$ &$\ket{^{1}(\textrm{S}_{0}\textrm{S}_{0})}$ &$\ket{^{1}(\textrm{C}\textrm{A})}$   &$\ket{^{1}(\textrm{A}\textrm{C})}$ 
			&$\ket{^{1}(\textrm{S}_{1}\textrm{S}_{0})}$
			&$\ket{^{1}(\textrm{S}_{0}\textrm{S}_{1})}$              &$\ket{^{1}(\textrm{T}_{1}\textrm{T}_{1})}$      &$\ket{^{1}(\textrm{DE})_{1}}$             &$\ket{^{1}(\textrm{DE})_{2}}$ \\
			\toprule
			$\bra{^{1}(\textrm{S}_{0}\textrm{S}_{0})}$ &0.0000 &  0.1958 & 0.1970 & 0.0434 & -0.0450 & 0.2569 & 0.6904 & -0.6903 \\
			$\bra{^{1}(\textrm{C}\textrm{A})}$   &0.1958 & 1.7763 & -0.1977 & -0.5714 & -0.5194 & 0.6497 & 0.0134 & -0.0692 \\
			$\bra{^{1}(\textrm{A}\textrm{C})}$ & 0.1970 & -0.1977 & 1.7768 & 0.5418 & 0.5497 & 0.6500 & 0.0694 & -0.0139 \\
			$\bra{^{1}(\textrm{S}_{1}\textrm{S}_{0})}$&0.0434 & -0.5714 & 0.5418 & 2.1066 & -0.1338  & -0.0090 & -0.0369 & -0.0531\\
			$\bra{^{1}(\textrm{S}_{0}\textrm{S}_{1})}$&-0.0450 & -0.5194 & 0.5497 & -0.1338 & 2.1238 & 0.0085 & -0.0518 & -0.0335 \\
			$\bra{^{1}(\textrm{T}_{1}\textrm{T}_{1})}$  & 0.2569 & 0.6497& 0.6500 & -0.0090 & 0.0085 & 2.5028 & -0.43317 & 0.4331 \\
			$\bra{^{1}(\textrm{DE})_{1}}$ &0.6904 & 0.0134 & 0.0694& -0.0369 & -0.0518 & -0.4331 & 2.7319 & 0.1062\\
			$\bra{^{1}(\textrm{DE})_{2}}$ &-0.6903 & -0.0692 & -0.0139 & -0.0531 & -0.0335 & 0.4331 & 0.1062 & 2.7319 \\
			\bottomrule
		\end{tabular}
		\label{tab:diab33}
	\end{center}
\end{table}

\begin{table}[!ht]
	\begin{center}
		\caption{The diabatic electronic energies and couplings (in eVs) between the diabatic states $\ket{^{1}(\textrm{S}_{0}\textrm{S}_{0})}$, $\ket{^{1}(\textrm{S}_{0}\textrm{S}_{1})}$, $\ket{^{1}(\textrm{S}_{0}\textrm{S}_{1})}$, $\ket{^{1}(\textrm{T}_{1}\textrm{T}_{1})}$, $\ket{^{1}(\textrm{C}\textrm{A})}$, $\ket{^{1}(\textrm{A}\textrm{C})}$, $\ket{^{1}(\textrm{DE})_{1}}$ and $\ket{^{1}(\textrm{DE})_{2}}$ for the system D[2,2] calculated at SA8-XMCQDPT(4,4)/cc-pVDZ level of theory.}
		\begin{tabular}{ccccccccc}
			\toprule
			$\mathcal{H}_{el}$ &$\ket{^{1}(\textrm{S}_{0}\textrm{S}_{0})}$ &$\ket{^{1}(\textrm{S}_{1}\textrm{S}_{0})}$   &$\ket{^{1}(\textrm{S}_{0}\textrm{S}_{1})}$ 
			&$\ket{^{1}(\textrm{T}_{1}\textrm{T}_{1})}$& $\ket{^{1}(\textrm{C}\textrm{A})}$
			&$\ket{^{1}(\textrm{A}\textrm{C})}$                &$\ket{^{1}(\textrm{DE})_{1}}$             &$\ket{^{1}(\textrm{DE})_{2}}$ \\
			\toprule
			$\bra{^{1}(\textrm{S}_{0}\textrm{S}_{0})}$ &0.0000 &  -0.0342 & 0.0341 & 0.0285 & 0.1755 & -0.1789 & -0.5284 & -0.5257 \\
			$\bra{^{1}(\textrm{S}_{1}\textrm{S}_{0})}$  &-0.0342 & 2.0048 & -0.1888 & -0.0120 & 0.1893 & 0.3893 & -0.0074 & -0.0208 \\
			$\bra{^{1}(\textrm{S}_{0}\textrm{S}_{1})}$ & 0.0341 & -0.1888 & 2.0221 & 0.0127 & 0.3834 & 0.1731 & 0.0182 & 0.0039 \\
			$\bra{^{1}(\textrm{T}_{1}\textrm{T}_{1})}$&0.0285 & -0.0120 & 0.0127 & 2.3346 & 0.2683 & -0.2674 & 0.0974& 0.0982\\
			$\bra{^{1}(\textrm{C}\textrm{A})}$& 0.1755 & 0.1893& 0.3834 & 0.2683 & 2.4812 & -0.0605 & -0.0652 & 0.0157 \\
			$\bra{^{1}(\textrm{A}\textrm{C})}$ &-0.1789 & 0.3893 & 0.1731 & -0.2674 & -0.0605 & 2.4828 & -0.0081 & 0.0665 \\
			$\bra{^{1}(\textrm{DE})_{1}}$ &-0.5284 & -0.0074 & 0.0182 & 0.0974 & -0.0652 & -0.0081 & 3.3546 & -0.1164\\
			$\bra{^{1}(\textrm{DE})_{2}}$ &-0.5257	& -0.0208 & 0.0039 & 0.0982 & 0.0157 & 0.0665 & -0.1164 & 3.3589 \\
			\bottomrule
		\end{tabular}
		\label{tab:diab22}
	\end{center}
\end{table}

\clearpage

\begin{table}[!ht]
	\begin{center}
		\caption{The diabatic electronic energies and couplings (in eV) between the diabatic states 
			$\ket{^{1}(\textrm{S}_{0}\textrm{S}_{0})}$, 
			$\ket{^{1}(\textrm{C}\textrm{A})}$, 
			$\ket{^{1}(\textrm{A}\textrm{C})}$, 
			$\ket{^{1}(\textrm{T}_{1}\textrm{T}_{1})}$, 
			$\ket{^{1}(\textrm{S}_{1}\textrm{S}_{0})}$, 
			$\ket{^{1}(\textrm{S}_{0}\textrm{S}_{1})}$, 
			$\ket{^{1}(\textrm{DE})_{1}}$ and 
			$\ket{^{1}(\textrm{DE})_{2}}$ for the system orthogonal-D[1,1] calculated at SA8-XMCQDPT(4,4)/cc-pVDZ level of theory.}
		
		\begin{tabular}{ccccccccc}
			\toprule
			$\mathcal{H}_{el}$ 
			& $\ket{^{1}(\textrm{S}_{0}\textrm{S}_{0})}$ 
			& $\ket{^{1}(\textrm{C}\textrm{A})}$ 
			& $\ket{^{1}(\textrm{A}\textrm{C})}$ 
			& $\ket{^{1}(\textrm{T}_{1}\textrm{T}_{1})}$ 
			& $\ket{^{1}(\textrm{S}_{1}\textrm{S}_{0})}$ 
			& $\ket{^{1}(\textrm{S}_{0}\textrm{S}_{1})}$ 
			& $\ket{^{1}(\textrm{DE})_{1}}$ 
			& $\ket{^{1}(\textrm{DE})_{2}}$ \\
			\midrule
			
			$\bra{^{1}(\textrm{S}_{0}\textrm{S}_{0})}$ 
			& 0.0000 & -0.0127 & 0.0161 & 0.0015 & 0.0118 & -0.0096 & 0.5509 & 0.5536 \\
			
			$\bra{^{1}(\textrm{C}\textrm{A})}$ 
			& -0.0127 & 2.0263 & -0.1339 & 0.0023 & 0.0064 & 0.0475 & 0.0150 & -0.0390 \\
			
			$\bra{^{1}(\textrm{A}\textrm{C})}$ 
			& 0.0161 & -0.1339 & 2.0329 & -0.0022 & 0.0406 & 0.0149 & 0.0374 & -0.0104 \\
			
			$\bra{^{1}(\textrm{T}_{1}\textrm{T}_{1})}$ 
			& 0.0015 & 0.0023 & -0.0022 & 2.3983 & -0.0057 & 0.0084 & -0.0012 & -0.0005 \\
			
			$\bra{^{1}(\textrm{S}_{1}\textrm{S}_{0})}$ 
			& 0.0118 & 0.0064 & 0.0406 & -0.0057 & 2.6275 & -0.0030 & 0.0197 & 0.0020 \\
			
			$\bra{^{1}(\textrm{S}_{0}\textrm{S}_{1})}$ 
			& -0.0096 & 0.0475 & 0.0149 & 0.0084 & -0.0030 & 2.6275 & -0.0004 & -0.0192 \\
			
			$\bra{^{1}(\textrm{DE})_{1}}$ 
			& 0.5509 & 0.0150 & 0.0374 & -0.0012 & 0.0197 & -0.0004 & 3.5875 & -0.0752 \\
			
			$\bra{^{1}(\textrm{DE})_{2}}$ 
			& 0.5536 & -0.0390 & -0.0104 & -0.0005 & 0.0020 & -0.0192 & -0.0752 & 3.5908 \\
			
			\bottomrule
		\end{tabular}
		\label{tab:diab11-90}
	\end{center}
\end{table}

\begin{table}[!ht]
	\begin{center}
		\caption{The diabatic electronic energies and couplings (in eVs) between the diabatic states $\ket{^{1}(\textrm{S}_{0}\textrm{S}_{0})}$, $\ket{^{1}(\textrm{S}_{0}\textrm{S}_{1})}$, $\ket{^{1}(\textrm{S}_{0}\textrm{S}_{1})}$, $\ket{^{1}(\textrm{T}_{1}\textrm{T}_{1})}$, $\ket{^{1}(\textrm{C}\textrm{A})}$, $\ket{^{1}(\textrm{A}\textrm{C})}$, $\ket{^{1}(\textrm{DE})_{1}}$ and $\ket{^{1}(\textrm{DE})_{2}}$ for the system orthogonal-D[1,3] calculated at SA8-XMCQDPT(4,4)/cc-pVDZ level of theory.}
		\begin{tabular}{ccccccccc}
			\toprule
			$\mathcal{H}_{el}$ &$\ket{^{1}(\textrm{S}_{0}\textrm{S}_{0})}$ &$\ket{^{1}(\textrm{S}_{1}\textrm{S}_{0})}$    &$\ket{^{1}(\textrm{S}_{0}\textrm{S}_{1})}$ 
			&$\ket{^{1}(\textrm{T}_{1}\textrm{T}_{1})}$
			&$\ket{^{1}(\textrm{C}\textrm{A})}$              &$\ket{^{1}(\textrm{A}\textrm{C})}$      &$\ket{^{1}(\textrm{DE})_{1}}$             &$\ket{^{1}(\textrm{DE})_{2}}$ \\
			\toprule
		$\bra{^{1}(\textrm{S}_{0}\textrm{S}_{0})}$ & 0.0000 &  0.0089 & -0.0551 & -0.0008 &  0.0129  & 0.0131 &  0.5149 & -0.5563 \\
		$\bra{^{1}(\textrm{S}_{1}\textrm{S}_{0})}$ &0.0089  & 1.9399   &0.1605   &0.0003  &-0.0292   &0.0410  &-0.0062  &-0.0195 \\
		$\bra{^{1}(\textrm{S}_{0}\textrm{S}_{1})}$&-0.0551 &  0.1605  & 2.0362  &-0.0041 &-0.0025  &-0.0064  &-0.0008  & 0.0451\\
		 $\bra{^{1}(\textrm{T}_{1}\textrm{T}_{1})}$&-0.0008 &  0.0003  &-0.0041  & 2.3651  &0.0269  &0.0165   &0.0015  &-0.0021\\
		$\bra{^{1}(\textrm{C}\textrm{A})}$  &0.0129  &-0.0292  &-0.0025   &0.0269  & 2.4349  &-0.0014   &0.0048  &-0.0253\\
		 $\bra{^{1}(\textrm{A}\textrm{C})}$&0.0131  & 0.0410  &-0.0064   &0.0165  &-0.0014   &2.5412   &0.0157  &-0.0206\\
		 $\bra{^{1}(\textrm{DE})_{1}}$ &0.5149  &-0.0062  &-0.0008   &0.0015  & 0.0048   &0.0157   &3.4414  & 0.0745\\
		 $\bra{^{1}(\textrm{DE})_{2}}$ &-0.5563 & -0.0195  & 0.0451  &-0.0021 & -0.0253  &-0.0206  & 0.0745  & 3.5643\\
		
			\bottomrule
		\end{tabular}
		\label{tab:diab13-90}
	\end{center}
\end{table}

\begin{table}[!ht]
	\begin{center}
		\caption{The diabatic electronic energies and couplings (in eVs) between the diabatic states $\ket{^{1}(\textrm{S}_{0}\textrm{S}_{0})}$, $\ket{^{1}(\textrm{S}_{0}\textrm{S}_{1})}$, $\ket{^{1}(\textrm{S}_{0}\textrm{S}_{1})}$, $\ket{^{1}(\textrm{T}_{1}\textrm{T}_{1})}$, $\ket{^{1}(\textrm{C}\textrm{A})}$, $\ket{^{1}(\textrm{A}\textrm{C})}$, $\ket{^{1}(\textrm{DE})_{1}}$ and $\ket{^{1}(\textrm{DE})_{2}}$ for the system orthogonal-D[3,3] calculated at SA8-XMCQDPT(4,4)/cc-pVDZ level of theory.}
		\begin{tabular}{ccccccccc}
			\toprule
			$\mathcal{H}_{el}$ &$\ket{^{1}(\textrm{S}_{0}\textrm{S}_{0})}$ &$\ket{^{1}(\textrm{S}_{1}\textrm{S}_{0})}$    &$\ket{^{1}(\textrm{S}_{0}\textrm{S}_{1})}$ 
			&$\ket{^{1}(\textrm{T}_{1}\textrm{T}_{1})}$
			&$\ket{^{1}(\textrm{C}\textrm{A})}$              &$\ket{^{1}(\textrm{A}\textrm{C})}$      &$\ket{^{1}(\textrm{DE})_{1}}$             &$\ket{^{1}(\textrm{DE})_{2}}$ \\
			\toprule
			$\bra{^{1}(\textrm{S}_{0}\textrm{S}_{0})}$ &  0.0000  & 0.0071  & 0.0071 &  0.0008  &-0.0470 & -0.0275 &  0.5326 & -0.5375 \\
			$\bra{^{1}(\textrm{S}_{1}\textrm{S}_{0})}$ &  0.0071 &  1.9532  & 0.1963 & -0.0031 & -0.0143  & 0.0470  & 0.0010 & -0.0041 \\
			$\bra{^{1}(\textrm{S}_{0}\textrm{S}_{1})}$&  0.0071  & 0.1963  & 1.9765  & 0.0021 &  0.0161 & -0.0499  & 0.0053 & -0.0004\\
			$\bra{^{1}(\textrm{T}_{1}\textrm{T}_{1})}$&  0.0008 & -0.0031 &  0.0021  & 2.3419 & -0.0491 & -0.0585  & 0.0026 & -0.0019 \\
			$\bra{^{1}(\textrm{C}\textrm{A})}$  & -0.0470 & -0.0143 &  0.0161 & -0.0491 &  2.3717 &  0.0006 &  0.0261 & -0.0161 \\
			$\bra{^{1}(\textrm{A}\textrm{C})}$& -0.0275 &  0.0470 & -0.0499 & -0.0585  & 0.0006 &  2.3817 &  0.0049 & -0.0098 \\
			$\bra{^{1}(\textrm{DE})_{1}}$ &  0.5326  & 0.0010 &  0.0053  & 0.0026 &  0.0261  & 0.0049  & 3.4281  & 0.0589 \\
			$\bra{^{1}(\textrm{DE})_{2}}$ & -0.5375 & -0.0041 & -0.0004 & -0.0019 & -0.0161 & -0.0098  & 0.0589  & 3.4314\\
			\bottomrule
		\end{tabular}
		\label{tab:diab33-90}
	\end{center}
\end{table}

\begin{table}[!ht]
	\begin{center}
		\caption{The diabatic electronic energies and couplings (in eVs) between the diabatic states $\ket{^{1}(\textrm{S}_{0}\textrm{S}_{0})}$, $\ket{^{1}(\textrm{S}_{0}\textrm{S}_{1})}$, $\ket{^{1}(\textrm{S}_{0}\textrm{S}_{1})}$, $\ket{^{1}(\textrm{T}_{1}\textrm{T}_{1})}$, $\ket{^{1}(\textrm{C}\textrm{A})}$, $\ket{^{1}(\textrm{A}\textrm{C})}$, $\ket{^{1}(\textrm{DE})_{1}}$ and $\ket{^{1}(\textrm{DE})_{2}}$ for the system orthogonal-D[2,2] calculated at SA8-XMCQDPT(4,4)/cc-pVDZ level of theory.}
		\begin{tabular}{ccccccccc}
			\toprule
			$\mathcal{H}_{el}$ &$\ket{^{1}(\textrm{S}_{0}\textrm{S}_{0})}$ &$\ket{^{1}(\textrm{S}_{1}\textrm{S}_{0})}$    &$\ket{^{1}(\textrm{S}_{0}\textrm{S}_{1})}$ 
			&$\ket{^{1}(\textrm{T}_{1}\textrm{T}_{1})}$
			&$\ket{^{1}(\textrm{C}\textrm{A})}$              &$\ket{^{1}(\textrm{A}\textrm{C})}$      &$\ket{^{1}(\textrm{DE})_{1}}$             &$\ket{^{1}(\textrm{DE})_{2}}$ \\
			\toprule
			$\bra{^{1}(\textrm{S}_{0}\textrm{S}_{0})}$ &  0.0000 & 0.0368 &	-0.0357 &	0.0021 &	0.0255 &	0.0236 &	0.5413 &	-0.5380 \\
			$\bra{^{1}(\textrm{S}_{1}\textrm{S}_{0})}$ & 0.0368	& 2.0263 &	-0.183	&0.0000 &	0.0669 &	-0.0152 &	0.0024 &	-0.0133	\\
			$\bra{^{1}(\textrm{S}_{0}\textrm{S}_{1})}$& -0.0357 &	-0.1830&	2.0396 &	0.0020 &	0.0327 &	-0.0608	 & -0.0154	& 0.0060 \\
			$\bra{^{1}(\textrm{T}_{1}\textrm{T}_{1})}$& 0.0021 & 0.0000	& 0.0020 &	2.4282 & 0.0764 &	0.0749 & 0.0025	& -0.0038\\
			$\bra{^{1}(\textrm{C}\textrm{A})}$  & 0.0255 &	0.0669 & 0.0327	 & 0.0764	& 2.8866 &	0.0043 &	-0.0024	 &0.0338\\
			$\bra{^{1}(\textrm{A}\textrm{C})}$& 0.0236	&-0.0152 &	-0.0608	& 0.0749 &	0.0043	& 2.8866 &	-0.0339 &	0.0033 \\
			$\bra{^{1}(\textrm{DE})_{1}}$ & 0.5413 &	0.0024 &	-0.0154 &	0.0025	&-0.0024 &	-0.0339 &	3.6008 &	0.0780 \\
			$\bra{^{1}(\textrm{DE})_{2}}$ & -0.5380 &	-0.0133	& 0.0060 &	-0.0038	& 0.0338 &	0.0033 &	0.0780 &	3.6008\\
			\bottomrule
		\end{tabular}
		\label{tab:diab22-90}
	\end{center}
\end{table}

\begin{table}[h]
	\centering
	\caption{Excitation energies ($\Delta E$, eV) computed using canonical orbitals and 4-fold DMOs for non-orthogonal investigated dimers.}
	\label{tab:excitation_energies-non-ortho}
	\begin{tabular}{c c c c c c c c c}
		\toprule
		State 
		& \multicolumn{2}{c}{D[1,1]} 
		& \multicolumn{2}{c}{D[1,3]}
		& \multicolumn{2}{c}{D[3,3]} 
		& \multicolumn{2}{c}{D[2,2]} \\
		\cmidrule(lr){2-3} \cmidrule(lr){4-5} \cmidrule(lr){6-7} \cmidrule(lr){8-9}
		& Canonical & \multirow{2}{*}{DMOs}  & Canonical & \multirow{2}{*}{DMOs} 	& Canonical & \multirow{2}{*}{DMOs} 	& Canonical & \multirow{2}{*}{DMOs} \\[0.3em]
		
		& & (4-fold) & & (4-fold)	&  &(4-fold)	&  &  (4-fold) \\
		\midrule
		S$_0$ & 0.000 & 0.000 & 0.000 & 0.000 & 0.000 & 0.000 & 0.000 & 0.000 \\
		S$_1$ & 1.6409 & 1.2781 & 1.6574 & 1.3893 & 1.8774 & 1.6790 & 1.7773 & 1.6642 \\
		S$_2$ & 1.6946 & 1.3268 & 1.8697 & 1.6715 & 1.9825 & 1.8029 & 2.2446 & 2.1511  \\
		S$_3$ & 2.6293 & 2.6466 & 2.4547 & 2.4484 & 2.4113 & 2.3964 & 2.4174 & 2.4026  \\
		S$_4$ & 3.2267 & 3.1493 & 2.6637 & 2.7133 & 2.7821 & 2.8428 & 2.8743 & 2.9434  \\
		S$_5$ & 3.3744 & 3.2381 & 3.1899 & 3.1645 & 2.8485 & 2.9183 & 3.0031 & 3.0783  \\
		S$_6$ & 3.4589 & 3.4684 & 3.3790 & 3.3017 & 3.6202 & 3.4841 & 3.6253 & 3.6079  \\
		S$_7$ & 3.7468 & 3.8189 & 3.7072 & 3.5969 & 3.6321 & 3.5289 & 3.7118 & 3.6635  \\
		\bottomrule
	\end{tabular}
\end{table}

\begin{table}[h]
	\centering
	\caption{Excitation energies ($\Delta E$, eV) computed using canonical orbitals and 4-fold DMOs for orthogonal investigated dimers.}
	\label{tab:excitation_energies-ortho}
	\begin{tabular}{c c c c c c c c c}
		\toprule
		State 
		& \multicolumn{2}{c}{D[1,1]} 
		& \multicolumn{2}{c}{D[1,3]}
		& \multicolumn{2}{c}{D[3,3]} 
		& \multicolumn{2}{c}{D[2,2]} \\
		\cmidrule(lr){2-3} \cmidrule(lr){4-5} \cmidrule(lr){6-7} \cmidrule(lr){8-9}
		& Canonical & \multirow{2}{*}{DMOs}  & Canonical & \multirow{2}{*}{DMOs} 	& Canonical & \multirow{2}{*}{DMOs} 	& Canonical & \multirow{2}{*}{DMOs} \\[0.3em]
		& & (4-fold) & & (4-fold)	&  &(4-fold)	&  &  (4-fold) \\
		\midrule
		S$_0$ & 0.000 & 0.000 & 0.000 & 0.000 & 0.000 & 0.000 & 0.000 & 0.000 \\
		S$_1$ & 1.9248 & 1.9220 & 1.9797 & 1.9796 & 2.0581 & 2.0560 & 2.0060 & 2.0027 \\
		S$_2$ & 2.3235 & 2.3233 & 2.3168 & 2.3156 & 2.3279 & 2.3266 & 2.3758 & 2.3742 \\
		S$_3$ & 2.4764 & 2.4447 & 2.5159 & 2.5153 & 2.5657 & 2.5650 & 2.5765 & 2.5661 \\
		S$_4$ & 2.5397 & 2.5440 & 2.6055 & 2.6077 & 2.7932 & 2.7945 & 3.0439 & 3.0474 \\
		S$_5$ & 2.5701 & 2.6043 & 2.7060 & 2.7059 & 2.7987 & 2.7999 & 3.0621 & 3.0753 \\
		S$_6$ & 3.6514 & 3.6506 & 3.6782 & 3.6782 & 3.8344 & 3.8333 & 3.8417 & 3.8404 \\
		S$_7$ & 3.6953 & 3.6957 & 3.8144 & 3.8146 & 3.8468 & 3.8479 & 3.8418 & 3.8429 \\
			\bottomrule
	\end{tabular}
\end{table}

\begin{figure}
	\caption{The ball and stick representation of optimized geometries of aza-BODIPY monomer and dimers.}
	\centering
	\includegraphics[width=0.8\textwidth]{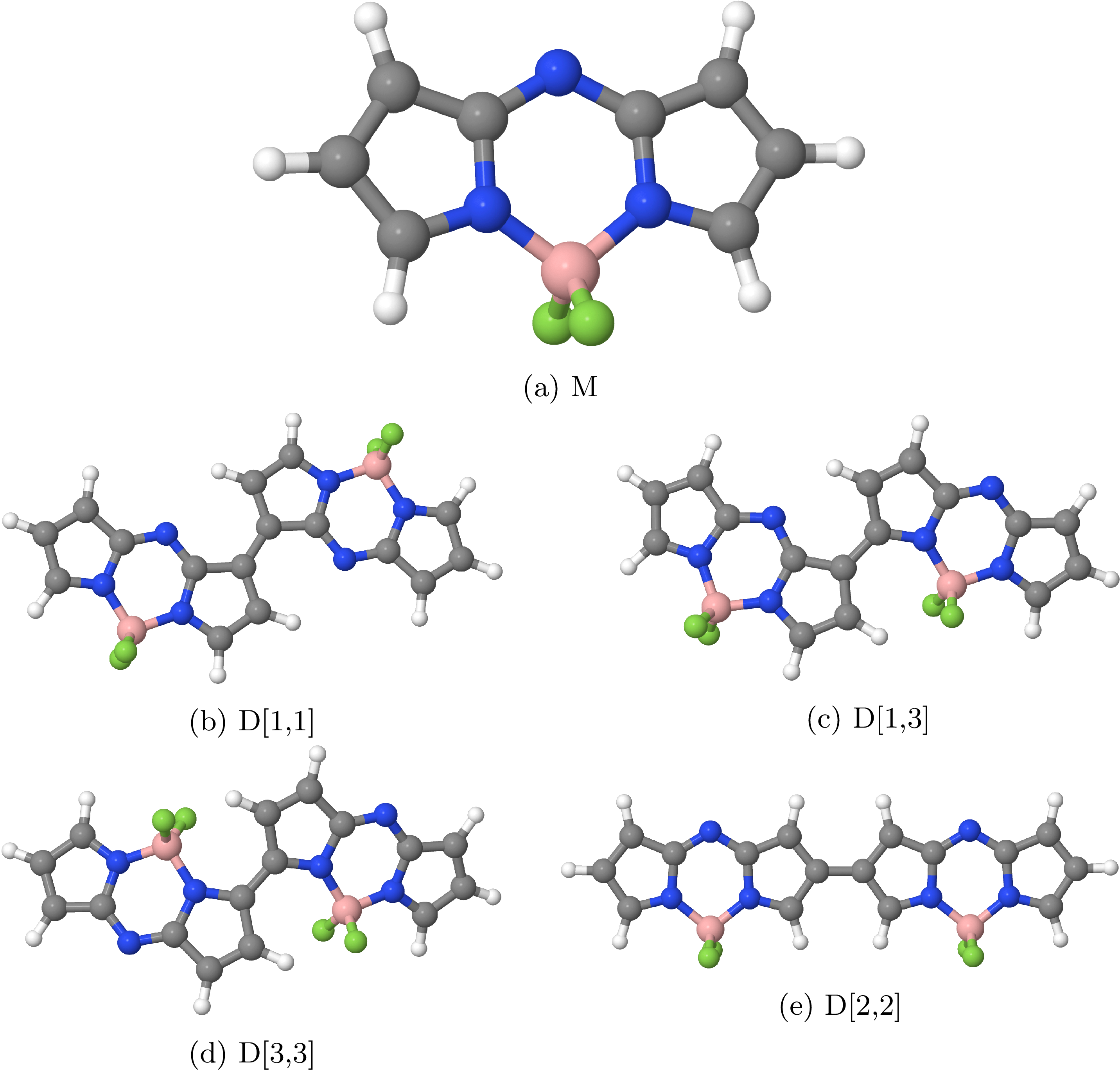}
	\label{fig:ball-stick}
\end{figure} 

\clearpage
\textbf{M}
\begin{verbatim}
20
Angstrom units
C         -6.33221       -0.60392        0.12392
C         -6.07991        1.58809        0.06570
C         -7.48026        1.33092        0.14809
C         -7.63247       -0.04331        0.18436
C         -4.15786        2.81081       -0.06307
C         -2.06745        2.10911       -0.16166
C         -2.02269        3.52535       -0.19135
C         -3.33072        3.97065       -0.12989
H         -6.03401       -1.64441        0.12840
H         -8.23926        2.09947        0.17459
H         -8.55278       -0.60460        0.24665
H         -1.25261        1.39729       -0.19170
H         -1.12412        4.12111       -0.25085
H         -3.70386        4.98466       -0.12927
N         -5.40559        0.36940        0.05316
N         -3.34146        1.68249       -0.08507
B         -3.84939        0.20043       -0.03105
F         -3.37125       -0.43204        1.10059
F         -3.49959       -0.47113       -1.18692
N         -5.48441        2.77490        0.00957

\end{verbatim}
\vskip15pt

\textbf{D[1,1]}
\begin{verbatim}
38
Angstrom units
N         -0.02317        2.99962        0.00001
C          0.72054        1.82868       -0.00006
C          2.72384        2.93847       -0.00022
N          2.12492        4.19414       -0.00011
C         -1.33580        2.65661       -0.00006
C         -0.19530        0.69285       -0.00003
C          4.15219        3.10829       -0.00043
C          3.11007        5.11921       -0.00047
C          4.38711        4.47947       -0.00059
C         -1.48830        1.24993       -0.00006
B          0.57132        4.45967        0.00039
F          0.19530        5.13267       -1.15449
F          0.19588        5.13181        1.15595
N          2.05302        1.77831       -0.00015
H          4.87008        2.28719       -0.00049
N          0.02317       -2.99962        0.00001
C         -0.72054       -1.82868       -0.00006
C         -2.72384       -2.93847       -0.00022
N         -2.12492       -4.19414       -0.00011
C          1.33580       -2.65661       -0.00006
C          0.19530       -0.69285       -0.00003
C         -4.15219       -3.10829       -0.00043
C         -3.11007       -5.11921       -0.00047
C         -4.38711       -4.47947       -0.00059
C          1.48830       -1.24993       -0.00006
B         -0.57132       -4.45967        0.00039
F         -0.19530       -5.13267       -1.15449
F         -0.19588       -5.13181        1.15595
N         -2.05302       -1.77831       -0.00015
H         -4.87008       -2.28719       -0.00049
H          5.34934        4.99244       -0.00079
H          2.87038        6.18450       -0.00051
H         -2.42496        0.69545       -0.00009
H          2.42496       -0.69545       -0.00009
H         -2.87038       -6.18450       -0.00051
H         -5.34934       -4.99244       -0.00079
H         -2.10573        3.43123       -0.00007
H          2.10573       -3.43123       -0.00007
\end{verbatim}
\vskip15pt

\textbf{D[1,3]}\\
$^*$In our previous paper, we had already provided the RICC2 optimized coordinates. Here, we present the MP2/cc-pVDZ optimized coordinates, as all subsequent calculations have been performed using these coordinates.
\begin{verbatim}
38
Angstrom units
C          0.70466        0.70990        0.36779
C          2.71765        1.67000        0.37146
C          1.75664        2.71573        0.61853
C          0.51188        2.12393        0.60724
C          4.78732        0.79532       -0.05894
C          5.31865       -1.29004       -0.58979
C          6.53669       -0.54583       -0.59490
C          6.21171        0.76589       -0.26185
H          2.02426        3.76151        0.77358
H          6.86340        1.63495       -0.16610
N          2.04958        0.45968        0.22991
N          4.28232       -0.48031       -0.27035
B          2.78497       -0.90336       -0.08572
F          2.69166       -1.77699        1.00385
F          2.29310       -1.47603       -1.25151
C         -3.62425        1.27924       -0.23355
C         -1.75155        0.01870        0.16057
C         -1.56993       -2.18510        0.32903
C         -0.32833       -0.28490        0.32349
C         -0.25055       -1.69068        0.41932
C         -4.39917        2.45758       -0.51579
C         -5.71326        2.03392       -0.68799
C         -5.72014        0.61732       -0.50617
H         -3.98617        3.46462       -0.58466
B         -4.03133       -1.29523        0.11198
N         -2.45975       -1.17181        0.17281
N         -4.47138        0.17632       -0.23641
F         -4.52553       -1.65929        1.35796
F         -4.40992       -2.17233       -0.89309
N          4.03658        1.85660        0.25415
N         -2.29962        1.21823       -0.03474
H          5.15269       -2.34886       -0.79794
H          7.52195       -0.95229       -0.82516
H         -0.46558        2.58268        0.74442
H         -6.55616       -0.08313       -0.55773
H         -6.58619        2.64369       -0.92284
H         -1.91758       -3.21930        0.37470
H          0.64522       -2.28599        0.57724
\end{verbatim}
\vskip15pt

\textbf{D[3,3]}
\begin{verbatim}
38
Angstrom units
N          1.83539        0.33977        0.31265
C          2.45246        1.57734        0.43166
C          4.52547        0.81903       -0.19514
N          4.05666       -0.46988       -0.41315
C          0.47801        0.52945        0.51224
C          1.45793        2.57743        0.74421
C          5.92378        0.86482       -0.53954
C          5.08954       -1.22327       -0.86570
C          6.26767       -0.41929       -0.96160
C          0.23824        1.93774        0.77142
B          2.63843       -0.97290       -0.00941
F          2.75964       -1.77186        1.15206
F          2.05165       -1.68729       -1.06279
N          3.75148        1.82966        0.22227
H          6.54484        1.75776       -0.47961
H          1.68249        3.63107        0.90493
N         -1.83539       -0.33985        0.31262
C         -2.45251       -1.57743        0.43156
C         -4.52554       -0.81900       -0.19522
N         -4.05665        0.46991       -0.41312
C         -0.47798       -0.52954        0.51222
C         -1.45795       -2.57752        0.74421
C         -5.92386       -0.86469       -0.53960
C         -5.08951        1.22339       -0.86561
C         -6.26769        0.41947       -0.96155
C         -0.23821       -1.93780        0.77153
B         -2.63838        0.97286       -0.00937
F         -2.75958        1.77178        1.15212
F         -2.05158        1.68730       -1.06272
N         -3.75156       -1.82966        0.22214
H         -6.54497       -1.75759       -0.47972
H         -1.68247       -3.63117        0.90498
H          7.23935       -0.76690       -1.31038
H          4.95002       -2.27996       -1.09663
H         -0.73580        2.36780        0.98707
H          0.73576       -2.36790        0.98736
H         -4.94994        2.28009       -1.09644
H         -7.23936        0.76718       -1.31027
\end{verbatim}
\vskip15pt

\textbf{D[2,2]}
\begin{verbatim}
38
Angstrom units
N          2.85772        0.29892       -0.00000
C          2.93194       -1.09692        0.00000
C          5.22163       -1.10625        0.00000
N          5.31748        0.28376        0.00000
C          1.55758        0.64109       -0.00000
C          1.60106       -1.62638       -0.00000
C          6.55077       -1.65769        0.00000
C          6.62915        0.60517        0.00000
C          7.42826       -0.57900        0.00000
C          0.72686       -0.53379       -0.00000
B          4.09768        1.27973       -0.00000
F          4.09628        2.04762       -1.15564
F          4.09628        2.04762        1.15563
N          4.07352       -1.79078        0.00000
H          6.77053       -2.72583        0.00000
N         -2.85772        0.29892       -0.00000
C         -2.93194       -1.09692        0.00000
C         -5.22163       -1.10625        0.00000
N         -5.31748        0.28376        0.00000
C         -1.55758        0.64109        0.00000
C         -1.60106       -1.62638       -0.00000
C         -6.55077       -1.65769        0.00000
C         -6.62915        0.60517        0.00000
C         -7.42826       -0.57900        0.00000
C         -0.72686       -0.53379       -0.00000
B         -4.09768        1.27973       -0.00000
F         -4.09628        2.04762        1.15563
F         -4.09628        2.04762       -1.15564
N         -4.07352       -1.79078        0.00000
H         -6.77053       -2.72583        0.00001
H          8.51830       -0.60359        0.00000
H          6.94393        1.65076        0.00000
H         -6.94393        1.65076        0.00000
H         -8.51830       -0.60359        0.00000
H          1.26275        1.69301       -0.00000
H         -1.26275        1.69301        0.00000
H         -1.36332       -2.69085       -0.00001
H          1.36332       -2.69085       -0.00001
\end{verbatim}
\vskip15pt